\documentclass[longbibliography,10pt,aps,pra,twocolumn,superscriptaddress]{revtex4-2}
\usepackage{amsmath}
\usepackage{graphicx}
\usepackage{amssymb}
\usepackage{graphicx,color}
\usepackage[caption=false]{subfig}
\usepackage{soul}
\usepackage{epsfig}
\usepackage{dcolumn}
\usepackage{bm}
\usepackage{lineno}
\usepackage{mathrsfs}
\usepackage{cases}
\usepackage{upgreek}
\usepackage{comment}
\usepackage{graphicx}
\usepackage{dcolumn}
\usepackage{bm}
\usepackage{braket}
\usepackage{amsmath}
\usepackage{graphicx}
\usepackage{amssymb}
\usepackage{color}
\usepackage{lineno}
\usepackage{etoolbox}
\usepackage{soul}
\usepackage[rgb,dvipsnames]{xcolor}
\usepackage{adjustbox}
\usepackage{float} 

\newcommand*\linenomathpatch[1]{%
  \cspreto{#1}{\linenomath}%
  \cspreto{#1*}{\linenomath}%
  \csappto{end#1}{\endlinenomath}%
  \csappto{end#1*}{\endlinenomath}%
}

\linenomathpatch{equation}
\linenomathpatch{gather}
\linenomathpatch{multline}
\linenomathpatch{align}
\linenomathpatch{alignat}
\linenomathpatch{flalign}

\newcommand{\ph}{\protect\phantom{(-1)}}

\newcommand{\REFone}[1]{\protect\textcolor{Black}{#1}}
\newcommand{\REFtwo}[1]{\protect\textcolor{Black}{#1}}
\newcommand{\REVone}[1]{\protect\textcolor{Black}{#1}}
\newcommand{\REVtwo}[1]{\protect\textcolor{Black}{#1}}
\newcommand{\COR}[1]{\protect\textcolor{Black}{#1}}
\newcommand{\CORnew}[1]{\protect\textcolor{Black}{#1}}
\newcommand{\CORnewnew}[1]{\protect\textcolor{Black}{#1}}

\begin{document}

\preprint{Laoutaris et al 2025}

\title{Projectile excitation to the $\mathbf{2s2p}$ $\mathbf{^{3}\!P}$ and $\mathbf{^{1}\!P}$ autoionizing states in swift collisions \\
of  He-like carbon and oxygen mixed-state $\mathbf{(1s^2, 1s2s\,^{3,1}S)}$  ion beams with helium}

\thanks{Paper has been published in Phys. Rev. A \textbf{111}, (2025) 062811 with doi: 10.1103/l6x2-cfs9},
\author{A. Laoutaris}
\affiliation{Department of Physics, University of Crete, GR-70013 Heraklion, Greece}%
\affiliation{Tandem Accelerator Laboratory, Institute of Nuclear and Particle Physics, NCSR “Demokritos”, GR-15310 Aghia~Paraskevi, Greece}%

\author{S. Passalidis}%
\affiliation{Sorbonne Universit\'{e}, CNRS, Laboratoire de Chimie Physique- Mati\`{e}re et Rayonnement, F-75005 Paris, France}%

\author{S. Nanos}
\altaffiliation{Present address:
Department of Physics, University of Crete, GR-70013 Heraklion, Greece}
\affiliation{Tandem Accelerator Laboratory, Institute of Nuclear and Particle Physics, NCSR “Demokritos”, GR-15310 Aghia~Paraskevi, Greece}%
\affiliation{Department of Physics, University of Ioannina, GR-45110 Ioannina, Greece}%

\author{A. Biniskos}
\altaffiliation{Present address: Institut {f\"{u}r} Kernphysik, J. W. {Goethe-Universit\"{a}t} Frankfurt am Main, 60438 Frankfurt am Main, Germany}
\affiliation{Department of Physics, University of Ioannina, GR-45110 Ioannina, Greece}%

\author{E.~P. Benis}%
\affiliation{Department of Physics, University of Ioannina, GR-45110 Ioannina, Greece}%

\author{A. Dubois}%
\affiliation{Sorbonne Universit\'{e}, CNRS, Laboratoire de Chimie Physique- Mati\`{e}re et Rayonnement, F-75005 Paris, France}%

\author{T.~J.~M. Zouros}%
\email[Contact author: ]{tzouros@physics.uoc.gr}
\affiliation{Department of Physics, University of Crete, GR-70013 Heraklion, Greece}%

\date{June 11, 2025}

\begin{abstract}
The production of the projectile $2s2p\,^{3}\!P$ and $2s2p\,^{1}\!P$ autoionizing states is investigated in 0.5-1.5 MeV/u collisions of He-like carbon and oxygen mixed-state three-component $(1s^2,1s2s\,^{3}\!S,1s2s\,^{1}\!S)$ ion beams with helium targets. The mixed-state beams are produced in the
stripping systems of the 5.5~MV Demokritos tandem accelerator. Using high-resolution  Auger  projectile electron spectroscopy, the normalized  Auger  electron yields are measured at $0^\circ$ relative to the beam direction.
In addition, a three-electron atomic orbital close-coupling approach, employing full configuration interaction and antisymmetrization of
the three-electron, two-center total wave function, is applied to calculate the production cross sections for these states from each of the three initial ion beam components.
Thereupon, the theoretical  Auger  yields are computed and found to be mostly smaller than experiment by factors ranging from near 1 to about 10.  Agreement, however, improves when larger $1s2s\,^{1}\!S$ fractions, not only based on spin statistics, are projected.
\REFone{Overall, we present a careful and sophisticated analysis with a thorough discussion of these results which show that our current understanding is still incomplete. Never-the-less, our non-perturbative excitation treatment — free from scaling parameters and renormalization — marks a significant advancement in modeling ultrafast multielectron dynamics within open-shell quantum systems.}
\end{abstract}

\maketitle


\section{Introduction}
\label{sec:introduction}
\REFone{Recently, we investigated} $1s\rightarrow2p$ projectile excitation, both experimentally and theoretically, in the production of $2s2p\,^3\!P$ states from initial $1s2s\,^3\!S$ metastable states in energetic (MeV/u) collisions of He-like carbon and oxygen ions with helium~\cite{lao24a}. %
Here, in a more comprehensive treatment, we extend these investigations to also include the production of the $2s2p\,^1\!P$ states from the $1s^2$ ground state, as well as from both $1s2s\,^3\!S$  and $1s2s\,^1\!S$ metastable states, all three initial states naturally found in He-like ion beams.
Excitation from such pre-excited initial states presents a real challenge to the modeling of such multielectron multi-open-shell dynamical quantum systems.

\REFone{High-energy two-electron projectile ions colliding with two-electron targets represent unique \textit{four}-electron collision systems for investigating few-electron quantum dynamics. While four-electron calculations are presently prohibitively costly in CPU and unavailable, we have recently successfully used a \textit{three}-electron \CORnew{close}-coupling approach to describe electron capture~\cite{mad20a,mad22a} and transfer-excitation~\cite{lao22a} in such systems, where the He target is described by a model potential, with one of the target electrons considered frozen~\cite{ben24a} (see also discussion in section III.C of Ref.~\cite{lao24a}).}
Such three-electron systems, while simple enough to allow for the identification of individual excitation processes and the calculation of their cross sections, are also complex enough to present a real challenge to \textit{ab initio} non-perturbative theoretical approaches~\cite{sis19a}.
Important applications include solar flares~\cite{gab72a}, calibration of existing and developing new X-ray line diagnostics~\cite{bei03b}, high temperature fusion and astrophysical plasmas~\cite{del06a}, as well as fusion plasma heating and diagnostics~\cite{fri91d}.

Swift (MeV/u), He-like ion beams provided by accelerators can deliver such two-electron projectiles. Tandem Van de Graaff accelerators, in particular, use their intrinsic beam up-charging stripper systems~\cite{zou25a} to generate such \REVtwo{multiply}-charged, low-$Z_p$ atomic number beams allowing for  collision energy $E_p$- and isoelectronic $Z_p$ -dependent studies that reveal intriguing and important systematic features of the collision dynamics~\cite{ben18b}.

The stripping process, in the case of such energetic He-like ion beams, gives rise not only to the $1s^2$ ground state, but also to the long-lived $1s2s\,^3\!S$ (for short ${^3\!S}$) and $1s2s\,^1\!S$ (for short ${^1\!S}$) states. The lifetimes of such first-row ion metastable states are in the range of $10^{-3}-10^{-7}$~s~\cite{ben18b} and therefore long enough to survive to the target.

This admixture of  metastable states is particularly rich in atomic physics information as it provides \textit{unique} access to both singly- and doubly-excited states otherwise inaccessible from just a \textit{pure} ground state beam~\cite{ben18b}. However, it presents the additional difficulty of having to accurately determine its fractional composition since production cross sections from these pre-excited states can be much larger than from the ground state. This knowledge is essential for precise quantitative comparisons between theory and experiment.

For the $1s2s\,^3\!S$ state, various measurement techniques have been used to date, indicating that for first-row atoms a significant fraction in the $^3\!S$ state, $f[^3\!S]\sim 10-30\%$~\cite{ped79a,dil84a,and92a,din96a,zam01a,ben02a,dmi03a} survives to the target. However, for the ${^1\!S}$ state - having a much shorter lifetime - the corresponding fraction
is smaller and has never been directly measured. Instead, its estimation, at $f[^1\!S]\sim0.1-3\%$~\cite{and92a} has been based on various stripper production models~\cite{zam01a,zam01b,and92a,ben02a,dmi03a,zou25a}, {\color{black}mostly assuming it is produced in a spin statistics ratio of $f[^1\!S]/f[^3\!S]=1/3$.}

Recently, there has been renewed interest in using mixed-state He-like ion beams in collisions with helium, driven by advancements in state-selective, \textit{ab initio}, non-perturbative close-coupling calculations~\cite{sis19a}. In particular, semi-classical three-electron atomic orbital close-coupling (3eAOCC) calculations, involving mixed-state He-like carbon and oxygen ions, have provided state-selective cross sections for processes such as single electron capture (SEC)~\cite{mad20a,mad22a}, transfer excitation (TE)~\cite{lao22a}, and projectile excitation~\cite{lao24a}, enhancing our understanding of multi-electronic interactions in multi-open-shell quantum systems under intense, ultrafast perturbations.

For low-$Z_p$ He-like ions, the production of doubly-excited autoionizing states has been effectively studied using  Auger  projectile spectroscopy. Zero-degree Auger projectile spectroscopy (ZAPS)~\cite{sto87a,zou97a}, which detects emitted  Auger  electrons at $\theta=0^\circ$ relative to the beam direction, has been particularly successful in providing \textit{state}-selective production cross sections. These  measurements offer well-defined initial and final states, thereby providing stringent tests of theory.

An important advantage of using ZAPS is that the $1s2s\,^3\!S$ metastable fraction can be measured directly from the Auger spectra \REVtwo{themselves, without relying on measurements from different experimental setups or ion beam conditions, where the $^3\!S$ content might vary.
In our ZAPS setup, this is achieved by recording two nearly identical Auger KLL spectra, but with ion beams containing significantly different $^3\!S$ fractions. The $^3\!S$ content is determined by the intensity of the $1s2s2p\,^4\!P$ Auger line, which is uniquely produced by single electron capture to the $^3\!S$ component. The $^3\!S$ fraction can be partially controlled by adjusting the stripping conditions: gas stripping generally results in substantially lower $^3\!S$ content~\cite{ben02a}.
To accurately determine the $^3\!S$ fraction, a second Auger line originating predominantly from the ground-state component is required. This is provided by the $1s2p^2\,^2\!D$ Auger line, which arises mainly from the $1s^2$ ground state via transfer-excitation processes~\cite{zam01a,zam01b}.
Using the single differential cross sections (SDCS) of these two Auger lines - represented by the normalized areas under the peaks - we can establish a system of four equations with four unknowns~\cite{ben02a,ben16b}. Solving this system yields the $^3\!S$ fraction in each of the two measurements, with the $1s^2$ fraction determined by the conservation relation $f[1s^2] = 1 - f[^3\!S]$.}

\REVtwo{Over the years, we have refined this approach, known as the ``two spectra measurement technique”, to provide accurate determinations of the $^3\!S$ and ground-state components within a two-component analysis framework~\cite{zam01a,zam01b,ben02a,ben16b}. In this two-component model, the small amount of $^1\!S$ content is neglected. However, as demonstrated in this paper, the production of the $2s2p\,^1\!P$ state strongly depends on the $^1\!S$ component. Consequently, we have recently extended our analysis to a three-component model that explicitly includes the $^1\!S$ fraction, enabling a self-consistent determination of all three \CORnew{components - $^3\!S$, $^1\!S$, and the ground state - in} the Auger spectra~\cite{zou25a}. This three-component model is used in the analysis of the results presented here.}

In particular, investigations of SEC in carbon resolved a long-standing spin-statistics problem, \REVtwo{i.e. showing that the production of \textit{identically} configured LS states differing only in their \textit{total} spins (e.g. the $1s2s2p\,^4\!P$ and $1s2s2p\,^2\!P$ states produced by single electron capture to the $^3\!S$) cannot be assumed to be populated according to their spin multiplicities, i.e. here, in the ratio of 4:2}~\cite{mad20a,mad22a}. Investigations of TE provided the first coherent treatment of dynamic electron-electron correlations, successfully describing resonance transfer-excitation (RTE) and revealed a new low-energy nonresonant one-step transfer-excitation mechanism~\cite{lao22a}.
{\color{black}Investigations of cusp-electron production using mixed-state He-like oxygen ion beams showed the $^3\!S$ component to play an important role which could be quantitatively well-described by \REFone{continuum}-distorted-wave \REFone{(CDW)} theories of electron-loss and electron-capture to the continuum~\cite{nan23b}.}
Very recently, investigations of $1s\rightarrow2p$ excitation in the production of $2s2p\,^3\!P$ states from just the $1s2s\,^3\!S$ state in collisions of carbon and oxygen ions with helium indicate that the conventional first Born picture of screening and antiscreening mechanisms might need revision~\cite{lao24a}. \REVtwo{In this approximation, the excitation (or loss) of the projectile electron arises from two fundamentally different interaction mechanisms: If the target remains in its ground state during the collision, the electrons effectively screen the target nucleus, thereby decreasing the projectile excitation (or loss) cross sections. If the target is excited or ionized as a result of the two-center electron-electron ($e$–$e$ or TCee) interactions, the target electrons contribute actively to projectile excitation (or loss) with a characteristic electron velocity dependence and threshold, below which this TCee interaction is negligible~\cite{mon94a,zou96b}.
These two mechanisms are commonly referred to as screening and antiscreening~\cite{mcg81a}, respectively, and are intrinsic to the Born collision picture. However, in close-coupling calculations, such electron-nucleus ($e$–$n$) and electron-electron ($e$–$e$) interactions are treated within a unified framework, making it difficult, in principle, to separate the contributions from each mechanism.
}

Here, we further pursue single and double excitation including
the production of both $2s2p\,^3\!P$ and $2s2p\,^1\!P$ states from all three initial ion beam components:
\begin{align}
\text{Z}^{q+}(1s2s\,^{3}\!S) + \text{He} \rightarrow &\,\text{Z}^{q+}(2s2p\,^{3,1}\!P) + \text{He(All),}\label{eq:2s2p31P_production_3S}\\
\text{Z}^{q+}(1s2s\,^{1}\!S) + \text{He} \rightarrow &\,\text{Z}^{q+}(2s2p\,^{3,1}\!P) + \text{He(All),}\label{eq:2s2p31P_production_1S}\\
\text{Z}^{q+}(1s^2\,^{1}\!S) + \text{He} \rightarrow &\,\text{Z}^{q+}(2s2p\,^{3,1}\!P) + \text{He(All),}\label{eq:2s2p31P_production_gs}\\
\phantom{\text{Z}^{q+}(1s^2\,^{1}\!S) + \text{He} \rightarrow } &\quad^|\!\!\!\rightarrow \text{Z}^{(q+1)+}(1s) + e_A^-(0^\circ), \label{eq:2s2p31P_0dgr_Auger_decay}
\end{align}
with the emitted  Auger  electrons, $e_A^-$, from the decay of the two $2s2p$ states {\color{black}[Eq.~(\ref{eq:2s2p31P_0dgr_Auger_decay})]} detected at the laboratory observation angle of $\theta=0^\circ$ relative to the ion beam using ZAPS. Normalized  Auger  yields are measured in the collision energy range of {0.5-1.5~MeV/u}, where $\text{Z}^{q+}$ denotes C$^{4+}$ or O$^{6+}$ ion projectiles. He(All), indicates that all resulting final helium target states, from processes including simultaneous target single excitation and ionization are considered in the calculations, since the final states of the target were not experimentally determined. Accompanying 3eAOCC calculations within a full configuration interaction approach provide the production cross sections. Thus, cross sections for both \textit{single} direct and exchange $1s\rightarrow2p$ excitation [Eqs.~(\ref{eq:2s2p31P_production_3S})-(\ref{eq:2s2p31P_production_1S})] and similarly for \textit{double} $(1s\rightarrow2s, 1s\rightarrow2p)$ excitation [Eq.~(\ref{eq:2s2p31P_production_gs}) direct for $^1\!P$ and exchange for $^3\!P$] are reported.

Historically, over the past 50 years, there has been much interest   in the excitation of atoms or ions in atomic collisions, as well as related work on electron impact excitation and photo-excitation.  \REFone{General} references were already given in Ref.~\cite{lao24a}.

Early high-resolution x-ray studies using He-like ions~\cite{mac73a,hop74a,hop76a,matt76a,schi77a,ter83a,rey88a,woh86a,ali91a,cha94a,ado95a} largely focused on the production of singly-excited $1s2p\,^3\!P$ and $1s2p\,^1\!P$ states. However, these measurements were often complicated by cascade effects~\cite{hop74a,matt76b,zuc85a}, which made interpretation challenging. Subsequent research examined doubly-excited states through high-resolution  Auger  spectroscopy~\cite{new84b,dil84a}, which are less affected by cascades due to low radiative branching ratios in first-row atoms. Notably, the first report on $2s2p\,^3\!P$ production from the $^3\!S$ state in energetic mixed-state F$^{7+}$ ions colliding with He and \CORnew{H$_2$ - compared to first Born cross-section calculations - was} presented in Ref.~\cite{lee91a}. Aside from our recent work on $2s2p\,^3\!P$ production from $1s2s\,^3\!S$ in He-like carbon and oxygen  ions colliding with He~\cite{lao24a}, with comparisons to 3eAOCC and first Born results, little else has appeared since.

In the following, experimental and theoretical considerations in the production of the $2s2p\,^3\!P$ and $2s2p\,^1\!P$ states are discussed in sections \ref{sec:experimental} and \ref{sec:theory}, respectively. Section~\ref{sec:results} provides a {\color{black}detailed
critical analysis} of theoretical and experimental cross-section results. Summary and conclusions are presented in section~\ref{sec:Conclusions}. The appendix includes  {\color{black} tables of our 3eAOCC production cross sections,} information on corrections due to SEC contaminants, fine-structure details related to the angular dependence of  Auger  emission at $\theta=0^\circ$, {\color{black}tables of the determined metastable fractions, the thereupon computed theoretical normalized Auger  yields compared to the measured  Auger  yields} and tables of known measured and calculated  Auger  energies used for energy calibration and state identification.
\begin{figure}[tbh]
\includegraphics[scale=0.31]{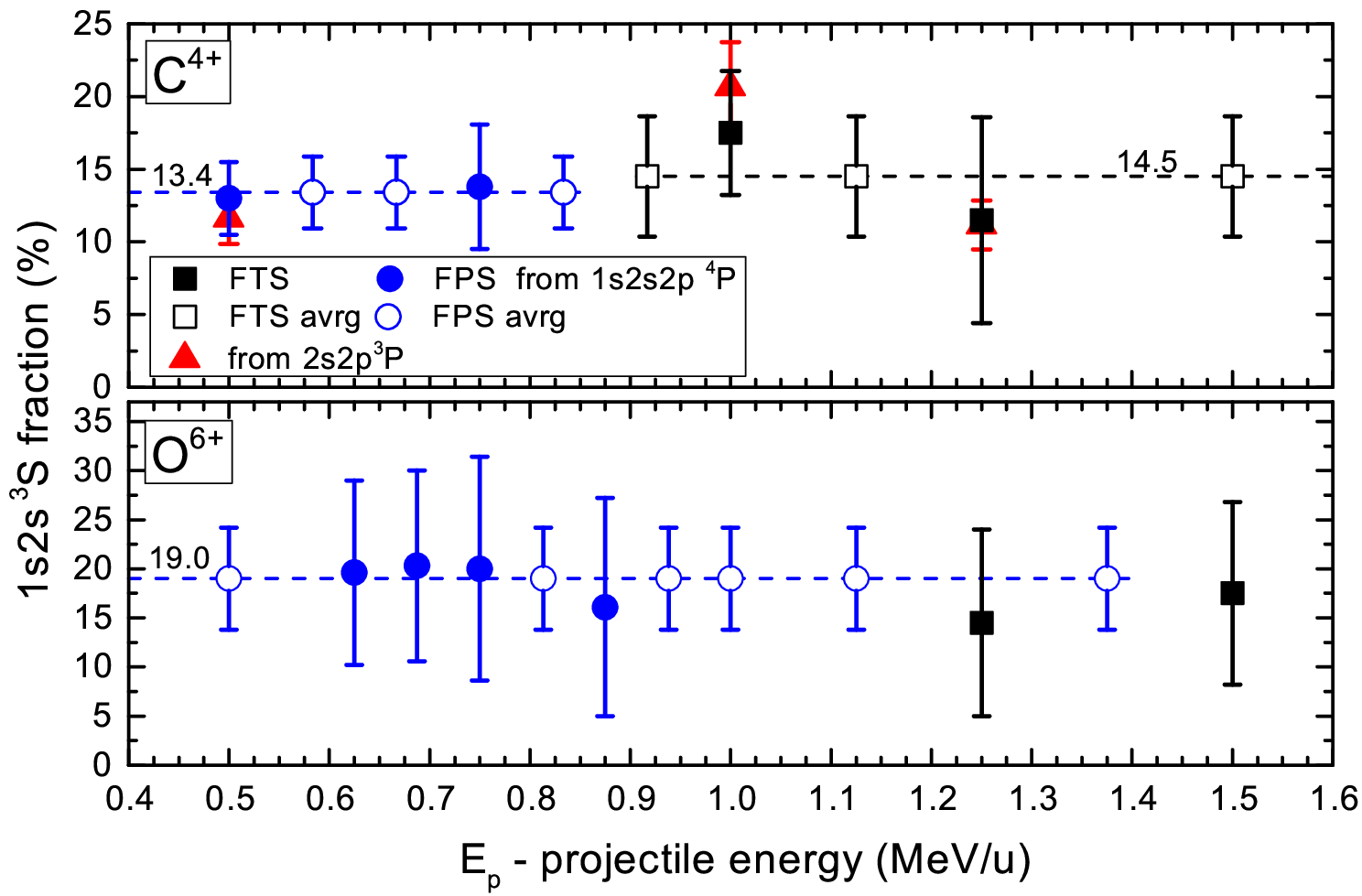}
\caption{Experimentally determined $1s2s\,^3\!S$ ion beam fractions as a function of projectile energy $E_p$ in MeV/u using the three-component model. (Top) C$^{4+}$, (Bottom) O$^{6+}$ projectile ions. Where available, fractions were determined from either the $1s2s2p\,^4\!P$ (circles and squares) or the $2s2p\,^3\!P$ (red triangles) states, both of which are dominantly produced from the $1s2s\,^3\!S$ component. Good consistency is observed between the two determinations. Squares are from FTS, \REFone{while circles from FPS 
measurements} as given in Tables~\ref{tb:C4He2s2pNY} and \ref{tb:O6He2s2pNY}. Open symbols refer to values estimated by the mean value of similarly stripped ions (dashed lines), whose values appear in parentheses in the tables.}
\label{fg:f3SC4O6}
\end{figure}

\begin{figure}[tbh]
\includegraphics[scale=0.31]{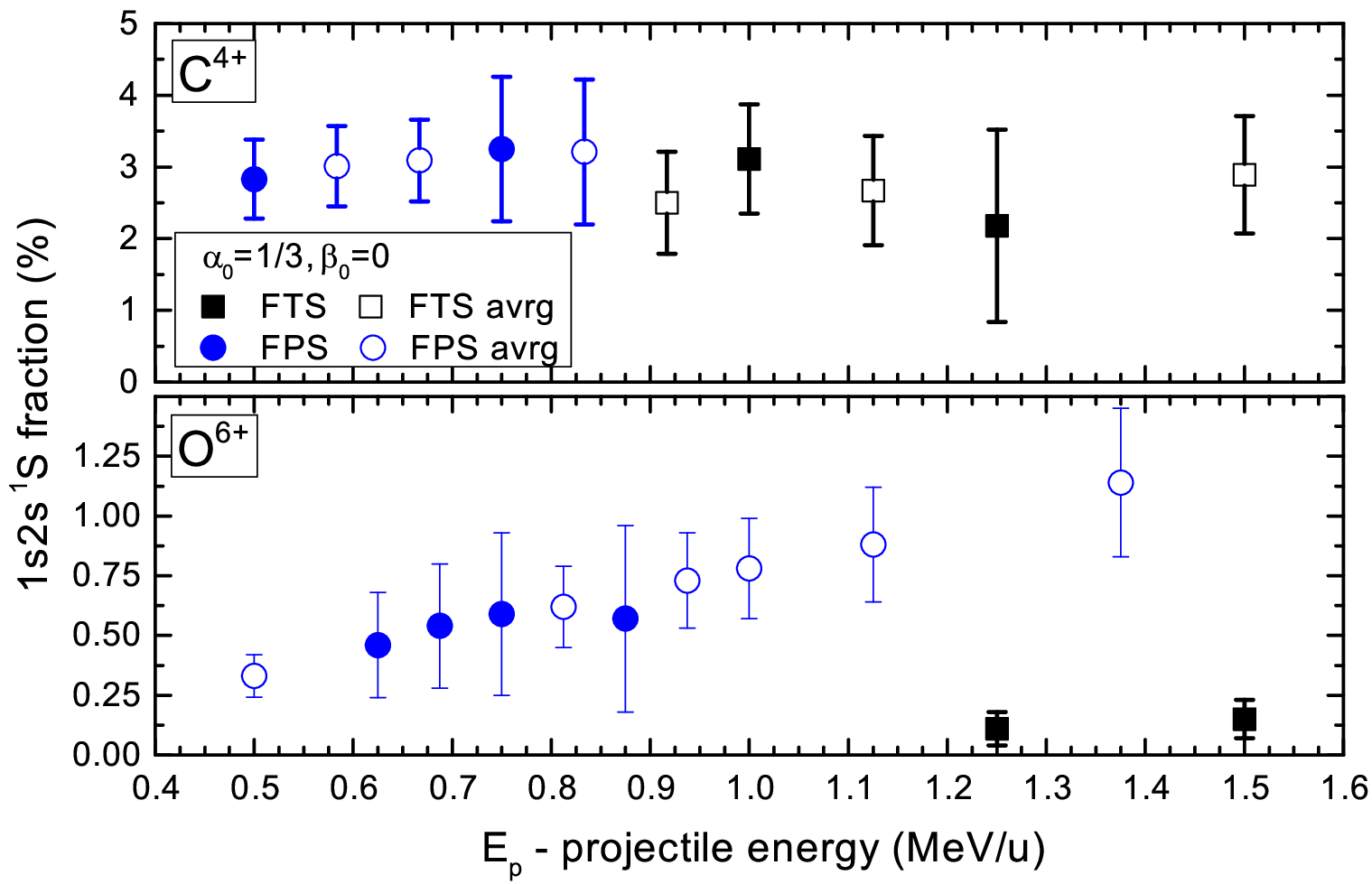}
\caption{Same as Fig.~\ref{fg:f3SC4O6}, but for $1s2s\,^1\!S$ (with $\alpha_0=1/3$ and $\beta_0=0$ - see text). For oxygen, the much smaller values of $f[^1\!S]$ and their larger spread reflect the $\sim$10 times shorter lifetime of the oxygen $^1\!S$ state compared to that of carbon. Since $f[^3\!S]$ is relatively constant over the same $E_p$ range the rise of $f[^1\!S]$ with $E_p$ reflects the decreasing time-of-flight $\Delta t_0$ in the negative exponential of Eq.~\ref{eq:bi}. The smaller values for FTS stripping in oxygen are due to the longer $\Delta t_0$ required from the terminal stripper.}
\label{fg:f1SC4O6}
\end{figure}

\section{Experiment}
\label{sec:experimental}
The measurements reported here were taken during the same beam time reported in Ref.~\cite{lao24a}. Experimental details of the setup and methodology was presented there so here only a very brief description is given. The experiment was conducted at the National Center
for Scientific Research (NCSR) ``Demokritos'' 5.5~MV
Tandem accelerator facility~\cite{har21a}, utilizing
our ZAPS setup  centered around a hemispherical electron spectrograph with a pre-retardation lens and a doubly-differentially pumped gas cell,
allowing for the detection of projectile  Auger  electrons with high efficiency and high energy resolution.
Existing spectroscopic information about the KLL  Auger  lines also measured here used in the  Auger  energy calibration and the $2s2p$  Auger  line identification is presented and compared to previous published results in tables found in appendix~\ref{apx:ecal}.

\subsection{Metastable fractions and their determination}
The $1s2s\,^3\!S_1$ and $1s2s\,^1\!S_0$ states decay to the ground state predominantly by $M1$ and two-photon ($2E1$) transitions, respectively. They are therefore metastable~\cite{tra24a} having relatively long lifetimes $\tau[^3\!S]$ and $\tau[^1\!S]$ {(\CORnewnew{we used the values $\tau[^3\!S]= 20.589 \times 10^{-3}$~s and $\tau[^1\!S]= 3.0322 \times 10^{-6}$~s for carbon, while for oxygen $\tau[^3\!S]= 95.60 \times 10^{-5}$~s and  $\tau[^1\!S]=4.3327\times 10^{-7}$~s -} see Table~I in \CORnew{the supplementary material} of Ref.~\cite{zou25a})}.
Both these beam components can survive to the target~\cite{ben18b} contributing, in general, to the production of the $2s2p\,^3\!P$ and $2s2p\,^1\!P$ states~\cite{ben16b}.
Using our \COR{two spectra measurement} technique~\cite{ben16b,mad19a}, we are able to accurately determine, \textit{in situ}, the $1s2s\,^3\!S$ beam fractional component $f[^3\!S]$, but not the $f[^1\!S]$ component.
However, since the production of the $2s2p\,^1\!P$ depends sensitively on the $1s2s\,^1\!S$ component this component has to also be considered. Here, this component is estimated using our recently published three-component model~\cite{zou25a}.

According to the three-component model~\cite{zou25a}, the three fractions \textit{at the target} are determined using our \COR{two spectra measurement} technique~\cite{ben16b,mad19a} by the following expressions:
\begin{align}
f^{[1]}[^3\!S] &= p\,f^{[2]}[^3\!S],\label{eq:f3S1_3c}\\
f^{[2]}[^3\!S] &= \frac{(1-\beta^{[1]})-d(1-\beta^{[2]})}{p(1+\alpha^{[1]})-d(1+\alpha^{[2]})},\label{eq:f3S2_3c}\\
f^{[i]}[^1\!S] &= \alpha^{[i]}f^{[i]}[^3\!S]+\beta^{[i]},\qquad\quad\,\text{for $i=1,2$,}\label{eq:f1Si_3c}\\
f^{[i]}[1s^2] &= 1-f^{[i]}[^3\!S]-f^{[i]}[^1\!S]\qquad\text{for $i=1,2$.}\label{eq:f1s2i_3c}
\end{align}
where $p$ and $d$ are the ratios of the $^4\!P$ and $^2\!D$  Auger  yields measured in each of the two  ($i=1$ and $i=2$) Auger  spectra~\cite{ben16b,zou25a} {\color{black}given by
\begin{align}
p \equiv \frac{dY^{[1]}_A[^4\!P]/d\Omega^\prime}{dY^{[2]}_A[^4\!P]/d\Omega^\prime}, \qquad
d \equiv \frac{dY^{[1]}_A[^2\!D]/d\Omega^\prime}{dY^{[2]}_A[^2\!D]/d\Omega^\prime}, \label{eq:pdsigma4P2D}
\end{align}
where \CORnew{$dY^{[i]}_A[X]/d\Omega^\prime$ represents} the measured normalized  Auger  yield~\cite{ben16b} for the  Auger  line $X$
in measurement $i$.}
The parameters $\alpha^{[i]},\beta^{[i]}$ are given by:
\begin{align}
\alpha^{[i]} & =\alpha_0^{[i]}\exp\left(\frac{\Delta t_0^{[i]}}{\tau[^3\!S]}\right)\exp\left(-\frac{\Delta t_0^{[i]}}{\tau[^1\!S]}\right)\label{eq:ai},\\
\beta^{[i]} &=\beta_0^{[i]}\exp\left(-\frac{\Delta t_0^{[i]}}{\tau[^1\!S]}\right),\label{eq:bi}
\intertext{where the metastable fractions at the point of their production (in the last stripper utilised - marked by the subscript 0) are assumed to be related by:}
f_0^{[i]}[{^1\!S}] &=  \alpha_0^{[i]} f_0^{[i]}[{^3\!S}] + \beta_0^{[i]}.\label{eq:f01Sa0f03Splusbeta0}
\end{align}
Here,  $\alpha_0^{[i]}=1/3$ according to spin statistics, while $\beta_0^{[i]}$  is a parameter  introduced
in Ref.~\cite{zou25a} to account for the production of $^1\!S$ in the stripper by singlet spin conserving excitation processes from the He-like ground state. Since there is no model yet to calculate $\beta_0$, we treat $\beta_0$ here as a free parameter with values in the range of 0-50\%.
The parameters $\alpha^{[i]}$ and $\beta^{[i]}$ then just propagate the ${^1\!S}$ fraction from the stripper, $f_0[{^1\!S}]$, to the target, $f[{^1\!S}]$, over the required time-of-flight, $\Delta t_0$, between the last stripper and the target~\cite{zou25a}.

In the \COR{two spectra} measurement technique both high and low $^3\!S$ fraction  Auger  spectra are used in the determination of $f[^3\!S]$~\cite{ben16b}. However, the normalized yields presented here are obtained from the high $^3\!S$ fraction ($i=1$) Auger spectrum which corresponds to larger $2s2p$ yields and therefore improved statistics. The determined high $f[^3\!S]$ fractions  are shown in Fig.~\ref{fg:f3SC4O6} for both carbon and oxygen computed according to Eqs.~(\ref{eq:f3S1_3c})-(\ref{eq:f3S2_3c}) and give the fractions at the target. Similarly, Fig.~\ref{fg:f1SC4O6} gives the $f[^1\!S]$ fractions derived from the $f[^3\!S]$ fractions according to Eq.~\ref{eq:f1Si_3c} for $\beta_0=0$.
For carbon, both $1s2s2p\,^4\!P$ and $2s2p\,^3\!P$ states were used in our technique since both states are predominantly produced from the $1s2s\,^3\!S$ component (a necessary requirement of the method). The good agreement underscores the consistency of the method.
For some energies, where only one spectrum was measured (usually the high $f[^3\!S]$ measurement), the metastable fraction was estimated and marked in the tables with parentheses and with open symbols in Fig.~\ref{fg:f3SC4O6}.
The fractions are also listed in Tables~\ref{tb:C4He2s2pNY} and \ref{tb:O6He2s2pNY} with stripping methods marked as gas terminal stripping (GTS), foil terminal stripping (FTS), gas post-stripping (GPS), foil post-stripping (FPS) and their combinations. These stripping methods are explained in more detail in Refs.~\cite{ben18b,zou25a}. {\color{black}Overall the $^3\!S$ metastable fraction over the energy range of the measurements remained rather constant around 13-14\% for carbon and 19-20\% for oxygen with the stripping methods used as seen in Fig.~\ref{fg:f3SC4O6}. The same is also seen in Fig.~\ref{fg:f1SC4O6}~(top) for the carbon $^1\!S$ fractions with a value of about 3\%.  However, this is not so for oxygen which due to its much shorter $^1\!S$ lifetime is much more sensitive to the stripper distance from the target and the speed of the ion beam~\cite{zou25a} as seen in Fig.~\ref{fg:f1SC4O6}~(bottom).}

\subsection{Zero-degree normalized Auger electron yields}
\label{subsec:NeY}
\begin{figure*}[tbh]
\includegraphics[scale=0.75]{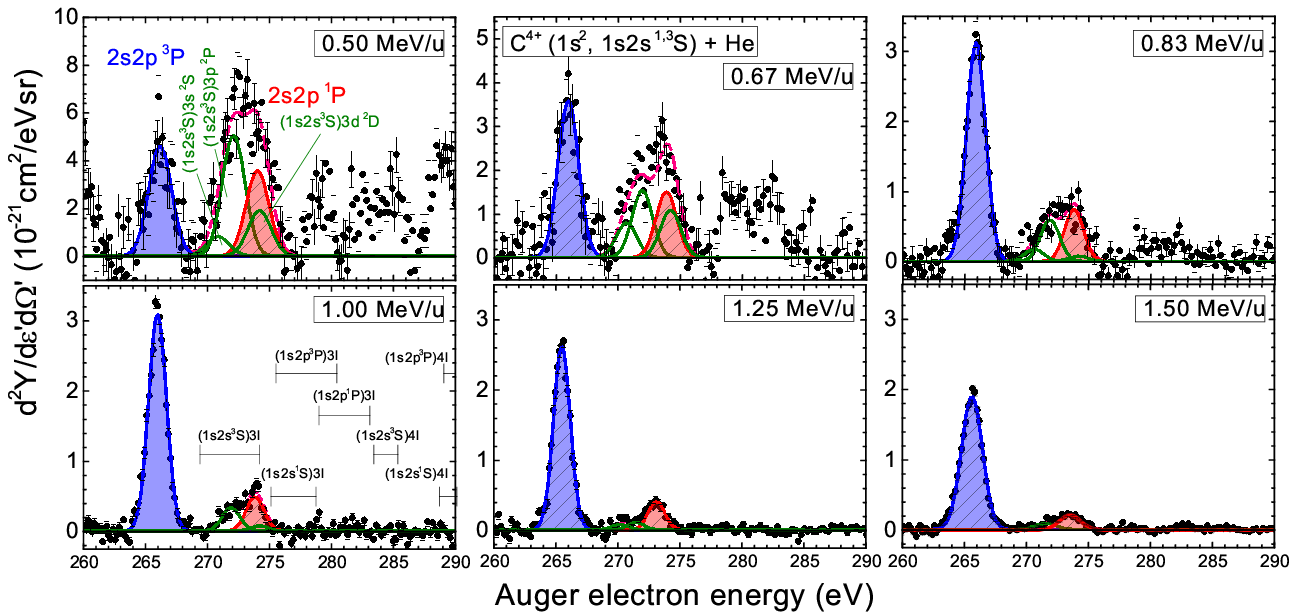}
\caption{\label{fg:C42s2p}Normalized ZAPS spectra after background subtraction and transformation to the projectile rest frame as a function of electron energy at the selected projectile energies \REFtwo{$E_p=0.50,0.67,0.83,1.00,1.25,1.50$~MeV/u} in collisions of the C$^{4+}(1s^2,\, 1s2s\,^{3,1}\!S)$ mixed-state ion beam with helium.
\REFtwo{The shaded areas indicate} both the $2s2p\,^{3}\!P$ (blue) and $2s2p\,^{1}\!P$ (pink) Auger lines.  The stripping method  \COR{used} and the extracted SDCS are listed in Table~\ref{tb:C4He2s2pNY}. The Gaussian fits in green correspond to the three near-lying Auger lines identified as the $(1s2s\,^3\!S)3l\,^2\!L$ with $l=0,1,2$ and $L=l$ due to $3l$ capture to the $1s2s\,^3\!S$
(see Table~\ref{tb:energiesCarbon}). Particularly the $(1s2s\,^3\!S)3d\,^2\!D$ line lies within the pink shaded area and cannot be resolved from the $2s2p\,^1\!P$. However, $3l$ capture drops rapidly with $E_p$, allowing for the $^1\!P$ line to be clearly identified above \COR{$E_p=1.25$~MeV/u}. \COR{In the leftmost bottom panel, the energy range of the higher-lying Auger line configurations that can contribute to the spectra in the top panels are indicated.}}
\end{figure*}

\begin{figure*}[tbh]
\includegraphics[scale=0.75]{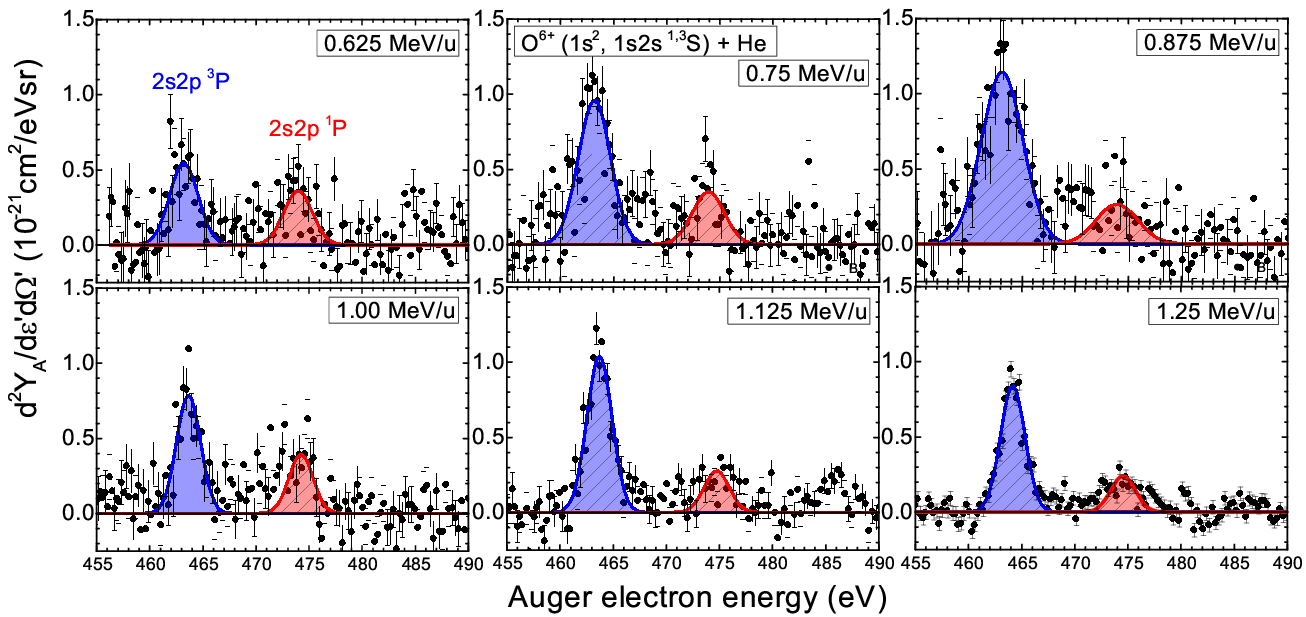}
\caption{\label{fg:O62s2p}Same as Fig.~\ref{fg:C42s2p}, but for the O$^{6+}(1s^2,\,1s2s\,^{3,1}\!S)$ mixed-state ion beam at the selected projectile energies \REFtwo{$E_p=0.625,0.75,0.875,1.00,1.125,1.25$~MeV/u}.
The stripping method  \COR{used} and the extracted SDCS are listed in Table~\ref{tb:O6He2s2pNY}. For oxygen there are no other troublesome Auger lines in between the two excitation lines as in the case of carbon.}
\end{figure*}

Normalized single differential Auger electron yields (for short normalized yields) at the observation angle $\theta$,  $dY_A(\theta)/d\Omega^\prime$ are obtained from normalized double differential electron yields $d^2Y_A/d\varepsilon^\prime d\Omega^\prime$ (after transformation to the rest frame of the projectile indicated by primed quantities) by extracting the area under the Auger line of interest, typically using peak fitting software or SIMION~\cite{sim14a} Monte Carlo simulations~\cite{nan23a} for improved accuracy. In Ref.~\cite{mad22a}, we have described in detail how these normalized Auger yields were obtained in the case of the $1s2l2l^\prime$ states produced by capture to the same mixed-state ion beams. In Figs.~\ref{fg:C42s2p} and \ref{fg:O62s2p}, the measured $\theta=0^\circ$ normalized double differential electron yields are shown with the fitted areas of the $2s2p$ $^3\!P$ and $^1\!P$ Auger lines indicated, from which the normalized yields $dY_A^\text{exp}/d\Omega^\prime$ for the corresponding states were obtained. These are also listed in Tables~\ref{tb:C4He2s2pNY} and \ref{tb:O6He2s2pNY} and plotted as symbols with error bars in Figs.~\ref{fg:NY0C4O62s2p3P} and \ref{fg:NY0C4O62s2p1P}.

\subsection{Line identification}
The $2s2p$ $^3\!P$ and $^1\!P$ Auger lines were identified in the normalized spectra shown in Figs.~\ref{fg:C42s2p} and \ref{fg:O62s2p}.
At the lowest collision energies, for carbon, the $^1\!P$ line lies in between other partially overlapping $(1s2s\,^3\!S)3l$ lines due to electron capture and can be difficult to identify. At the highest projectile energies, the two lines are the only ones left as capture has become negligible. To help in the unambiguous identification, we have provided in Table~\ref{tb:energiesCarbon} and Table~\ref{tb:energiesOxygen} some indicative experimental and theoretical Auger energies of the various lines in the spectra.
As seen from these tables the energies of the carbon $2s2p$ $^3\!P$ and $^1\!P$ Auger lines \COR{appear} never to have been measured, while for the oxygen lines there \COR{appear} to be only two such measurements one from collisions~\cite{bru79a} and the other from dielectronic recombination~\cite{kil90a}. Our own measured values are given in the first column of these tables.
The well-known $1s2l2l^\prime$ Auger lines~\CORnew{\cite{aza23a}} used to calibrate our spectra are also tabulated. The NIST reference energy levels (see Table~\ref{tb:energy_levels}) were used to determine the theoretical Auger energies where needed.
The relevant energy \COR{level} diagrams for both carbon and oxygen are shown in Fig.~\ref{fg:Absolute_BE_C3O5_C4O6_Auger}.

\section{Theory}
\label{sec:theory}
In Ref.~\cite{lao24a}, we presented 3eAOCC {\color{black}and first Born} calculations for the production of $2s2p\,^3\!P$ states from the $1s2s\,^3\!S$ component [Eq.~(\ref{eq:2s2p31P_production_3S})] for both carbon and oxygen He-like ion beams.
Here, we present only 3eAOCC results for the production of $2s2p\,^1\!P$ states from all three beam components [(Eqs.~(\ref{eq:2s2p31P_production_3S})-(\ref{eq:2s2p31P_production_gs})].
In addition, we also present 3eAOCC results for the production of the $2s2p\,^3\!P$ states, including the other two initial components [Eqs.~(\ref{eq:2s2p31P_production_1S}) and (\ref{eq:2s2p31P_production_gs})]\COR{, not treated in Ref.~\cite{lao24a}}.

\subsection{3eAOCC calculations}
The 3eAOCC method has been described in detail previously in  \cite{sis11a,gao18a,sis19a,iga24b}  and already used in our  single electron capture~\cite{mad20a,mad21a,mad22a} and transfer-excitation~\cite{lao22a} investigations in  C$^{4+}$-He (and H$_2$) MeV collisions.

The 3eAOCC uses a semiclassical close-coupling approach based on a time dependent expansion of the scattering states onto sets of asymptotic states, i.e. states of the two isolated target and projectile partners of the collision. As in Ref.~\cite{lao24a}, the two collision systems
are described here using a \textit{three}-active electrons representation which allows for the  accurate description of C$^{4+}$ and O$^{6+}$ after excitation including   spatial and spin  components (but neglecting spin-orbit coupling), as well as the final state of the target. Details of the 3eAOCC calculation for the process of Eq.~(\ref{eq:2s2p31P_production_3S})
were provided in \cite{lao24a}. Here, the calculations using the same {\color{black}basis sets} are also used to obtain production cross sections for the $2s2p\,^1\!P$ state, as well as for the production of both $2s2p\,^3\!P$ and $2s2p\,^1\!P$  states from the other two components, not addressed in Ref.~\cite{lao24a}.

For each of the three initial states of the He-like ion beam an independent 3eAOCC calculation was performed which included the production of both $2s2p\,^3\!P$ and $2s2p\,^1\!P$ states.
For the target, one of the He electrons is considered frozen so that the interactions between the He$^+$ target core and the three active electrons  is described by a model potential (see Table III in~\cite{mad22a}). For the static (state and basis sets  construction) and dynamical (collision) stages of the calculations, all Coulombic interactions and bi-electronic couplings  were taken into account within a full configuration interaction scheme.

In the previous works  we used  very large basis sets to describe simultaneously  one-electron processes (transfer, excitation and, in a more limited way, ionization) and two-electron processes (mainly transfer-excitation and double excitation). The present results therefore stem from the same computations for C$^{4+}$ projectile: the same sets of Gaussian-Type Orbitals (GTO) for the genuine representation of the helium and carbon states (see Table II in~\cite{mad22a}) were used. For oxygen projectile, we have an equivalent representation of the O$^{6+}$ and O$^{5+}$ states, with a set of 22 GTOs, 10 for $\ell$=0 and 3 $\times$ 4 $\ell$ =1 symmetries, see Table~\ref{tb:GTOO}. With these GTO sets, the  helium ground state is bound by 0.901~a.u. (to be compared to the NIST value of 0.9035698802~a.u.~\CORnew{\cite{kra24a}}) and the energies of the states for the C$^{4+}$ and O$^{6+}$ ions under consideration in the present work are shown in Table~\ref{tb:energy_levels}. They are compared to reference values, with agreement  better than $\sim$0.9\% for carbon and $\sim$0.5\% for oxygen.

\begin{table}[htb]
\caption{ Orbital angular momentum quantum numbers $\ell$ and exponents $\alpha$ of the GTOs $\mathcal{G}(r) = N r^\ell \exp(- \alpha r^2 )$ for  oxygen ions. The notation 3.00[-1] denotes $3.00\times 10^{-1}$. Note that the number of GTOs  is 22 considering the multiplicity of 3 for each of the $\ell$ = 1 orbitals. }
\begin{tabular}{|llc||llc|}
\hline
$\ell$  && $\alpha$ & $\ell$  && $\alpha$ \\
\hline
0  && 3.00[-1]      & 0  && \!\!1.30[2] \\
0  && 7.50[-1]      & 0  && \!\!3.06[2] \\
0  && \,\,1.77\ph   & 0  && \!\!1.73[3] \\
0  && \,\,4.18\ph   & 1  && 7.22[-1] \\
0  && \,\,9.86\ph   & 1  && \,\,2.08\ph  \\
0  && \!\!2.33[1]   & 1  && \,\,6.80\ph  \\
0   && \!\!5.49[1]  & 1  && \!\!2.67[1] \\
\hline
\end{tabular}
\label{tb:GTOO}
\end{table}

\begin{table}[htb]
\caption{Energies (in a.u.) of states under direct consideration for
C$^{4+}$ and O$^{6+}$ ions. The present values are compared with the ones listed in  NIST unless otherwise indicated.}
\begin{tabular}{|c|cc||cc|}
\hline\\[-3.2mm]
& \multicolumn{2}{c||}{C$^{4+}$} &\multicolumn{2}{c|}{O$^{6+}$}  \\
\hline
state  & present & NIST\footnotemark[1] & present & NIST\footnotemark[1]  \\
\hline\\[-3.2mm]
 1s$^2\ ^1\!S$  & -32.219    & \!\!-32.409                          & -59.130  & -59.193 \\
 1s2s$\ ^3\!S$  & -21.314    & \,\,-21.430\footnotemark[2]      & -38.533	& -38.578  \\
 1s2s$\ ^1\!S$  & -21.114    & -21.223                          & -38.246 & -38.287 \\
 2s2p$\ ^3\!P$  & \,\,-8.196 & \,\,\,\,\,-8.234\footnotemark[2] & -14.949 & -14.971 \\
 2s2p$\ ^1\!P$  & \,\,-7.868 & \,\,\,\,\,-7.939\footnotemark[3] & -14.498 & -14.565 \\
\hline
\end{tabular}
\footnotetext[1]{NIST~\CORnew{\cite{kra24a}}}
\footnotetext[2]{M\"{u}ller \textit{et al.}  2018~\cite{mul18b}.}
\footnotetext[3]{van der Hart and Hansen 1993~\cite{van93a}.}
\label{tb:energy_levels}
\end{table}

For C$^{4+}$+He collisions, to solve the time-dependent Schr\"odinger equation,  the time-dependent expansion of the scattering state spans the same Hilbert space as in~\cite{mad20a}, i.e. with a total 1794 3-electron bound, autoionizing and continuum  states  (799 of type C$^{4+}~\times$~He and 995 of type C$^{3+}$) for doublet spin symmetry (respectively 802, 380 and 422 for quartet). For O$^{6+}$+He collisions, the basis set includes 1357 three-electron~states, with 694 of O$^{6+}~\times$~He and O$^{5+}$ types, for doublet spin symmetry (respectively 598, 322 and 276 for quartet).

The cross sections stemming from the close-coupling computations and shown in the following are inclusive cross sections, i.e. cross sections for excitation to C$^{4+}$(2s2p) and O$^{6+}$(2s2p), whatever the final state of the helium target. This is mandatory since  (i) the target is not analyzed experimentally after collision and (ii) our calculations prove that  He excitation and ionization are very important channels, especially for initial metastable (1s2s$ \ ^{1,3}\!S$) \CORnew{He}-like ions.

Since the He-like Z$^{q+}$ ions are in a mixture of the $1s^2$ ground state and the two long-lived $1s2s\,^3\!S$ and $1s2s\,^1\!S$ states, three independent calculations had to be performed one for each initial state as in the processes of Eqs.~(\ref{eq:2s2p31P_production_3S})-(\ref{eq:2s2p31P_production_gs}).
Here, He(All) signifies that all final states of the He target were considered in the calculation, including the He$(1s)$, He$(nl)$, and even ionization, i.e. He$^+$.

{\color{black}For the production of the $2s2p\,^{1,3}\!P$ states, the cross sections for the $M_L=0$ component,
$\sigma_j(M_L=0)$, as well as the total (sum over $M_L$), $\sigma_j^\text{tot}$,}
are listed in Tables~\ref{tb:C4He1_2s2pcsM} and \ref{tb:O6He1_2s2pcsM} {\color{black}as noted}, for carbon and oxygen ion beams, respectively, where $j$ signifies one of the three initial ion states, i.e. $1s^2$, $^3\!S$ and $^1\!S$. In Figs.~\ref{fg:2s2p3PC4O6CSlog} and \ref{fg:2s2p1PC4O6CSlog}, the $E_p$ energy dependence of the cross sections is shown.

\subsection{\label{subsec:AAD}Auger angular distributions and single differential cross sections}
In an ion-atom collision the produced doubly-excited $SLJ$ projectile state  may Auger decay to a final $S_fL_fJ_f$ state
\begin{align}
(SLJ) \rightarrow (S_fL_fJ_f) + e^-_A(\theta_e^\prime,\varepsilon_A;\, s=1/2, \ell, j)\label{eq:Augerdecay}
\end{align}
emitting an Auger electron $e^-_A$ at angle $\theta_e^\prime$ (the prime refers to the projectile rest frame) with respect to the initial beam direction with  energy $\varepsilon_A$ and $\ell$ orbital-  and $j$ total- angular momenta.
The Auger SDCS  are then angular distributions
expressed as a sum over \textit{even} (due to parity conservation) Legendre polynomials $P_k(\cos\theta_e^\prime)$  given by (see \cite{mad22a} and references therein):
\begin{align}
\frac{d\sigma^j_A}{d\Omega^\prime}(\theta_e^\prime)
&= \overline{\xi}\,\frac{\sigma_j^\text{tot}}{4\pi}\,\left[1 +
\sum_{k=2,4\ldots}a^j_k\,P_k(\cos\theta_e^\prime)\right],\label{eq:sdcsAugertheta}
\end{align}
where the index $j=1s^2,^3\!\!S,^1\!\!S$, refers to the three different initial components of the ion beam and $\sigma_j$ the production cross sections from each of these components to the $2s2p\,^{3,1}\!P$ states.
The coefficients $a^j_k$ can be theoretically computed in various approximations, $\sigma_j^\text{tot}$ is the total state production cross section, while $\overline{\xi}$ is the mean Auger yield given in Tables~\ref{tb:C4He2s2pNY}-\ref{tb:O6He2s2pNY} for the two states.

For unresolved $LSJ$ multiplets one has to sum over the various $J$ levels in various formulations depending on whether the fine structure is in principle resolvable or not. Furthermore, the Auger electron might have more than one allowed $\ell$ or $j$ angular momenta (see Eq.~\ref{eq:Augerdecay}), in which case, further complications arise since the different partial $(l,j)$-waves can interfere. Examples of calculations in the $LSJ$ intermediate coupling approximation are given in Refs.~\cite{che92a,sur08a,fri12a} and for $LS$ coupling in Refs.~\cite{kab94a,mehl80a}.

In particular, for the $P$ states ($L=1$) of interest here and for $k=2$ in Eq.~(\ref{eq:sdcsAugertheta}), the coefficient $a_2$ is given by:
\begin{align}
a^j_2 & = A^j_2 D_2,\label{eq:a2}
\intertext{with the anisotropy coefficient $A^j_2$ given by (see Table I of Ref.~\cite{mehl80a}): }
A^j_2 & = 2\,\frac{\sigma_j[M_L=0]-\sigma_j(M_L=1)}{\sigma_j^\text{tot}}\quad(j=1s^2,{^3\!S},{^1\!S})\label{eq:A2}\\
\sigma_j^\text{tot} & = \sigma_j(M_L=0)+2\sigma_j(M_L=1),\label{eq:sigmatot}
\end{align}
and the dealignment factor $D_2$ (which accounts for the average loss of orbital alignment into spin alignment) is given  by Eq.~(\ref{eq:D2}) \COR{in appendix~\ref{apx:sec_D2}}. The partial production cross sections $\sigma_j(M_L)$
depend on the magnetic quantum number $M_L$ and
are computed in the 3eAOCC approach for each of the three initial beam components. They are listed for the production of the $2s2p\,^3\!P$ and $^1\!P$ states in Tables~\ref{tb:C4He1_2s2pcsM} and \ref{tb:O6He1_2s2pcsM} for collisions of carbon and oxygen ions with
{\color{black} He} target as already discussed in the Theory section. The anisotropy parameter is seen to take values from $A_2=2$, when $\sigma(M_L=1)=0$ to $A_2=-1$ when $\sigma(M_L=0)=0$ and thus is an indicator for alignment. And of course, when all partial cross sections are equal, then $A_2=0$ and we have isotropy. The anisotropy parameter is plotted in Fig.~\ref{fg:A2C4O63P1P} for the two collision systems and states.

Evaluating Eq.~(\ref{eq:sdcsAugertheta}) at the laboratory observation angle $\theta=0^\circ$ (for which $\theta_e^\prime=0^\circ$ or $180^\circ$ - see Eq.~(27) in Ref.~\cite{mad22a}), we then obtain for the Auger SDCS:
\small
\begin{align}
\frac{d\sigma^j_A}{d\Omega^\prime}(0^\circ)
&= \overline{\xi}\,\frac{(1+2D_2)\sigma_j(M_L=0)+2(1-D_2)\sigma_j(M_L=1)}{4\,\pi}.\label{eq:sdcsAugertheta0D2}
\end{align}
\normalsize
For no dealignment (completely overlapping resonances, i.e. $\varepsilon_{J,J^\prime}=0$ in Eq.~(\ref{eq:varepsilonJJp}) appendix~\ref{apx:sec_D2}), $D_2=1$~\cite{mehl80a}, and we get the well-known $LS$-coupling result:
\begin{align}
\frac{d\sigma^j_A}{d\Omega^\prime}(0^\circ)
&= \overline{\xi}\,\frac{3\,\sigma_j(M_L=0)}{4\,\pi},\qquad(D_2=1)\label{eq:sdcsAugertheta0}
\end{align}
while if all partial cross sections are equal, i.e. $\sigma_j\equiv\sigma_j(M_L=0)=\sigma_j(M_L=1)\left(=\sigma_j[M_L=-1]\right)$, then $A^j_2=0$ [see Eq.~(\ref{eq:A2})]
and we have the case of isotropy as expected (independent of dealignment):
\begin{align}
\frac{d\sigma^j_A}{d\Omega^\prime}(0^\circ)
&= \overline{\xi}\,\frac{\sigma_j^\text{tot}}{4\,\pi},\qquad(\text{isotropic})\label{eq:sdcsAugertheta0isotropic}
\end{align}
which is seen to also correspond to the case of $D_2=0$.
We note that for maximum dealignment, i.e. cases of extreme spin-orbit coupling encountered in much heavier projectiles, i.e. $\varepsilon_{J,J^\prime}>>1$ (non-overlapping resonances), Eq.~(\ref{eq:D2}) then gives $D_2=5/18 = 0.2778$~(see Eq.~(23) of Ref.~\cite{mehl80a}). The relative overlap between the three resonances and its effect on the value of $D_2$ is shown schematically in Fig.~\ref{fg:C4O62s2p3PJD2} in appendix~\ref{apx:sec_D2}.
Its effect on the normalized yield compared to that for $D_2=1$ or isotropy is rather small.

\subsection{Normalized Auger yields}
Comparisons to the measured (normalized)  $\theta=0^\circ$ Auger yield, $dY_A^\text{exp}(\theta=0^\circ)/d\Omega^\prime$ require the computation of the corresponding \textit{total} theoretical normalized yields. These are calculated as the sum of the \textit{partial} normalized yields {\color{black}(also known as apparent cross section~\cite{mul25a})} from each one of the three $j$ initial states:
\begin{align}
\frac{dY_A^\text{tot}}{d\Omega^\prime} (\theta=0^\circ) = \sum_j \frac{dY_A^j}{d\Omega^\prime} (\theta=0^\circ) \label{eq:sumfisdcsi}
\end{align}
with $dY_A^j/d\Omega^\prime$ {\color{black} given by:}
\begin{align}
\frac{dY_A^j}{d\Omega^\prime}
(\theta=0^\circ) = f[j]\,\frac{d\sigma^j_A}{d\Omega^\prime}(\theta=0^\circ)\quad(j=1s^2,{^3\!S},{^1\!S})\label{eq:dYi}
\end{align}
where $f[j]$ are the three fractional components of the mixed-state ion beam and \CORnew{$d\sigma^j_A(\theta=0^\circ)/d\Omega^\prime$} are the computed SDCSs according to Eqs.~(\ref{eq:sdcsAugertheta0D2})-(\ref{eq:sdcsAugertheta0isotropic}) discussed above and dependent on the 3eAOCC partial cross sections via the alignment parameter $A_2$ given by Eq.~(\ref{eq:A2}). The computed values of $dY_A^\text{tot}/d\Omega^\prime$ are listed in Tables~\ref{tb:C4He2s2pNY} and \ref{tb:O6He2s2pNY} and shown in Fig.~\ref{fg:NY0C4O62s2p3P} and \ref{fg:NY0C4O62s2p1P}.

\section{Results and discussion}
\label{sec:results}
In this section we present our measurements and theoretical results in both figures and tables and discuss the observed features.
In all subsections, except the last, we assume that the metastable fractions are related just by spin-statistics as assumed in the past, i.e. $f_0[^1\!S]=(1/3) f_0[^3\!S]$, 
or $\beta_0=0$. In the last section, \ref{subsec:results.betaneq0}, we explore non-zero values for $\beta_0$.

\subsection{$2s2p\,^3\!P$ and $2s2p\,^1\!P$ production}
\subsubsection{3eAOCC production cross sections}
\begin{figure}[tbh]
\includegraphics[scale=0.47,angle=0]{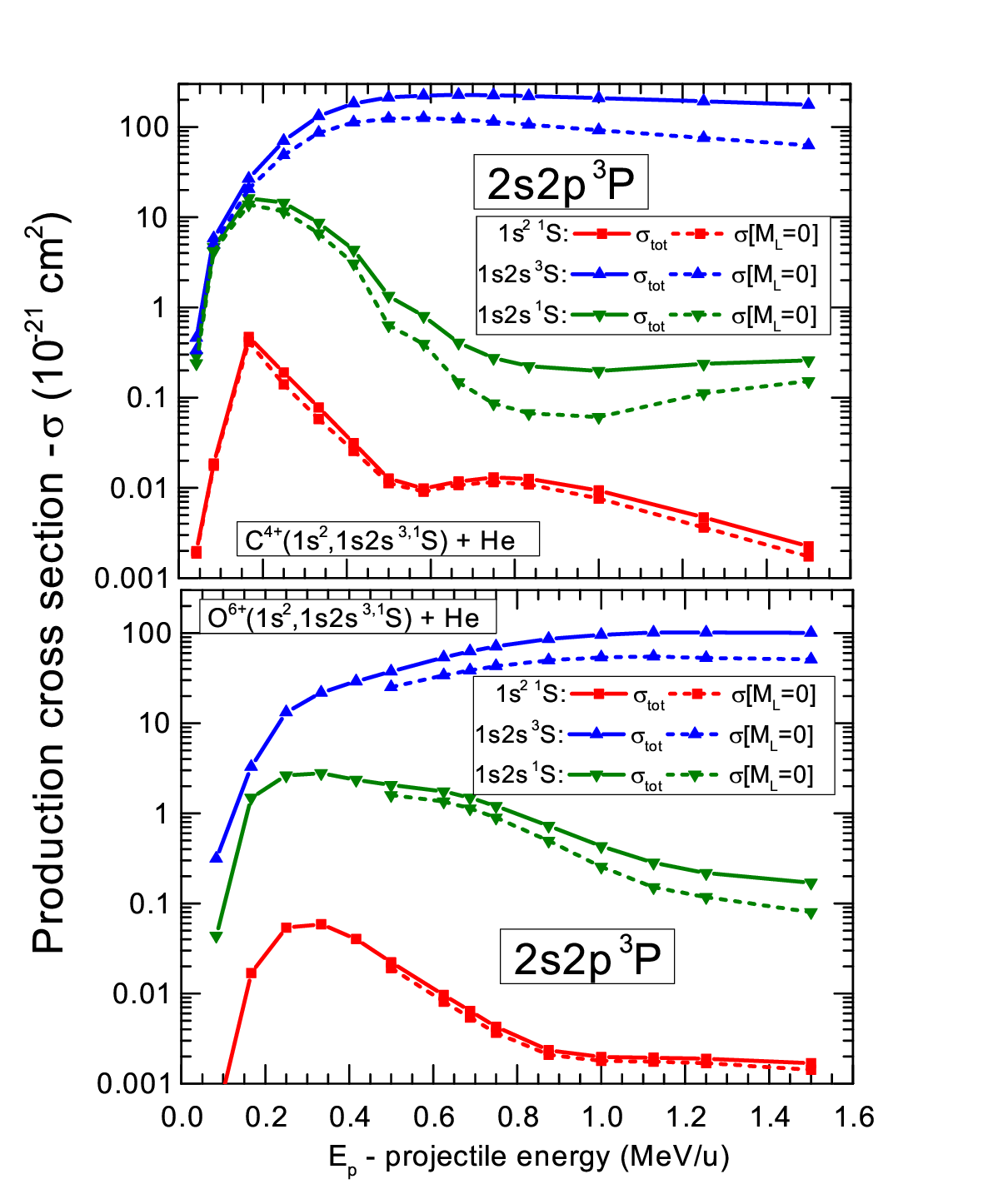}
\caption{\label{fg:2s2p3PC4O6CSlog}3eAOCC $(2s2p\,^3\!P)$ production cross sections as a function of projectile energy $E_p$ for C$^{4+}$ (top) and O$^{6+}$ (bottom) from each of the three different initial ion beam components in collisions with He: The $(1s2s\,^3\!S)$ state (blue lines with triangles), the $(1s2s\,^1\!S)$ state (green lines with inverted triangles) and the $(1s^2)$ ground state (red lines with squares). The full lines correspond to total cross sections (sum over all partial cross sections), while the dashed lines to just the $M_L=0$ partial cross sections, $\sigma(M_L=0)$ {\color{black}(where shown)},  as also listed in Table~\ref{tb:C4He1_2s2pcsM} and Table~\ref{tb:O6He1_2s2pcsM}.
}
\end{figure}

\begin{figure}[tbh]
\includegraphics[scale=0.47,angle=0]{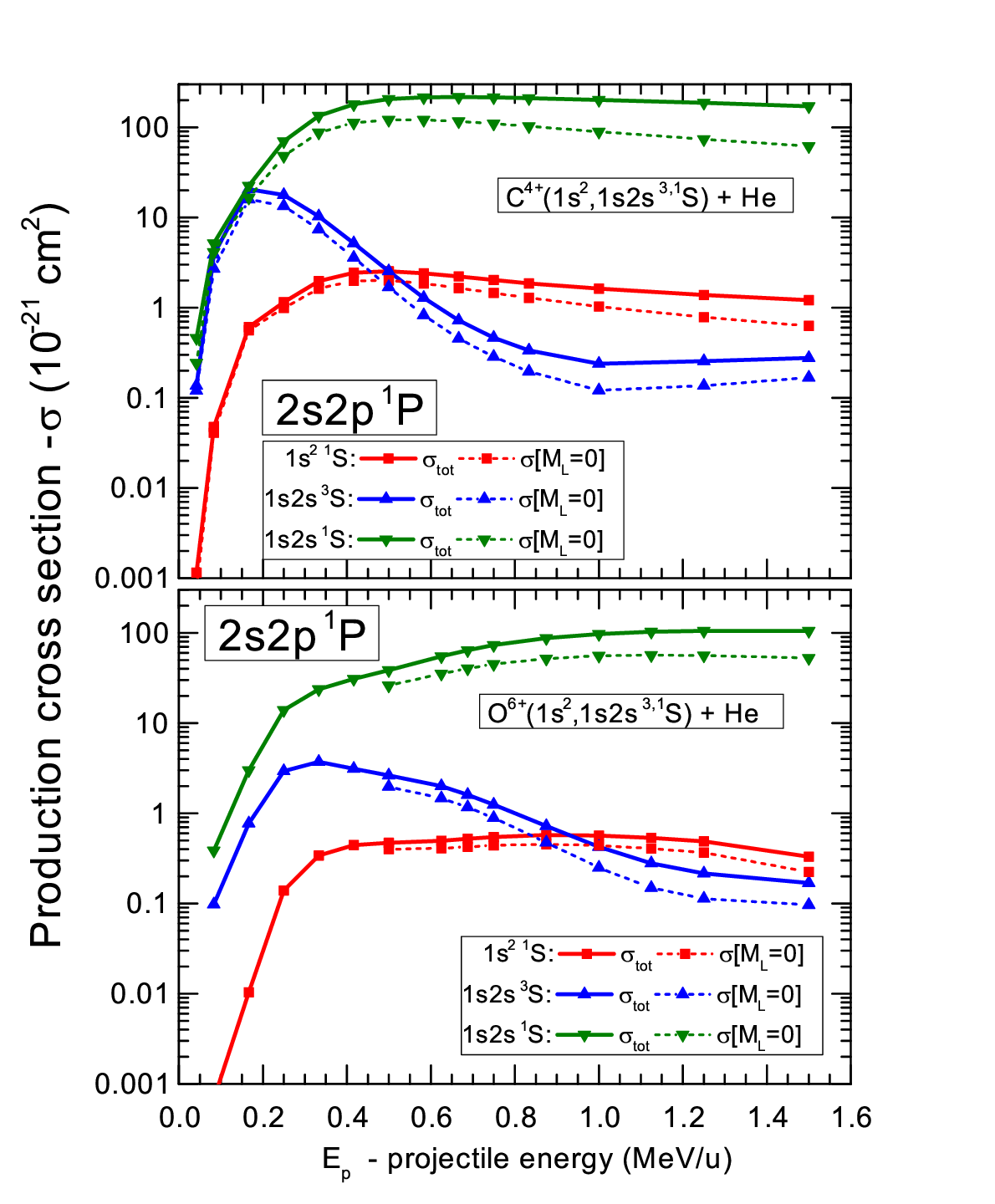}
\caption{\label{fg:2s2p1PC4O6CSlog}Same as Fig.~\ref{fg:2s2p3PC4O6CSlog}, but for the production of the $2s2p\,^1\!P$ state. Again, direct $1s\rightarrow 2p$ excitation, but from the $1s2s\,^1\!S$ state (green lines with inverted triangles) is seen to be the dominant excitation mode, followed by $1s\rightarrow 2p$ excitation with exchange (blue lines with triangles) and double excitation (red lines with squares). Double excitation is seen to become larger than  excitation with exchange as the collision energy $E_p$ increases and the time allowed for spin-exchange is correspondingly reduced.}
\end{figure}

In Tables~\ref{tb:C4He1_2s2pcsM} and \ref{tb:O6He1_2s2pcsM}  (see Appendix~\ref{ap:3eAOCCxsections}) the 3eAOCC cross sections for the production of the $2s2p\,^{3,1}\!P$ states from each of the three initial states are tabulated as a function of collision energy \CORnew{$E_p$. These results are shown in Figs.~\ref{fg:2s2p3PC4O6CSlog} and \ref{fg:2s2p1PC4O6CSlog} and discussed in detail.}

\subsubsection{Single excitation}
As can be seen, single excitation without exchange [\CORnewnew{blue triangles in Fig.~\ref{fg:2s2p3PC4O6CSlog}} - see Eq.~(\ref{eq:2s2p31P_production_3S}) for $^3\!P$ and \CORnewnew{green triangles in Fig.~\ref{fg:2s2p1PC4O6CSlog}} - see Eq.~(\ref{eq:2s2p31P_production_1S}) for $^1\!P$] is by almost two orders of magnitude the dominant production mechanism (\REFone{above} $\sim$0.5~MeV/u for carbon and \REFone{above} $\sim$0.94~MeV/u for oxygen) exhibiting the well-known excitation $E_p$ dependence: a low energy threshold followed by an increasing cross section eventually dropping off slowly with increasing energy $E_p$.
For these triplet to triplet and singlet to singlet excitations \REFone{their excitation energies} are very similar (see Fig.~\ref{fg:Absolute_BE_C3O5_C4O6_Auger}) which might explain their very similar energy dependence.
However, excitation with spin-exchange [\CORnewnew{green triangles in Fig.~\ref{fg:2s2p3PC4O6CSlog}} - see Eq.~(\ref{eq:2s2p31P_production_1S}) for $^3\!P$ and \CORnewnew{blue triangles in Fig.~\ref{fg:2s2p1PC4O6CSlog}} - see Eq.~(\ref{eq:2s2p31P_production_3S}) for $^1\!P$],
while having a similar low $E_p$ behavior, falls off much more rapidly than direct excitation.
Thus, it \REFone{appears} spin-exchange is much more probable at the lowest collision energies, where more time is available for the spin-exchange to occur.

\subsubsection{Double-excitation}
Overall, as seen in Figs.~\ref{fg:2s2p3PC4O6CSlog} and \ref{fg:2s2p1PC4O6CSlog}, the double-excitation process (red squares - {\color{black}excitation from the initial projectile ground state}) is by far the weakest, followed by single excitation with exchange (green inverted triangles), while direct single excitation (blue triangles) is seen to be the strongest.
\REFone{These general features are seen to apply to both carbon and oxygen.}
%
Interestingly though,  for the $2s2p\,^1\!P$ and $E_p$ energies larger than \REFtwo{$\sim$0.5~MeV/u} for carbon and \REFtwo{$\sim$0.94~MeV/u} for oxygen, double-excitation without spin-exchange \REFone{appears} to become more efficient than single-excitation with exchange. And of course, double-excitation with spin-exchange needed in the production of $2s2p\,^3\!P$ from the ground-state is seen to be the weakest process.

Finally, the $\sigma(M_L=0)$ partial cross section is seen to follow very closely the total cross section in its energy dependence. More on the difference between the $M_L=0$ and the $M_L=1$ partial cross sections can also be gained from the anisotropy parameter $A^j_2$ discussed next.

\subsubsection{\label{subsec:A2}Anisotropy parameters $A_2$}
The anisotropy coefficient $A_2$ gives important information about the alignment of the states due to excitation. From Eq.~(\ref{eq:A2}) it is clear that minimum alignment is attained when the $M_L=0$ and $M_L=1$ partial cross sections are equal in which case $A_2=0$ resulting also in the isotropic distribution of the Auger emission. Extreme alignment occurs when one of the two partial cross sections is zero. Then, $A_2$ is positive with $A^j_2=2$, when $\sigma_j(M_L=1)=0$ or negative with $A^j_2=-1$, when $\sigma_j(M_L=0)=0$.

\begin{figure}[tbh]
\includegraphics[scale=0.47,angle=0]{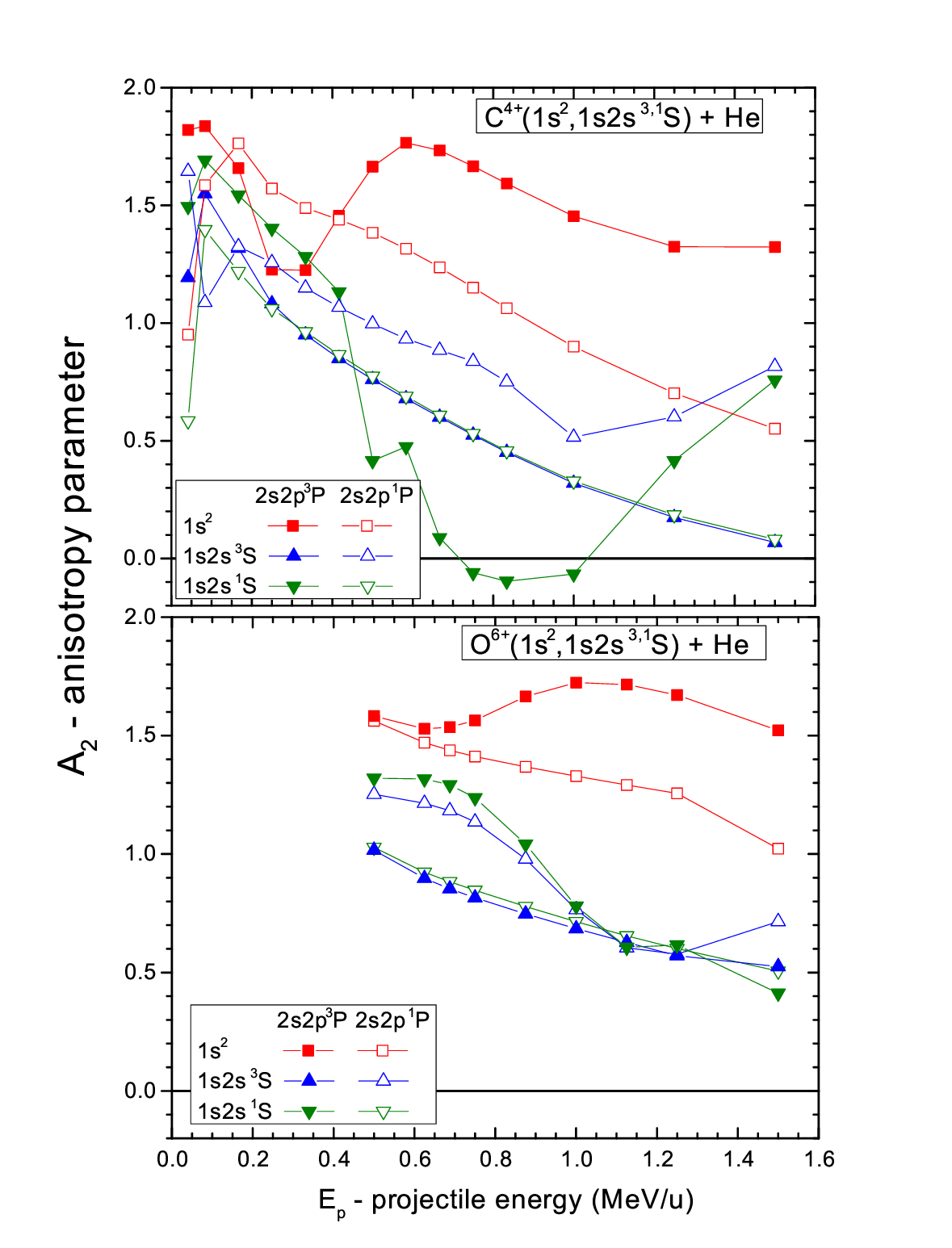}
\caption{\label{fg:A2C4O63P1P} Anisotropy parameter $A_2$ [see Eq.~(\ref{eq:A2})] for $2s2p\,^3\!P$ (filled symbols) and $2s2p\,^1\!P$ (open symbols) as a function of projectile energy $E_p$ for each one of the three initial states in collisions of C$^{4+}$ (top) and O$^{6+}$ (bottom) with He. Direct excitation for both $^3\!P$ (blue triangles) and $^1\!P$ (green inverted triangles) is seen to show very similar behavior. {\color{black}For the oxygen energy points below 0.5~MeV/u, which are outside the range of the measurements, no partial cross sections were computed and therefore no $A_2$ values are shown.}}
\end{figure}

In Fig.~\ref{fg:A2C4O63P1P}
the anisotropy parameter $A^j_2$ is plotted as a function of collision energy $E_p$ for excitation from each of the three initial beam components $j=1s^2\,^1\!S,1s2s\,^3\!S,1s2s\,^1\!S$ for both $2s2p\,^3\!P$ and $2s2p\,^1\!P$ for carbon and oxygen, respectively. All three initial state excitation processes \COR{appear} to be preferentially populated in the $M_L=0$ state with the process of double-excitation with spin-exchange being the most strongly aligned, both for carbon, but particularly for oxygen. The exception \REFone{appears} to be the excitation of the $2s2p\,^3\!P$ state from the $1s2s\,^1\!S$ state of carbon for which $A_2$ drops strongly, even attaining negative values, in the energy range of \REFtwo{0.5--1.5~MeV/u}. Interestingly, in the same rough energy range $A_2$ for $2s2p\,^3\!P$ excitation from the ground state \REFone{appears} to take on its most positive values approaching the maximum of 2. Oxygen \REFone{appears to demonstrate} a similar energy dependence, but with much less variation.

\subsubsection{\label{subsec:rations3Pto1P} $2s2p$ excitation ratio -  $\sigma({^3\!P})/\sigma({^1\!P})$}
\begin{figure}
\includegraphics[scale=0.47,angle=0]{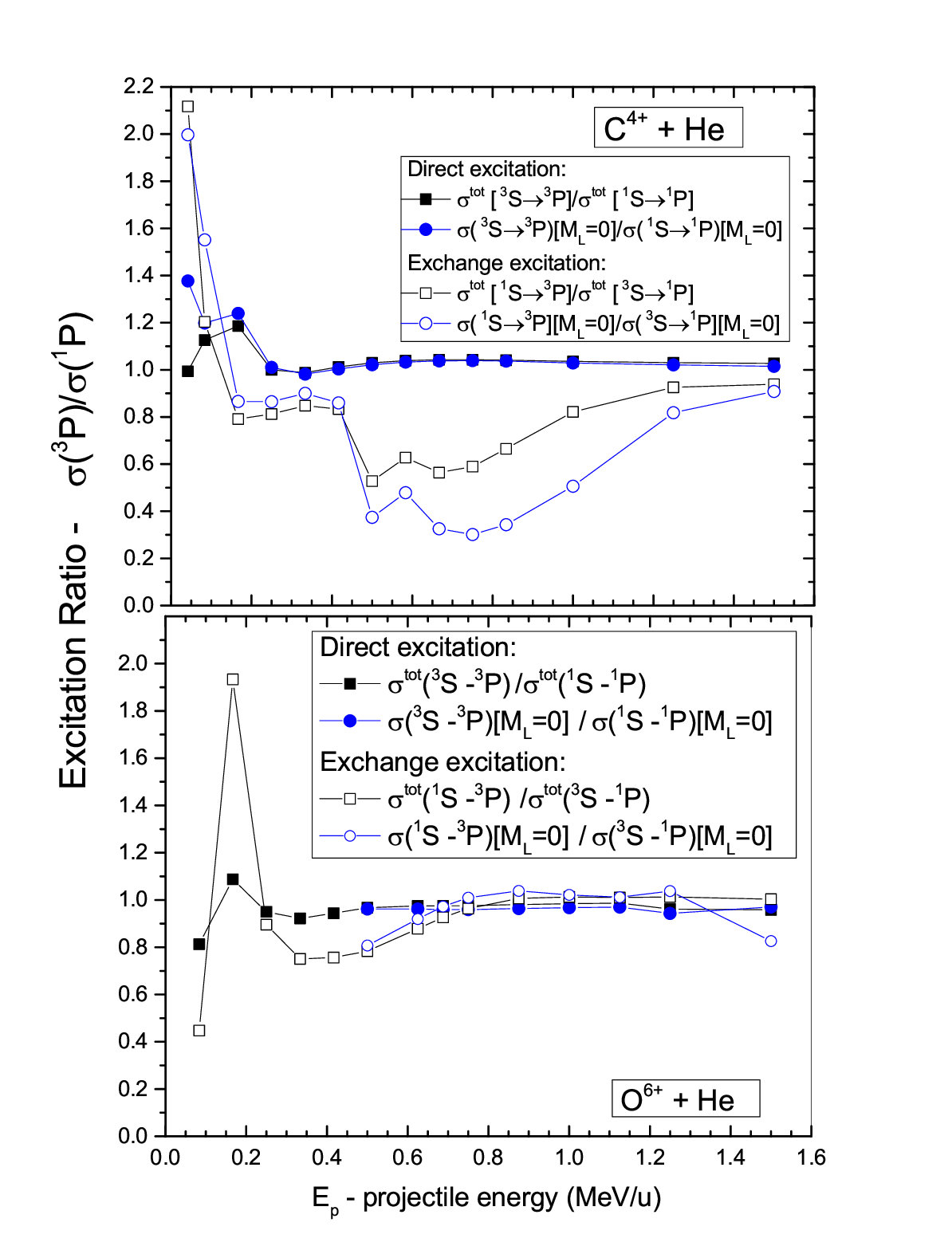}
\caption{\label{fg:ratiocsC4O62s2p3Pto1P}Ratio of $2s2p\,^3\!P$ to $2s2p\,^1\!P$ cross sections for C$^{4+}$ (top) and O$^{6+}$ (bottom). Direct excitation: Total cross sections (black squares), $M_L=0$ partial cross sections (blue circles). Exchange excitation: Total cross sections (black open squares), $M_L=0$ partial cross sections (blue open circles).}
\end{figure}

In Fig.~\ref{fg:ratiocsC4O62s2p3Pto1P}
the computed cross section ratios for direct and exchange single excitation for carbon and oxygen are shown, respectively.

For the process of direct excitation [Eq.~(\ref{eq:2s2p31P_production_3S}) for $^3\!P$ and Eq.~(\ref{eq:2s2p31P_production_1S}) for $^1\!P$], this ratio involves the production cross sections of $^3\!S$ to $^3\!P$ and of $^1\!S$ to $^1\!P$. This ratio is found to be nearly 1 for carbon and slightly below 1 for oxygen, remaining almost constant across the full range of projectile energies, except at {\color{black} the lowest energy points for both ions}. This near-identical behavior suggests a similar underlying excitation mechanism. The $M_L=0$ ratios show a similar trend as well.

However, for the process of exchange excitation [Eq.~(\ref{eq:2s2p31P_production_3S}) for $^1\!P$ and Eq.~(\ref{eq:2s2p31P_production_1S}) for $^3\!P$] the behavior is very different with this ratio dropping much below 1 {\color{black}above $\sim$0.4~MeV/u, while increasing back to near 1 with increasing collision energy, and also surpassing 1 at the very low collision energies}. This might indicate a different excitation mechanism with the production of the $2s2p\,^1\!P$ via spin-exchange being more probable than that for $2s2p\,^3\!P$.
However, for oxygen this \CORnew{does not} \COR{appear} to be the case since both ratios for direct and exchange excitation are seen to have almost identical behavior as a function of $E_p${\color{black}, except at the lowest collision energies}.

\subsubsection{\label{subsec:impact_parameter}Impact parameter dependence}
\begin{figure*}
\includegraphics[scale=0.6,angle=0]{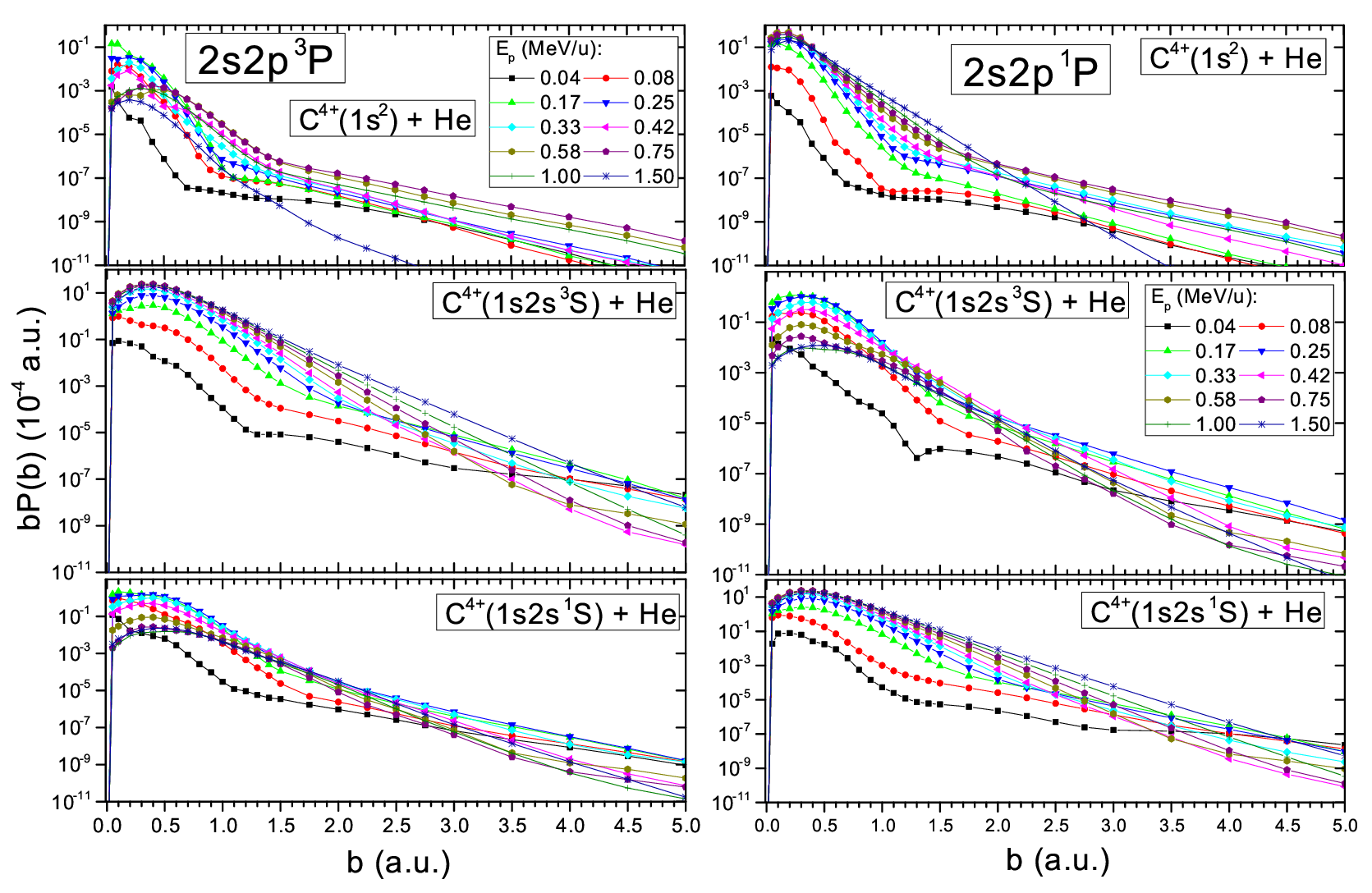}
\caption{\label{fg:2s2p3P1PC4bPblin}Reduced probabilities $bP(b)$ plotted as a function of impact parameter $b$ for selected characteristic projectile energies $E_p$. (Left) $2s2p\,^3\!P$ production, (Right) $2s2p\,^1\!P$ production. (Top) From the  C$^{4+}(1s^2)$ ground state, (Middle) From the C$^{4+}(1s2s\,^3\!S)$, and (Bottom) From the C$^{4+}(1s2s\,^1\!S)$ metastable states. Calculations are for the total probabilities (sum over all $M_L$ and \REVtwo{all final target states)}.}
\end{figure*}
\REVtwo{In Fig.~\ref{fg:2s2p3P1PC4bPblin}, the reduced probabilities $ bP(b) $ are shown on a logarithmic scale for the production of the $ 2s2p\,^3\!P $ (left) and $ 2s2p\,^1\!P $ (right) states from each of the three initial states ($1s^2$, $^3\!S$ and $^1\!S$). The calculations extend up to a maximum impact parameter of 5.0 a.u., but beyond approximately 1.4 a.u., the probabilities become negligible and do not significantly contribute to the cross sections.}

\REVtwo{No clear evidence of a large impact parameter structure is found that would indicate the onset of a TCee interaction.
TCee excitation might be expected in spin-exchange excitation processes, as observed in the production of the $ 1s2s2p\,^4\!P $ state from the $ 1s^22s $ initial state in Li-like O$^{5+}$ and F$^{6+}$ ion collisions with He and H$_2$~\cite{zou89b}.
However, no such signature is evident in the impact parameter dependence shown in Fig.~\ref{fg:2s2p3P1PC4bPblin}.}

\REVtwo{This observation is also in accord  with the absence of any clear sign of TCee excitation thresholds, as reflected in the $E_p$ energy dependence of the corresponding cross sections presented in Figs.~\ref{fg:2s2p3PC4O6CSlog} and \ref{fg:2s2p1PC4O6CSlog}, and also previously noted in Ref.~\cite{lao24a}.}

\REVtwo{Moreover, Fig.~\ref{fg:2s2p3P1PC4bPblin} reveals a significant reduction - by at least one order of magnitude - in the probabilities for spin-exchange excitation compared to spin-conserved (direct) excitation. This difference explains the correspondingly smaller cross sections for spin-exchange processes shown in Figs.~\ref{fg:2s2p3PC4O6CSlog} and \ref{fg:2s2p1PC4O6CSlog}. The spin-exchange excitation probabilities are confined to a narrower range of impact parameters, indicating that these processes require more violent (smaller impact parameter) collisions to occur. This effect is especially pronounced for low-velocity collisions involving the projectile’s compact $1s^2$ ground state, whose smaller spatial extent compared to that of metastable states
further limits the overlap with target electrons. Similar behavior is observed in O$^{6+}$ collisions, underscoring the generality of this mechanism.}

\REVtwo{The clear absence of a TCee signature at large impact parameters contrasts with observations in electron loss (bound-free transitions). There, TCee contributions can be significant at larger impact parameters (soft collisions) due to the extended spatial overlap between projectile and target electronic clouds~\cite{mon93b}. In contrast, excitation processes (bound-bound transitions) involve spatial overlaps confined to smaller regions, requiring
both $e$–$e$ and $e$–$n$ interactions to occur within the same impact parameter range. This obscures the distinct TCee signature observed in electron loss~\cite{mon92a,mon93b}. }

\REVtwo{To further investigate this behavior, studies across an isoelectronic series of ions involving coincidence measurements between the scattered projectile and
the recoil target would be valuable.}

\subsection{Normalized Auger yields}
\begin{figure}
\includegraphics[scale=0.47,angle=0]{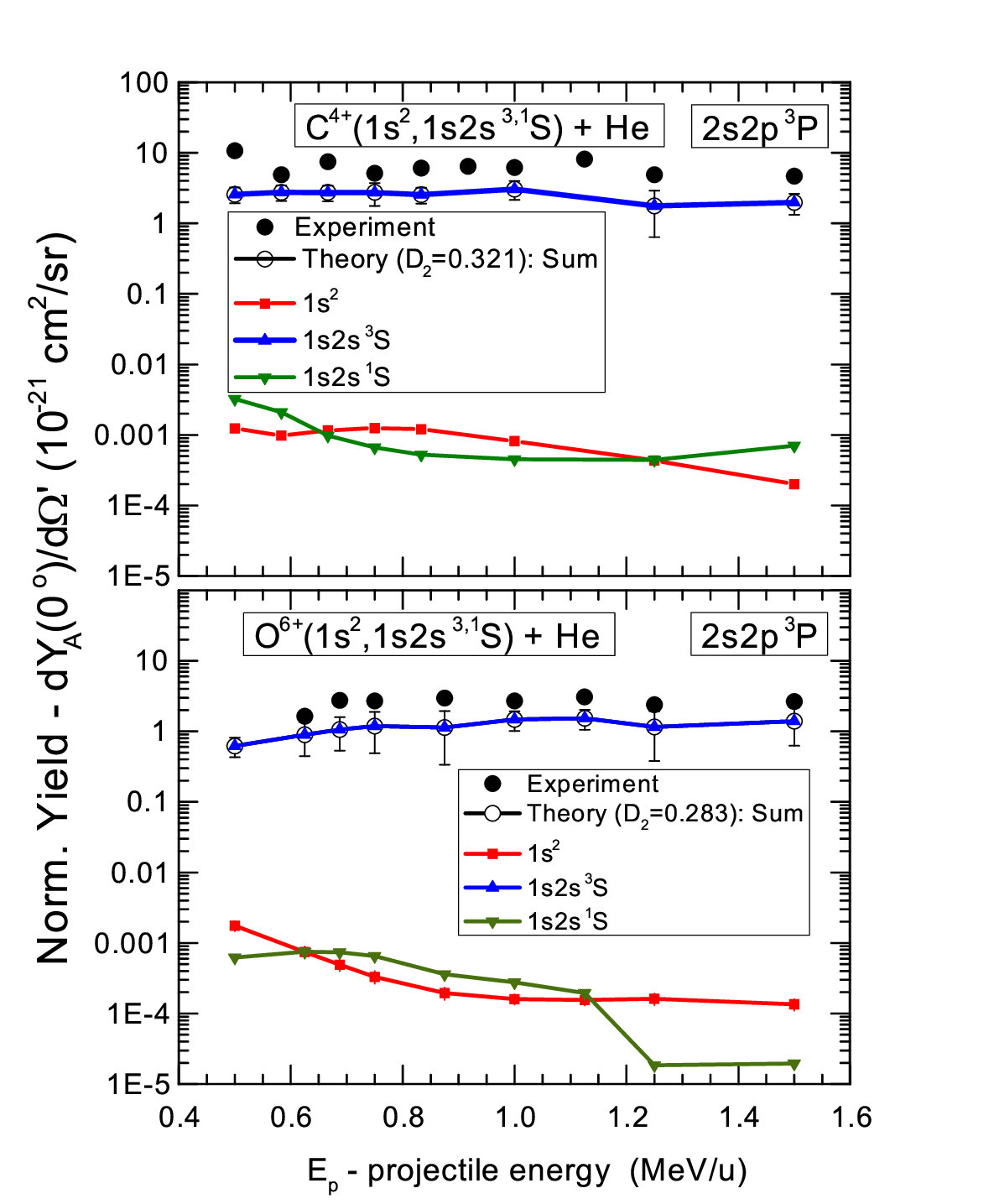}
\caption{\label{fg:NY0C4O62s2p3P}Zero-degree  \CORnew{normalized Auger} yields for the production of the $2s2p\,^3\!P$ from each of the three ion beam components with $D_2=0.321$ for C$^{4+}$ (top) and $D_2=0.283$ for O$^{6+}$ (bottom). Black circles: Measured normalized yields. Contributions from the $1s2s\,^3\!S$ beam component (blue line with triangles) dominate as the sum of the three contributions (black line with circles) hides behind the contribution of the $1s2s\,^3\!S$ beam component. The uncertainty in the theoretical results includes the uncertainties in the beam component fractions (see text) and the 3eAOCC cross sections ($\sim$15\%)  added in quadrature. Error bars shown on the experimental values include only the statistical \REFtwo{uncertainties which are smaller and are mostly hidden under the size of the symbols}.
}
\end{figure}

\begin{figure}
\includegraphics[scale=0.47,angle=0]{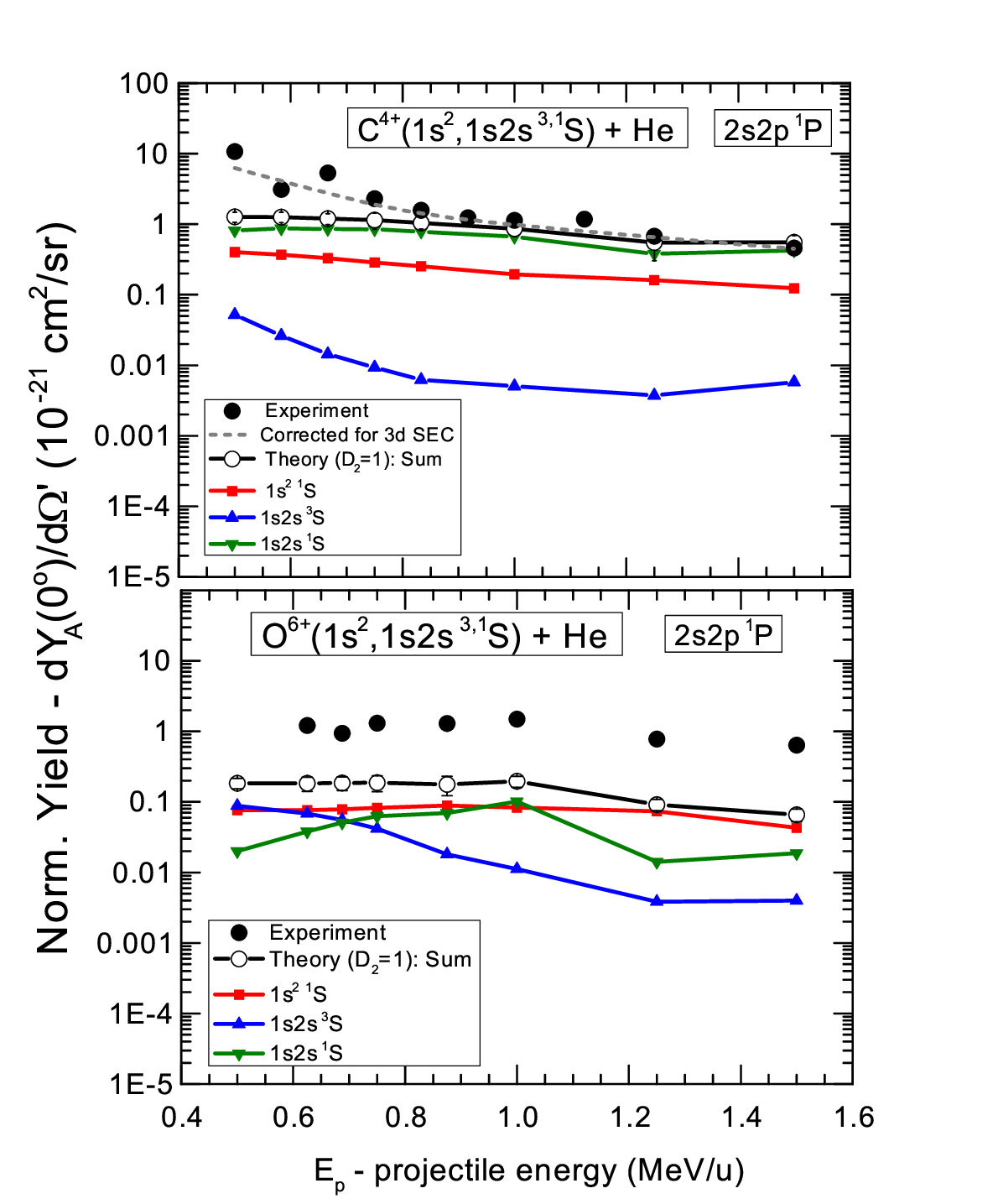}
\caption{\label{fg:NY0C4O62s2p1P}Same as Fig.~\ref{fg:NY0C4O62s2p3P}, but for the $2s2p\,^1\!P$ with $D_2=1$. In the case of carbon (top) contributions from the $1s2s\,^1\!S$ are seen to be the most important, but
the ground state contributions cannot be neglected as they
are only about a factor 3-5 smaller.  \CORnew{Gray dashed line} (top): Estimated contribution of the blended $(1s2s\,^3\!S)3d\,^2\!D$ Auger line due to $3d$ SEC to the $1s2s\,^3\!S$ component has been subtracted (see Appendix~\ref{apx:SEC_to_the_1s2s3d} for more details).
}
\end{figure}

\subsubsection{\label{sec:comp_theo_exp}Comparison to experiment for $\beta_0=0$}
In Figs.~\ref{fg:NY0C4O62s2p3P} and \ref{fg:NY0C4O62s2p1P} the computed normalized Auger yields from each initial state [Eq.~(\ref{eq:dYi})] and their sum $dY_A^\text{tot}(0^\circ)/d\Omega^\prime$  [Eq.~(\ref{eq:sumfisdcsi})] {
for $\beta_0=0$} are compared to the measured normalized Auger yields $dY_A^\text{exp}(0^\circ)/d\Omega^\prime$ for the $2s2p\,^{3}\!P$ and $2s2p\,^{1}\!P$ states. In the case of the carbon $2s2p\,^1\!P$ production [Fig.~\ref{fg:NY0C4O62s2p1P} (top)], the estimated contribution of the $(1s2s\,^3\!S)3d\,^2\!D$ Auger line [produced by single electron capture (SEC) - which could not be resolved - see appendix~\ref{apx:SEC_to_the_1s2s3d}] was subtracted, slightly improving the overall agreement between theory and experiment at the lowest energies.
Numerical results are also listed
in Tables~\ref{tb:C4He2s2pNY} and \ref{tb:O6He2s2pNY} together with the determined fractional beam components.

The calculated total normalized Auger yields $dY_A^\text{tot}(0^\circ)/d\Omega^\prime$ \CORnewnew{(black lines with open circles)} for C$^{4+}(2s2p\,^3\!P)$ are seen in Fig.~\ref{fg:NY0C4O62s2p3P} (top) to be roughly within a factor of 2 of the measured normalized Auger yields and in good overall agreement as to their $E_p$-dependence with what was already reported in Ref.~\cite{lao24a}, where the approximation  in Eq.~(3) \REFone{of Ref.~\cite{lao24a}) was used there,} instead of the full contribution given in Eq.~(\ref{eq:sumfisdcsi})\REFone{, here.} Similarly, for C$^{4+}(2s2p\,^1\!P)$, excellent agreement is found for $E_p>0.8$~MeV/u, but an increasing discrepancy is observed with decreasing $E_p$, even after subtraction of the contamination due to the $(1s2s\,^3\!S)3d\,^2\!D$ state.

The normalized Auger yield calculations for carbon excitation show that the production of the $^3\!P$ is dominantly due to \textit{direct} single excitation from the $1s2s\,^3\!S$ beam component [Eq.~(\ref{eq:2s2p31P_production_3S})] by almost three orders of magnitude justifying the use of the approximate Eq.~(3) in Ref.~\cite{lao24a}. Similarly, the production of the $^1\!P$ state by direct single excitation [Eq.~(\ref{eq:2s2p31P_production_1S})] is also seen to dominate, even though the beam fraction $f[^1\!S]$ is smaller (see Table~\ref{tb:C4He2s2pNY}). However, now the ground state contributions are also seen to be important as they are roughly only about a factor of 2 smaller, mainly due to the about 20 times larger ground state fraction. %

A similar picture is also seen to hold for the oxygen $2s2p\,^3\!P$ state with agreement being slightly better than for carbon as seen in Fig.~\ref{fg:NY0C4O62s2p3P} (bottom). However, for the oxygen $2s2p\,^1\!P$ state, the agreement with experiment is seen quite a bit worse, with experiment being larger by factors of more than 2-5, as seen in Fig.~\ref{fg:NY0C4O62s2p1P} (bottom). In the case of oxygen, the $1s2s\,^1\!S$ component is seen to be much weaker (less ions survive to the target due to the much shorter lifetime), at only about 0.10-0.79\% (see Table~\ref{tb:O6He2s2pNY}). Thus, as also seen for carbon, production from the ground state is now relatively enhanced because of the much larger ground state fraction. Finally, also similarly to carbon, contributions from excitation with exchange (i.e. from the $^3\!S$) is seen to be more than an order of magnitude smaller than from the ground state, particularly at the highest projectile energies.

It is notable that the oxygen $f[^1\!S]$ fractions for $\beta_0=0$ remains very small, primarily due to its significantly shorter lifetime. Consequently, even a substantial increase in its value — by a factor of about 10 — would have minimal impact on the other two components of the He-like ion beam, yet would significantly help in narrowing the observed discrepancy between theory and experiment for $2s2p\,^1\!P$ production. This could occur for $\beta_0$ values larger than 0, as shown in Ref.~\cite{zou25a}.

Our three-component model with $\beta_0=0$ assumes that $^1\!S$ production is tied to $^3\!S$ in a 1:3 ratio, as dictated by statistical considerations~\cite{and92a}. This approach has been applied previously in studies of low-energy capture to He$^+(1s)$ forming metastable He$(1s2l\,^{3,1}\!l)$ states~\cite{ols73a, mcc78a}, capture into He-like $1s2s\,^3\!S$ ions in carbon yielding $1s2l2l'$ states~\cite{mad20a, mad22a}, and dielectronic recombination (DR) measurements of He-like ions with electron coolers~\cite{and92a}. The assumption has also been used in $1s$ loss studies for Be-like carbon and oxygen resulting in $1s2l2l'$ states~\cite{lee92a} and in Li-like low-$Z_p$ ions for $1s2s\,^{3,1}\!S$ states~\cite{ben02a}. However, while plausible, this assumption may warrant further scrutiny as recently spin statistics has been found not always to apply~\cite{rza06a,mad20a,zhu24a}, highlighting the need for additional investigation.
In section~\ref{subsec:results.betaneq0} we explore three-\COR{component} model results with $\beta_0>0$.

\subsubsection{Differences in the Auger angular distributions}
The comparison with experiment \CORnew{is} shown in the previous normalized yield figures only for the computed values of the anisotropy parameter corrected by the dealignment parameter $D_2$. In the case of the $^1\!P$ there is no such correction since this state has no fine structure splitting and therefore $D_2=1$. Typically, most calculations make rather rough estimates of the angular distributions either assuming isotropy and/or ignoring alignment. In Fig.~\ref{fg:C4O62s2p3PJD2} the normalized yields for the three different cases are compared. The largest differences are seen to be of the order of 40\% between isotropic and $D_2=0.321$ and slightly bigger in the case of oxygen. Overall, using the correct $D_2$ value \REFone{appears} to improve the discrepancy between theory and experiment.

\subsubsection{Validity of a one-electron model for the He target for projectile excitation}
As already mentioned our production cross sections for excitation calculated in the 3eAOCC approach assumed a one-electron model for the He target. Thus, the question arises whether it would be more correct to assume further an independent electron approximation (IEA) for the He target and multiply our results by 2 (see also discussion in Ref.~\cite{lao24a}), as was done in the case of SEC~\cite{mad22a} and transfer-excitation~\cite{lao22a}. This correction \COR{appeared} to be justified within an \REFtwo{Oppenheimer-Brinkman-Kramers (OBK) simplified model and therefore was adopted in that work (see section IV.D of Ref.~\cite{lao22a})}. In the case of excitation, the IEA is rather confusing.
The terms appearing in the three-electron (3e) OBK  formulation would also  be found in a four-electron (4e) formulation and  indeed also multiplied by 2 (when shared). However, there are many additional terms appearing in the 4e OBK which are clearly not negligible. Therefore, we feel it is not legitimate to apply a multiplication factor of 2 in this case.
Here, we have tried to include all relevant factors (Auger angular distributions with corrections for fine structure effects and alignment, partial cross sections with dependence on $M_L$, fractional composition of the He-like ion beam) and in our 3eAOCC treatment all couplings to other non-negligible states, as well as contributions from all three initial states.

In addition, it should be reminded that in our 3eAOCC treatment both the interaction with the target nucleus, as well as with the target electron are treated on the same footing and thus included coherently for the first time.
In the past, the factor of 2 has been applied to theoretical calculations of two-center electron-electron interactions (also called electron-electron excitation (eeE)~\cite{zou89b,zou94b,zou96b} or electron impact excitation (EIE)~\cite{gum19a})
performed within the impulse approximation. These eeE results were then added incoherently to the excitation due to the interaction with the target nucleus (referred to as electron-nucleus excitation (enE)~\cite{zou89b} or proton impact excitation (PIE)~\cite{gum19a}).
Clearly, a {\color{black}four-electron AOCC treatment (4eAOCC)} would be more appropriate, but for the time being is just too difficult and time consuming making it impractical for the present.

\REFone{Finally, it is interesting to speculate on the contribution of the target excited and ionized states to the total projectile excitation given the existing discrepancy between theory and experiment. This was \CORnew{already addressed} in section IV.D of Ref.~\cite{lao24a}, where the separate contributions from the helium ground state [He(gs)] and the helium excited plus ionized states [He(exc+ion)] were shown in Fig. 4 of Ref.~\cite{lao24a}. Including more excitation states and positive pseudostates on the target to evaluate further their contribution would be very CPU demanding and presently not viable. However, we do not expect the total projectile excitation cross sections to change much since the observed discrepancy between theory and experiment is seen to be rather constant over most of the collision energy range considered, while contributions from target excitation and ionization were found to be dependent on impact energy and also on the projectile initial state.}

\subsubsection{Cascade contributions}
The possibility of cascade contributions was already discussed~\cite{lao24a} and found negligible in the case of $2s2p\,^3\!P$ excitation, basically due to the low radiative branching ratios for cascade feeding by dipole transitions from higher lying $2snl\,^3\!L$ states which can also be excited. Similarly, higher lying $2snl\,^1\!L$ states can also be excited and can similarly be expected to have small radiative branching ratios. Thus, $2s2p\,^1\!P$ excitation can also be expected to have minimal contributions from cascades. Of course, radiative branching ratios increase roughly as $Z_p^4$ so excitation of higher $Z_p$ projectiles would be increasingly prone to such radiative cascade feeding.

\subsubsection{\label{subsec:results.betaneq0}Comparison to experiment for $\beta_0 > 0$}

\begin{figure}
\includegraphics[scale=0.47,angle=0]{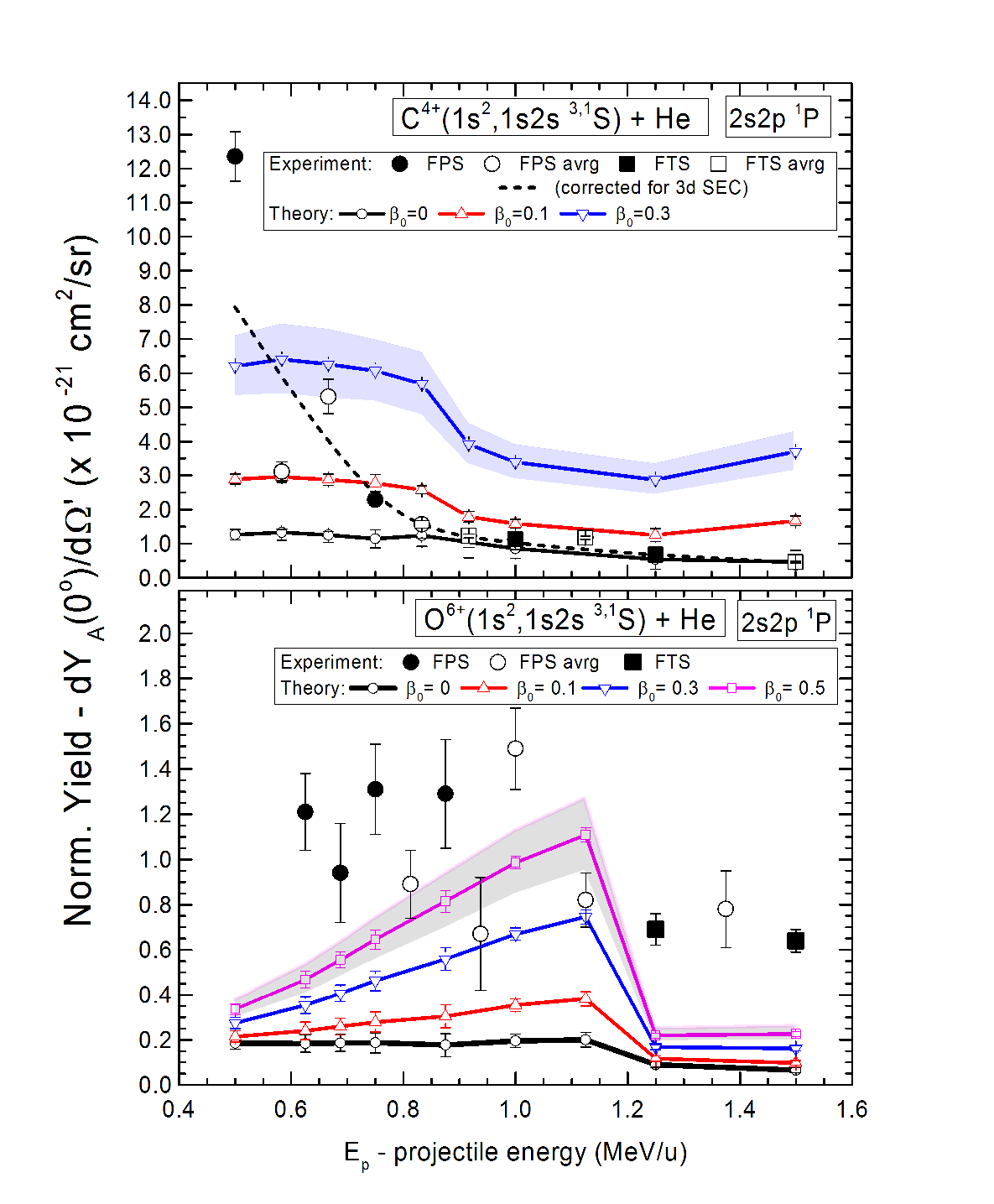}
\caption{\label{fg:NY0C4O62s2p1P_beta0neq0}Comparison of experimental (same as in Fig.~\ref{fg:NY0C4O62s2p1P}, but with last stripper indicated) and theoretical ($D_2=1$ - sum of all three contributions) normalized Auger yields for the production of the $2s2p\,^1\!P$ state in C$^{4+}$ (top) and O$^{6+}$ (bottom) in collisions with helium as a function of projectile energy.
For carbon the dashed line indicates the estimated experimental results after subtraction of $1s2s3d\,^2\!D$ contributions.
The theory has been weighted by the three-component model fractions according to Eq.~\ref{eq:sumfisdcsi} with $f[^1\!S]$ parameterized by $\beta_0$ [see Eqs.~(\ref{eq:f1Si_3c}) and (\ref{eq:f01Sa0f03Splusbeta0})]. The error bars on both experiment and theory are purely statistical. The shaded zones also include a maximum theoretical uncertainty of 15\% in the 3eAOCC calculation of the production cross sections (indicated only for $\beta_0=0.3$ for C$^{4+}$ and $\beta_0=0.5$ for O$^{6+}$). The $\beta_0=0$ results are the same as shown in Fig.~\ref{fg:NY0C4O62s2p1P}. Increasing the value of $\beta_0$ is seen to close the gap between theory and experiment for oxygen, while it does not \COR{appear} to help much in the case of carbon.
}
\end{figure}

The large discrepancy between theory and experiment observed for $\beta_0 = 0$, particularly for the $2s2p\,^1\!P$ level, suggests the need for comparisons using larger values of $f[^1\!S]$, which can be controlled by non-zero values of $\beta_0$. In Ref.~\cite{zou25a}, we demonstrated that using $\beta_0 > 0$ values, while remaining consistent with previous studies, improves this agreement.

In Fig.~\ref{fg:NY0C4O62s2p1P_beta0neq0}, theoretical normalized Auger yields are calculated now also using $\beta_0$ values ranging from 0–30\% for carbon and 0–50\% for oxygen.
As $\beta_0$ increases, $f[^1\!S]$ fractions grow significantly, while $f[^3\!S]$ and $f[1s^2]$ decrease only slightly, by just a few percent~\cite{zou25a}. Therefore, Fig.~\ref{fg:NY0C4O62s2p1P_beta0neq0} presents (on a linear scale) the effect of $\beta_0$ on the final summed yield, $dY_A^\text{tot}/d\Omega^\prime$.

As discussed in Ref.~\cite{zou25a}, no measurements of the $f[^1\!S]$ fraction currently exist, and all value estimates rely on statistical assumptions, such as Eq.~(\ref{eq:f01Sa0f03Splusbeta0}) with $\alpha_0 = 1/3$ and $\beta_0 = 0$ (e.g., see~\cite{and92b}). Non-zero values of $\beta_0$, which imply a larger-than-expected $^1\!S$ fraction, were considered for the first time in Ref.~\cite{zou25a}.

These results suggest that a singlet spin-conserving excitation process from the He-like ground state, specifically $1s^2\,^1\!S\rightarrow 1s2s\,^1\!S$, becomes increasingly favorable for elements with higher $Z_p$. Although the underlying mechanism remains unclear, further isoelectronic studies on this excitation channel would be beneficial for \CORnew{elucidating such a process}.

\section{Summary and Conclusions}
\label{sec:Conclusions}
We have presented theoretical and experimental results for the production of $2s2p\,^3\!P$ and $2s2p\,^1\!P$ states
in 0.5-1.5 MeV/u collisions of He-like mixed-state ($1s^2$, $1s2s\,^3\!S$ and $1s2s\,^1\!S$) carbon and oxygen ion beams with He. A nonperturbative, three-electron treatment was used to calculate the cross sections for the production of these doubly excited states from each of the three possible initial ionic states.
In parallel, the production of these states was also experimentally determined using high-resolution Auger projectile spectrography at $\theta=0^\circ$ with respect to the beam direction. The $1s2s\,^3\!S$ metastable component was also determined experimentally,
while the $1s2s\,^1\!S$ component was assumed to be statistically produced in the ratio of 3:1 according to the $1s2s$ $^3\!S$ to $^1\!S$ spin \COR{multiplicities} and was included in a more complete three-component analysis. The effects of dealignment due to fine structure splitting were also included in the Auger angular distributions at the observation angle of $\theta=0^\circ$. Thus, using this three-component fractional composition the $\theta=0^\circ$ theoretical normalized  yields were determined and compared to the measured mixed-stated normalized  yields.

In all cases, the measured yields were higher than the theoretical predictions. For carbon, the yields exceeded theory by factors ranging from 1.9 to 4.1 for the $2s2p\,^3\!P$ state, and from 0.84 to 6.3 for the $2s2p\,^1\!P$ state. For oxygen, these factors ranged from 1.8 to 2.6 for the $2s2p\,^3\!P$ state, and from 4.1 to 9.7 for the $2s2p\,^1\!P$ state, with the disagreement being significantly larger for the $2s2p\,^1\!P$ state compared to \COR{that for} the $2s2p\,^3\!P$.

An alternative interpretation of the \REFone{above-mentioned disagreements between theory and experiment} could be due to a novel, unknown, mechanism not described in the present close-coupling  calculations, involving eventually, both target electrons, while presently the activity of only one is taken into account  in our  three-electron approach. However, it's hard to imagine how such a mechanism could account for the \CORnew{large up to factor of $\sim$10 disagreement}. Future 4eAOCC calculations with \COR{true two-electron} He targets or measurements using a one-electron target such as atomic hydrogen \REVone{for comparison with 3eAOCC calculations,} should be able to shed more light on this issue and therefore would be of interest.
The larger discrepancy for the $2s2p\,^1\!P$ states may also result from an underestimation of the $^1\!S$ fraction in our three-component model with $\beta_0 = 0$. \CORnew{Increasing the value of $\beta_0$ increases the $^1\!S$ fraction resulting in improved} agreement between theory and experiment, particularly for oxygen. This suggests that the $^1\!S$ fraction might be higher than anticipated in the $\beta_0 = 0$ models.

\REVtwo{In the present study, no clear signature of two-center bi-electronic (TCee) repulsion was observed, in contrast to older investigations involving other collision systems and processes. While tentative explanations for this difference are proposed, it is evident that more systematic studies - particularly along isoelectronic sequences - are needed to clarify these new findings and to resolve the observed disagreement with experiment.} In addition, more work on better defining the amount of the $f[^1\!S]$ fraction in He-like ion beams either theoretically or experimentally would clearly also be very helpful.

\begin{acknowledgments}
We would like to thank the personnel of the “Demokritos”
Tandem for their help with the measurements. We acknowledge partial support of this work by the project “CALIBRA/EYIE” (MIS 5002799) which is implemented under the Action “Reinforcement of the Research and Innovation Infrastructure”, funded by the Operational Programme “Competitiveness, Entrepreneurship and Innovation” (NSRF 2014-2020) and co-financed by Greece and the European Union (European Regional Development Fund).
\end{acknowledgments}
\newpage

\onecolumngrid
\appendix
\addcontentsline{toc}{section}{Appendix}
\section{3eAOCC excitation cross sections}
\label{ap:3eAOCCxsections}
\begin{table}[tbh]
\setlength{\tabcolsep}{5pt}
\centering
{\caption{\label{tb:C4He1_2s2pcsM} Calculated cross sections
for the production of the $2s2p\,^3\!P$ and $2s2p\,^1\!P$ states in collisions of mixed-state $^{12}\text{C}^{4+}(1s^2,1s2s\,^{3,1}\!\!S)$ ion beams with He
as a function of projectile energy $E_p$.
Listed from left to right are the projectile velocity $V_p$ and energy $E_p$, the 3eAOCC partial cross section for $M_L=0$, $\sigma(M_L=0)$, and the total production cross sections $\sigma^\text{tot} =\sigma(M_L=0)+2\sigma(M_L=1)$.
An uncertainty of about 15\% is assigned to all computed cross sections (see text). The notation 4.31[-$\!$1] denotes $4.31\times10^{-1}$.}
\begin{tabular}{cclccccccccc}
\hline\hline\\[-3mm]
&&&\space&\multicolumn{8}{c}{$\,\text{C}^{4+}(1s^2,1s2s\,^{3,1}\!\!S) + \text{He} \rightarrow \text{C}^{4+}(2s2p\,^{3,1}\!P)+ \text{He(All)}$}\\
\cline{5-12}\\[-3mm]
&&&\space&\multicolumn{2}{c}{C$^{4+}(1s^2)$}&\space&    \multicolumn{2}{c}{C$^{4+}(1s2s\,^3\!S)$\footnotemark[2]}&\space&\multicolumn{2}{c}{C$^{4+}(1s2s\,^1\!S)$}\\
\cline{5-6}\cline{8-9}\cline{11-12}\\[-3mm]
$V_p$\,\,\footnotemark[1] & \multicolumn{2}{c}{$E_p$}
&\space& $\,\sigma${\tiny(M$_L\!\!=\!0$)} &  $\sigma^\text{tot}$    &\space&  $\,\sigma${\tiny(M$_L\!\!=\!0$)} &  $\sigma^\text{tot}$
&\space&  $\,\sigma${\tiny(M$_L\!\!=\!0$)} &  $\sigma^\text{tot}$\\[0.3mm]
\cline{2-3}\cline{5-12}\\[-3mm]
(a.u.)&(MeV)&\multicolumn{1}{l}{(MeV/u)}&\space& \multicolumn{7}{c}{[$\times10^{-21}$~cm$^2$]}\\ \hline \\[-2.9mm]
&&&&\multicolumn{8}{l}{\,\underline{$2s2p\,^3\!P$:}} \\
1.291 & 0.5&0.0417&\space& 1.87[-3]   & 1.99[-3]    &\space&\,3.36[-1]   & 4.59[-1]      &\space& 2.40[-1]       & 2.89[-1] \\
1.826 & 1  &0.0833&\space& 1.76[-2]   & 1.86[-2]    &\space&\,\,\,4.95\ph&\,\,5.82\ph    &\space&\,\,4.19\ph     & \,\,4.67\ph\\
2.582 & 2  &0.1667&\space& 4.16[-1]   & 4.70[-1]    &\space& 2.06[1]     & 2.67[1]       &\space&\!\!1.38[1]     & 1.63[1]  \\
3.162 & 3  &0.2500&\space& 1.41[-1]   & 1.89[-1]    &\space& 4.87[1]     & 7.02[1]       &\space&\!\!1.16[1]     & 1.45[1]  \\
3.652 & 4  &0.3333&\space& 5.77[-2]   & 7.79[-2]    &\space& 8.61[1]     & 1.32[2]       &\space&\,6.62\ph       & \,\,\,8.70\ph  \\
4.082 & 5  &0.4167&\space& 2.57[-2]   & 3.13[-2]    &\space& 1.13[2]     & 1.83[2]       &\space&\,3.08\ph       & \,\,4.33\ph  \\
4.472 & 6  &0.5000&\space& 1.13[-2]   & 1.27[-2]    &\space& 1.24[2]     & 2.11[2]       &\space& 6.32[-1]       & \,1.34\ph  \\
4.830 & 7  &0.5833&\space& 9.10[-3]   & 9.88[-3]    &\space& 1.25[2]     & 2.24[2]       &\space& 3.95[-1]       & 8.04[-1] \\
5.164 & 8  &0.6667&\space& 1.07[-2]   & 1.18[-2]    &\space& 1.21[2]     & 2.27[2]       &\space& 1.48[-1]       & 4.07[-1] \\
5.477 & 9  &0.7500&\space& 1.16[-2]   & 1.30[-2]    &\space& 1.14[2]     & 2.24[2]       &\space& 8.60[-2]       & 2.75[-1] \\
5.774 & 10 &0.8333&\space& 1.09[-2]   & 1.26[-2]    &\space& 1.06[2]     & 2.20[2]       &\space& 6.72[-2]       & 2.23[-1] \\
6.325 & 12 &1.000 &\space& 7.63[-3]   & 9.32[-3]    &\space& 9.20[1]     & 2.09[2]       &\space& 6.12[-2]       & 1.96[-1] \\
7.071 & 15 &1.250 &\space& 3.64[-3]   & 4.69[-3]    &\space& 7.53[1]     & 1.92[2]       &\space& 1.12[-1]       & 2.36[-1] \\
7.746 & 18 &1.500 &\space& 1.74[-3]   & 2.25[-3]    &\space& 6.26[1]     & 1.76[2]       &\space& 1.52[-1]       & 2.59[-1] \\
\hline\\[-2.9mm]
&&&&\multicolumn{8}{l}{\,\underline{$2s2p\,^1\!P$:}} \\
1.291 & 0.5&0.0417&\space& 7.48[-4]   & 1.15[-3]    &\space& 1.20[-1]     & 1.36[-1]     &\space&\COR{\,2.44[-1]}&\COR{\!4.62[-1]} \\
1.826 & 1  &0.0833&\space& 4.08[-2]   & 4.73[-2]    &\space& \,\,2.70\ph  & \,\,3.88\ph  &\space&\,\,4.13\ph     &\,\,5.17\ph \\
2.582 & 2  &0.1667&\space& 5.61[-1]   & 6.09[-1]    &\space& \!\!1.60[1]  & \!\!2.06[1]  &\space& \!\!1.66[1]    & \!2.25[1] \\
3.162 & 3  &0.2500&\space& 9.88[-1]   & \,\,1.15\ph &\space& \!\!1.34[1]  & \!\!1.78[1]  &\space& \!\!4.82[1]    & \!7.02[1] \\
3.652 & 4  &0.3333&\space&\,\,1.63\ph & \,\,1.96\ph &\space& \,\,7.36\ph  & \!\!1.03[1]  &\space& \!\!8.77[2]    & 1.34[2] \\
4.082 & 5  &0.4167&\space&\,\,1.98\ph & \,\,2.44\ph &\space& \,\,3.58\ph  & \,\,5.19\ph  &\space& \!\!1.12[2]    & 1.80[2] \\
4.472 & 6  &0.500 &\space&\,\,2.01\ph & \,\,2.52\ph &\space& \,\,1.69\ph  & \,\,2.54\ph  &\space& \!\!1.21[2]    & 2.05[2] \\
4.830 & 7  &0.5833&\space&\,\,1.86\ph & \,\,2.41\ph &\space& 8.25[-1]     & \,\,1.28\ph  &\space& \!\!1.21[2]    & 2.15[2] \\
5.164 & 8  &0.6667&\space&\,\,1.65\ph & \,\,2.22\ph &\space& 4.53[-1]     & 7.22[-1]     &\space& \!\!1.17[2]    & 2.17[2] \\
5.477 & 9  &0.7500&\space& \,\,1.45\ph& \,\,2.03\ph &\space& 2.85[-1]     & 4.66[-1]     &\space& \!\!1.10[2]    & 2.15[2] \\
5.774 & 10 &0.8333&\space& \,\,1.28\ph& \,\,1.86\ph &\space& 1.96[-1]     & 3.36[-1]     &\space& \!\!1.03[2]    & 2.11[2] \\
6.325 & 12 &1.000 &\space& \,\,1.03\ph& \,\,1.62\ph &\space& 1.21[-1]     & 2.39[-1]     &\space& \!\!8.94[1]    & 2.02[2] \\
7.071 & 15 &1.250 &\space& 7.85[-1]   & \,\,1.38\ph &\space& 1.36[-1]     & 2.55[-1]     &\space& \!\!7.38[1]    & 1.87[2] \\
7.746 & 18 &1.500 &\space& 6.27[-1]   & \,\,1.21\ph &\space& 1.67[-1]     & 2.76[-1]     &\space& \!\!6.17[1]    & 1.71[2] \\
\hline\hline
\end{tabular}
\footnotetext[1]{$V_p \text{(a.u.)} \approx 2\,\sqrt{10\,E_p\text{(MeV)}/M_p\text{(u)}}$.}
\footnotetext[2]{From the
{\color{black}$1s2s\,^3\!\!S$
initial beam component the partial cross sections $\sigma(M_L)$ for the production of the $2s2p\,^3\!P$} are the mean of the two contributions from the doublet and quartet total spin of the collision partners $\mathbf{S}_\text{tot}$, i.e.    $\sigma[^3\!S](M_L) = 0.5\sigma[^3\!S](M_L,\mathbf{S}_\text{tot}=3/2) + 0.5\sigma[^3\!S](M_L,\mathbf{S}_\text{tot}=1/2)$, when using a one-electron model for the He target (see text).}
}
\end{table}

\begin{table}[tbh]
\setlength{\tabcolsep}{5pt}
\centering
{
\caption{\label{tb:O6He1_2s2pcsM} Same as Table~\ref{tb:C4He1_2s2pcsM}, but for $^{16}\text{O}^{6+}(1s^2,1s2s\,^{3,1}\!\!S)$ ion beams. {\color{black}Entries indicated by $-$ denote that no result was calculated.}}
\begin{tabular}{cclccccccccc}
\hline\hline\\[-3mm]
&&&\space&\multicolumn{8}{c}{$\,\text{O}^{6+}(1s^2,1s2s\,^{3,1}\!\!S) + \text{He}\rightarrow \text{O}^{6+}(2s2p\,^{3,1}\!P)+ \text{He(All)}$}\\
\cline{5-12}\\[-3mm]
&&&\space&\multicolumn{2}{c}{O$^{6+}(1s^2)$}&\space&    \multicolumn{2}{c}{O$^{6+}(1s2s\,^3\!S)$\footnotemark[2]}&\space&\multicolumn{2}{c}{O$^{6+}(1s2s\,^1\!S)$}\\
\cline{5-6}\cline{8-9}\cline{11-12}\\[-3mm]
$V_p$\,\,\footnotemark[1] & \multicolumn{2}{c}{$E_p$}
&\space& $\,\sigma${\tiny(M$_L\!\!=\!0$)} &  $\sigma^\text{tot}$    &\space&  $\,\sigma${\tiny(M$_L\!\!=\!0$)} &  $\sigma^\text{tot}$
&\space&  $\,\sigma${\tiny(M$_L\!\!=\!0$)} &  $\sigma^\text{tot}$\\[0.3mm]
\cline{2-3}\cline{5-12}\\[-3mm]
(a.u.)&(MeV)&\multicolumn{1}{l}{(MeV/u)}&\space& \multicolumn{7}{c}{[$\times10^{-21}$~cm$^2$]}\\ \hline \\[-2.9mm]
&&&&\multicolumn{8}{l}{\,\underline{$2s2p\,^3\!P$:}} \\
1.826 & 1.33&0.0834    &\space& -        & 3.88[-4] &\space& -        &\,3.14\ph &\space& -       & 4.36[-2] \\
2.582 & 2.67&0.1667    &\space& -        & 1.69[-2] &\space& -        &\,3.27\ph &\space& -       & \,1.49\ph  \\
3.162 & 4.00&0.2500    &\space& -        & 5.36[-2] &\space& -        & 1.32[1]  &\space& -       & \,2.63\ph  \\
3.652 & 5.33&0.3333    &\space& -        & 5.88[-2] &\space& -        & 2.19[1]  &\space& -       & \,2.78\ph  \\
4.082 & 6.67&0.4166    &\space& -        & 4.02[-2] &\space& -        & 2.91[1]  &\space& -       & \,2.36\ph \\
4.472 & 8   &0.5000    &\space& 1.91[-2] & 2.22[-2] &\space& 2.51[1]  & 3.74[1]  &\space&\,1.59\ph& \,2.06\ph \\
5.000 & 10  &0.6250    &\space& 8.11[-3] & 9.63[-3] &\space& 3.39[1]  & 5.37[1]  &\space&\,1.35\ph& \,1.75\ph \\
5.244 & 11  &0.6875    &\space& 5.42[-3] & 6.42[-3] &\space& 3.87[1]  & 6.26[1]  &\space&\,1.14\ph& \,1.49\ph \\
5.477 & 12  &0.7500    &\space& 3.67[-3] & 4.29[-3] &\space& 4.32[1]  & 7.13[1]  &\space& 8.97[-1]& \,1.20\ph \\
5.916 & 14  &0.8750    &\space& 2.09[-3] & 2.35[-3] &\space& 5.01[1]  & 8.60[1]  &\space& 4.94[-1]& 7.26[-1] \\
6.325 & 16  &1.000     &\space& 1.79[-3] & 1.98[-3] &\space& 5.38[1]  & 9.58[1]  &\space& 2.55[-1]& 4.29[-1] \\
6.709 & 18  &1.125     &\space& 1.76[-3] & 1.94[-3] &\space& 5.50[1]  & 1.01[2]  &\space& 1.51[-1]& 2.82[-1] \\
7.071 & 20  &1.250     &\space& 1.68[-3] & 1.89[-3] &\space& 5.29[1]  & 1.01[2]  &\space& 1.17[-1]& 2.18[-1] \\
7.746 & 24  &1.500     &\space& 1.42[-3] & 1.69[-3] &\space& 5.11[1]  & 1.04[2]  &\space& 7.98[-2]& 1.70[-1] \\
\hline\\[-2.9mm]
&&&&\multicolumn{8}{l}{\,\underline{$2s2p\,^1\!P$:}} \\
1.826 & 1.33&0.0834    &\space& -        & 6.83[-4] &\space& -        & 9.75[-2] &\space& -       & 3.86[-1] \\
2.582 & 2.67&0.1667    &\space& -        & 1.04[-2] &\space& -        & 7.71[-1] &\space& -       & \,3.01\ph  \\
3.162 & 4.00&0.2500    &\space& -        & 1.40[-1] &\space& -        &\,2.93\ph &\space& -       & 1.39[1] \\
3.652 & 5.33&0.3333    &\space& -        & 3.40[-1] &\space& -        &\,3.70\ph &\space& -       & 2.37[1] \\
4.082 & 6.67&0.4166    &\space& -        & 4.45[-1] &\space& -        &\,3.12\ph &\space& -       & 3.09[1] \\
4.472 & 8   &0.5000    &\space& 4.01[-1] & 4.69[-1] &\space& \,1.97\ph&\,2.63\ph &\space& 2.61[1] & 3.86[1] \\
5.000 & 10  &0.6250    &\space& 4.08[-1] & 4.96[-1] &\space& \,1.47\ph&\,1.99\ph &\space& 3.53[1] & 5.51[1] \\
5.244 & 11  &0.6875    &\space& 4.23[-1] & 5.21[-1] &\space& \,1.17\ph&\,1.60\ph &\space& 4.03[1] & 6.42[1] \\
5.477 & 12  &0.7500    &\space& 4.40[-1] & 5.47[-1] &\space& 8.89[-1] &\,1.25\ph &\space& 4.50[1] & 7.31[1] \\
5.916 & 14  &0.8750    &\space& 4.54[-1] & 5.75[-1] &\space& 4.76[-1] & 7.21[-1] &\space& 5.20[1] & 8.77[1] \\
6.325 & 16  &1.000     &\space& 4.39[-1] & 5.66[-1] &\space& 2.49[-1] & 4.24[-1] &\space& 5.56[1] & 9.73[1] \\
6.709 & 18  &1.125     &\space& 4.06[-1] & 5.32[-1] &\space& 1.49[-1] & 2.79[-1] &\space& 5.67[1] & 1.03[2] \\
7.071 & 20  &1.250     &\space& 3.67[-1] & 4.88[-1] &\space& 1.13[-1] & 2.15[-1] &\space& 5.61[1] & 1.05[2] \\
7.746 & 24  &1.500     &\space& 2.23[-1] & 3.30[-1] &\space& 9.66[-2] & 1.69[-1] &\space& 5.27[1] & 1.05[2] \\
\hline\hline
\end{tabular}
}
\end{table}

\newpage
\section{Correction of the C$^{4+}(2s2p\,^1\!P)$ normalized yields due to SEC contribution from $(1s2s\,^3\!S)3d\,^2\!D$}
\label{apx:SEC_to_the_1s2s3d}
The $(1s2s\,^3\!S)3d\,^2\!D$ contribution were estimated from our $(1s2s\,^3\!S)3d\,^4\!D$ SEC cross sections (they were assumed to be roughly equal - see Table VII in Ref.~\cite{mad22a}) calculated in a one-electron AOCC (1eAOCC) treatment since our SEC 3eAOCC calculations included only $l=0$ and $l=1$ orbitals~\cite{mad22a}. Furthermore, an isotropic emission was assumed from this state. These 1eAOCC SEC cross sections  $\sigma[(1s2s\,^3\!S)3d\,^2\!D]$ were also multiplied by 2 to further account for the two He electrons, a correction that was shown in previous work~\cite{mad22a} to be in principle justifiable in the case of SEC. {\color{black}Their contribution to the $2s2p\,^1\!P$ normalized yields were thus computed as $2f[^3\!S]\overline{\xi}[(1s2s\,^3\!S)3d\,^2\!D]\sigma[(1s2s\,^3\!S)3d\,^2\!D]/(4\pi)$ with $\overline{\xi}[(1s2s\,^3\!S)3d\,^2\!D]=0.86$~\cite{dev93b}. They are depicted as Gaussians with the corresponding area in  Fig.~\ref{fg:C42s2p} and were subtracted and shown as the dashed line in Figs.~\ref{fg:NY0C4O62s2p1P} (top) and \ref{fg:NY0C4O62s2p1P_beta0neq0} (top).}
These $(1s2s\,^3\!S)3d\,^2\!D$ results can be considered as rough estimates.  Never-the-less they do \COR{appear} to correctly indicate that SEC to this state drops rapidly with increasing projectile energy $E_p$, in agreement with the observed drop in intensity of the other $(1s2s\,^3\!S)3s\,^2\!S$ and $(1s2s\,^3\!S)3p\,^2\!P$
Auger lines {\color{black}seen in Fig.~\ref{fg:C42s2p}} with increasing $E_p$.

\section{$D_2$ dealignment factor and effect of fine structure}
\label{apx:sec_D2}
The dealignment factor $D_2$ appearing in the Auger angular distributions [Eq.~(\ref{eq:a2})] accounts for the average loss of orbital alignment into spin alignment in states having fine structure and is given  by (see Eq.~(20) in Ref.~\cite{mehl80a}):
\begin{align}
D_2 & = \sum_{J,J^\prime=0,1,2}\frac{(2J+1)(2J^\prime+1)}{3}\frac{\left\{\begin{array}{ccc}
                                                                     J & J^\prime & 2 \\
                                                                     1 & 1 & 1
                                                                   \end{array}\right\}^2}{(1+\varepsilon^2_{JJ^\prime})}\label{eq:D2}\\
\varepsilon_{JJ^\prime} & = \frac{\Delta E(J,J^\prime)}{\Gamma(J,J^\prime)}\label{eq:varepsilonJJp}
\end{align}
where $\Delta E(J,J^\prime)$ and $\Gamma(J,J^\prime)$ are defined as:
\begin{align}
\Delta E(J,J^\prime) & = |BE[J]-BE[J^\prime]|\label{eq:DeltaEJJp}\\
\Gamma(J,J^\prime) & = \frac{\Gamma[J]+\Gamma[J^\prime]}{2}.\label{eq:GammaJJp}
\end{align}
Here, $BE[J]$ and $\Gamma[J]$ are the energies and linewidths of the state $2s2p\,^3\!P_J$ with $J=0,1,2$ given in Table~\ref{tb:C42s2p3Pfinestructure}. The parameters  $\varepsilon_{JJ^\prime}$ and $\Gamma_{JJ^\prime}$ have been computed in Table~\ref{tb:D2dealignmentfactor} leading to dealignment factors $D_2=0.321$ and 0.283, for the $2s2p\,^3\!P$ states of carbon and oxygen, respectively.
\begin{table*}[tbh]
\centering
\caption{\label{tb:C42s2p3Pfinestructure}Fine structure details of carbon and oxygen $2s2p\,^3\!P_J$ levels. Entries indicated by $-$ denote that no result was acquired. Lorentzians with the tabulated parameters are depicted in Fig.~\ref{fg:C4O62s2p3PJD2}.
}
\begin{tabular}{cccccccccc}
\hline\hline
Ion & State           & $J$ &$E_\text{res}$ & BE\footnotemark[2]
& $\varepsilon_A$\footnotemark[4] &$\Gamma$\footnotemark[5] & $A_a$\footnotemark[5] & $A_x$\footnotemark[5]
& $\xi$\footnotemark[6]\\
    &            &   & (eV)      & (eV)
     & (eV) & (meV)                             &(s$^{-1}$)                 & (s$^{-1}$)
&                                                 \\
\hline\\[-2.6mm]
C$^{4+}$ & $2s2p\,^3\!P_0$ & 0 & 359.0493\footnotemark[1]  & -224.07174\footnotemark[2] & 265.921 & 9.694\footnotemark[7] & 1.402[13]\footnotemark[7] & 7.058[11]\footnotemark[7] & 0.9521  \\
& $2s2p\,^3\!P_1$ & 1 & 359.0649\footnotemark[1]  & -224.05620\footnotemark[2] & 265.937 & 9.517\footnotemark[7] & 1.375[13]\footnotemark[7] & 7.058[11]\footnotemark[7] & 0.9512  \\
& $2s2p\,^3\!P_2$ & 2 & 359.0992\footnotemark[1]  & -224.02191\footnotemark[2] & 265.971 & 9.360\footnotemark[7] & 1.352[13]\footnotemark[7] & 7.059[11]\footnotemark[7] & 0.9504  \\
& $2s2p\,^3\!P$   & mean &                        & 
& 265.954 & 9.449 &           &           & 0.9508  \\
\hline\\[-2.6mm]
O$^{6+}$  & $2s2p\,^3\!P_0$ & 0 & -                         & -407.60768\footnotemark[3] & 463.802 & 10.81\footnotemark[8] & 1.454[13]\footnotemark[8] & 1.885[12]\footnotemark[9] & 0.8852\\
& $2s2p\,^3\!P_1$ & 1 & -                         & -407.55595\footnotemark[3] & 463.854 & 10.19\footnotemark[8]
& 1.409[13]\footnotemark[8] & 1.381[12]\footnotemark[9]& 0.9107\\
& $2s2p\,^3\!P_2$ & 2 & -                         & -407.44308\footnotemark[3] & 463.967 & 10.47\footnotemark[8]
& 1.366[13]\footnotemark[8] & 2.245[12]\footnotemark[9]  & 0.8589\\
& $2s2p\,^3\!P$   & mean &                        & 
& 463.911 &  10.41\footnotemark[8] &           &           & 0.8791\\
\hline\hline
\end{tabular}
\footnotetext[1]{Resonance energy of the parent ion measured in the photoionization of C$^{4+}(1s2s\,^3\!S)$~\cite{mul18b}, i.e. $\gamma + \text{C}^{4+}(1s2s\,^3\!S) \rightarrow  \text{C}^{4+}(2s2p\,^3\!P) \rightarrow \text{C}^{5+}(1s) + e^-_A$.}
\footnotetext[2]{Absolute binding energy computed from $E_\text{res}$ as: $BE[J] = BE(^3\!S)+E_\text{res}[J]$, where $BE(^3\!S)=-583.12107$~eV is the binding energy of C$^{4+}(1s2s\,^3\!S)$ (given as $-21.430284\cdot27.21014177$~eV  in Table I~\cite{mul18b}), slightly different from the NIST value given in our Table~\ref{tb:energy_levels}.}
\footnotetext[3]{Absolute binding energy computed in eV from energies $E$(au) in Table IV of Zaytsev \textit{et al.}  ~\cite{zay19a} using 1 au = 27.21014177.}
\footnotetext[4]{Auger energy, $\varepsilon_A[J] = BE(1s)-BE[J]$, where $BE(1s)$ is the binding energy of the $(1s)$ configuration~(see Table~\ref{tb:energy_levels}). Center-of-gravity Auger energy computed as $\overline{\varepsilon}_A = \sum_J(2J+1)\varepsilon_A[J]/\sum_J(2J+1)$.}
\footnotetext[5]{$\Gamma[J]$ natural linewidth, $A_a$ Auger  and $A_x$  radiative rates.  Mean total width computed as $\overline{\Gamma} = \sum_J(2J+1)\Gamma[J]/\sum_J(2J+1)$.}
\footnotetext[6]{Auger yield $\xi[J]$.  Mean Auger yield computed as $\overline{\xi}=\sum_J(2J+1)\xi[J]/\sum_J(2J+1)$.}
\footnotetext[7]{M\"{u}ller \textit{et al.}  ~\cite{mul18b} Table I using complex rotation (CR) and
many-body perturbation theory (MBPT).}
\footnotetext[8]{Zaytsev \textit{et al.}  ~\cite{zay19a} Table IV using the complex-scaled configuration-interaction approach within the framework of the Dirac-Coulomb-Breit Hamiltonian.}
\footnotetext[9]{Manai\textit{et al.}  ~\cite{man22a} using AMBiT code~\cite{kah19a} (Particle–hole configuration interaction with many-body perturbation theory (CI+MBPT) for fully relativistic calculations of atomic energy levels).}
\end{table*}
\begin{table}[htb]
\centering
\caption{\label{tb:D2dealignmentfactor}Fine structure parameters used in the computation of the dealignment factor $D_2$ for the carbon and oxygen $2s2p\,^3\!P$ states.}
\begin{tabular}{ccccccc}
\hline\hline
Ionic State & $J$ & $J^\prime$ &$\Delta E(J,J^\prime)$\footnotemark[1] &$\Gamma(J,J^\prime)$\footnotemark[2] & $\varepsilon_{J,J^\prime}$\footnotemark[3] & $D_2$\footnotemark[4]  \\
&    &            &  (meV)                                & (meV)                               &               & \\
\hline\\[-2.46mm]
C$^{4+}(2s2p\,^3\!P_J)$ & 0   & 1          & 15.5                    &  9.606              & 1.62                   & \\
                      & 1   & 2          & 34.3                    &  9.439              & 3.64                   & \\
                      & 0   & 2          & 49.8                    &  9.528              & 5.24                   & \\
                      &     &            &                         &                     &                  & 0.321\\
   \hline\\[-2.46mm]
O$^{6+}(2s2p\,^3\!P_J)$ & 0   & 1          & 51.73                   &  10.50              & 4.92 & \\
                      & 1   & 2          & 112.9                   &  10.33              & 10.9 &\\
                      & 0   & 2          & 164.6                   &  10.64              & 15.5 &\\
                      &     &            &                         &                     &      & 0.283\\
\hline\hline
\end{tabular}
\footnotetext[1]{Fine structure energy splitting $\Delta E(J,J^\prime) = |BE[J] - BE[J^\prime]|$ with binding energies $BE[J]$ from Table~\ref{tb:C42s2p3Pfinestructure}.}
\footnotetext[2]{Mean adjacent widths~\cite{kab94a}, $\Gamma(J,J^\prime) \equiv (\Gamma[J]+\Gamma[J^\prime])/2$ with natural widths $\Gamma[J]$ from Table~\ref{tb:C42s2p3Pfinestructure}.}
\footnotetext[3]{Mehlhorn and Taulbjerg~\cite{mehl80a} overlap parameter
$\varepsilon_{J,J^\prime} \equiv \Delta E(J,J^\prime)/\Gamma(J,J^\prime)$.}
\footnotetext[4]{Mehlhorn and Taulbjerg~\cite{mehl80a} dealignment parameter $D_k$, given for $k=2$ by Eq.~(\ref{eq:D2}).}
\end{table}
\hspace*{-3mm}
\begin{figure*}[tbh]
\includegraphics[scale=0.65,angle=0]{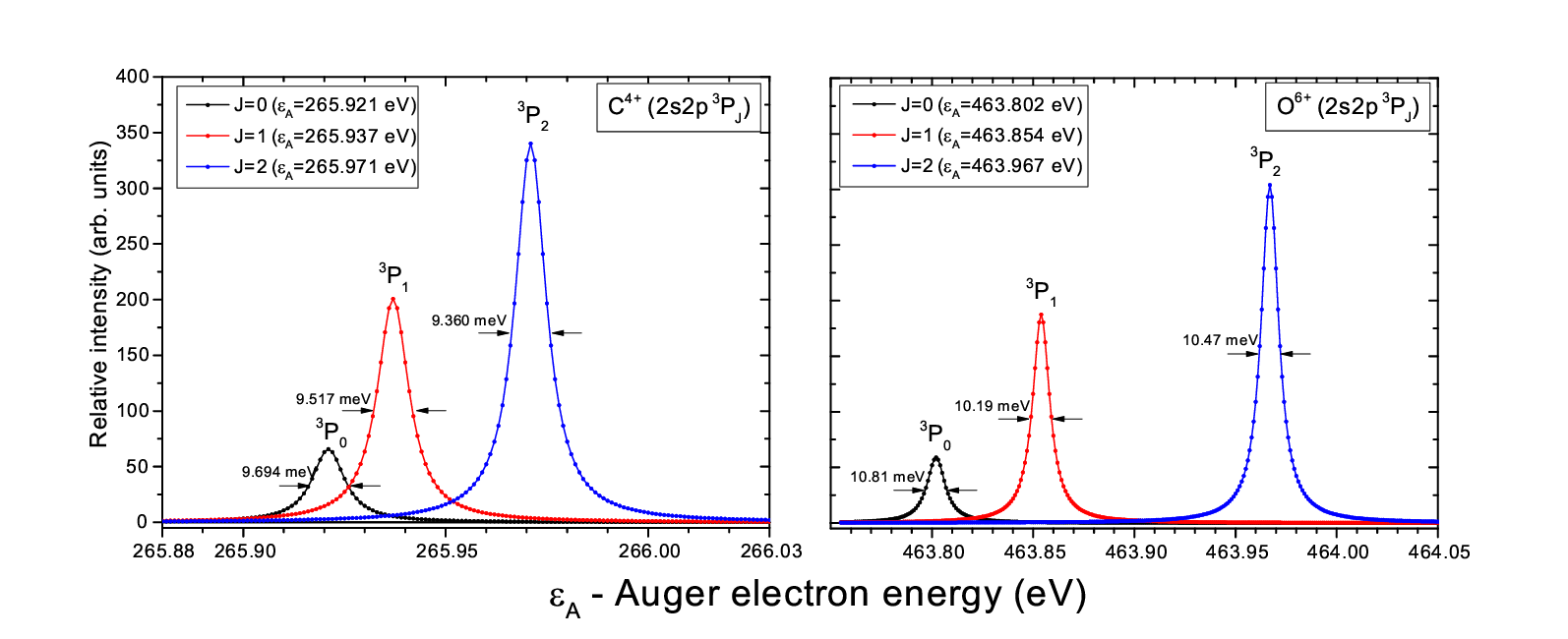}\\[-8mm]
\includegraphics[scale=0.65,angle=0]{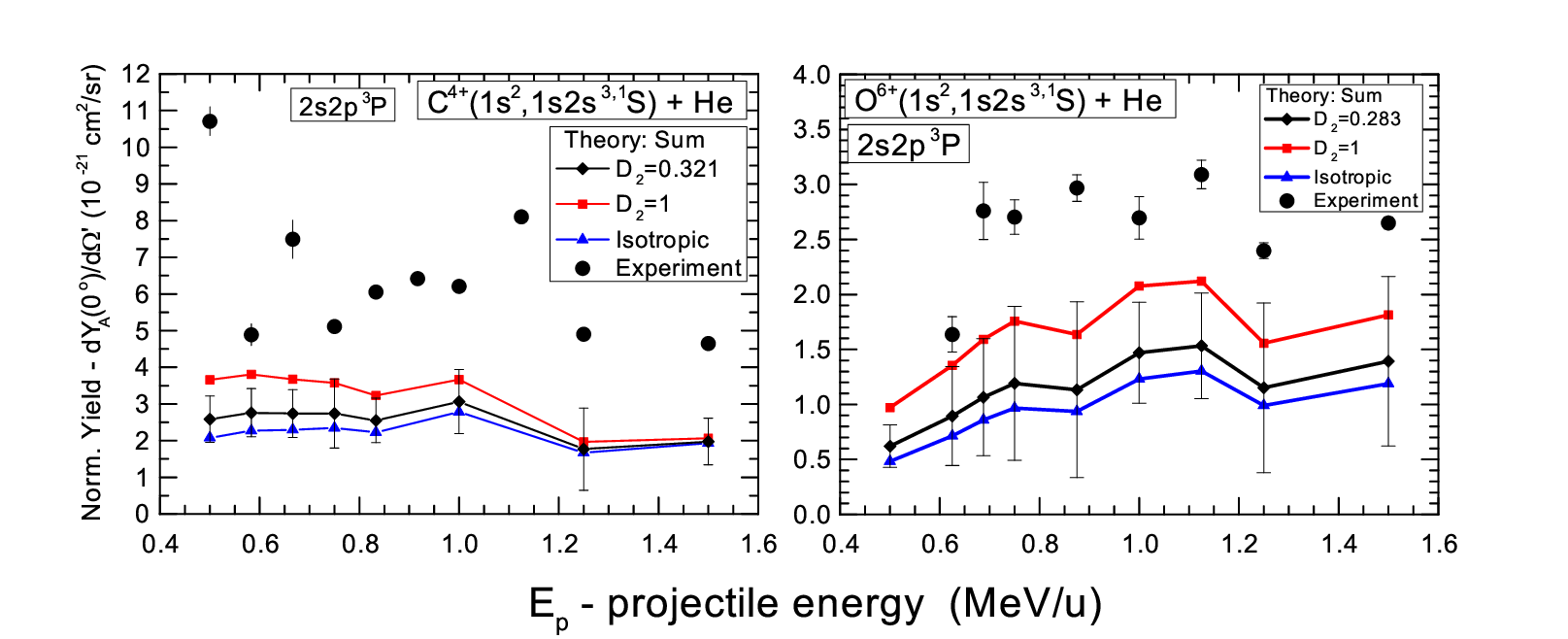}
\caption{\label{fg:C4O62s2p3PJD2}Top: Fine structure of carbon (left) and oxygen (right) $2s2p\,^3\!P_{0,1,2}$ resonances (Lorentzians) using parameters from Table~\ref{tb:C42s2p3Pfinestructure} leads to a dealignment coefficient of $D_2=0.321$ for carbon and $D_2=0.283$ for oxygen. Bottom: Comparison of the three different calculated normalized yields to the experimental data. The degree of relative overlap of the three $J$ levels (top) affects the value of the $D_2$ dealignment coefficient calculated using Eq.~(\ref{eq:D2}).
A value of $D_2=1$ corresponds to maximum overlap or minimal fine structure level separation. A value of $D_2=0$ corresponds to isotropy.
}
\end{figure*}

\clearpage
\section{Normalized Auger yields - Theory and Experiment}
\label{apx:NY}

In Tables~\ref{tb:C4He2s2pNY} and \ref{tb:O6He2s2pNY} the determined fractional beam components and the thereupon computed normalized Auger yields are compared to the experimental yields.

\begin{table*}[tbh]
\centering
\caption{\label{tb:C4He2s2pNY} Theoretical and experimental results
for the production of the $2s2p\,^3\!P$ and $2s2p\,^1\!P$ states in collisions of mixed-state $^{12}\text{C}^{4+}(1s^2,1s2s\,^{3,1}\!\!S)$ ion beams with He as a function of projectile energy $E_p$.
Listed from left to right are the projectile velocity $V_p$ and energy $E_p$, the ion beam stripper combinations used in the measurements (see Sec.~\ref{sec:experimental} for explanations), the fractional composition of the three ion beam $f[1s^2],f[^3\!S],f[^1\!S]$ (for $i=1$), the parameter $\alpha$ [see Eq.~(\ref{eq:ai})], the
sum of the $0^\circ$ normalized  yield contributions from each of component $dY_A^\text{tot}/d\Omega^\prime$ [see Eq.~(\ref{eq:sumfisdcsi})]
and the experimentally determined $0^\circ$ normalized Auger electron yields, $dY_A^\text{exp}/d\Omega^\prime$, respectively. Uncertainties in $dY_A^\text{exp}/d\Omega^\prime$ include just the statistical error, while uncertainties in the $dY_A^\text{tot}/d\Omega^\prime$ include both the computational uncertainty of $\sim$15\% and the listed experimental uncertainties in the ion beam fractions added in quadrature. Entries indicated by $-$ denote that no result was calculated.}
\small
\begin{tabular}{cccccc|ccc||c|c|c||c}
\hline\hline\\[-3mm]
\multicolumn{9}{c||}{$\text{C}^{4+}(1s^2,1s2s\,^{3,1}\!\!S) + \text{He} \rightarrow \text{C}^{4+}(2s2p\,^{3,1}\!P)+ \text{He(All)}$}
&\multicolumn{4}{c}{$0^\circ$ Normalized Auger yields}\\\hline\\[-3mm]
&&&&&&\multicolumn{3}{c||}{Ion beam fractions\footnotemark[1]}&\multicolumn{3}{c||}{Theory\footnotemark[2] - ${\frac{dY_A^{\text{tot}}}{d\Omega^\prime}(\theta=0^\circ)}$}& \multicolumn{1}{c}{Exp.}\\\hline\\[-3mm]
$V_p$ & \multicolumn{2}{c}{$E_p$} & Stripping\footnotemark[3]& $\Delta t_0$\footnotemark[4]& $\alpha^{[1]}$ & $f[1s^2]$\footnotemark[5]& $f[^3\!S]$\footnotemark[6]& $f[^1\!S]$\footnotemark[7]
&\multicolumn{1}{c|}{Isotropic} &\multicolumn{1}{c|}{$D_2$=1}& \multicolumn{1}{c||}{$D_2$=0.321}
& {${\frac{dY_A^\text{exp}}{d\Omega^\prime}}$}\\[0.3mm]
\cline{1-1}\cline{2-3}\cline{5-5}\cline{6-8}\cline{9-13}\\[-2.6mm]
(a.u.) & (MeV) & (MeV/u)             &    Method               & [$\times10^{-6}$s] & (Eq.~\ref{eq:ai}) & \multicolumn{3}{c||}{(\%)}  &\multicolumn{4}{c}{[$\times10^{-21}$~cm$^2$/sr]}\\ \hline \\[-3mm]
&&&&&&&&&\multicolumn{4}{l}{\underline{$2s2p\,^3\!P$ ($\overline{\xi}=0.951$\CORnew{\footnotemark[8]}):}} \\
4.472 & 6 & 0.500 & GTS-FPS &1.287 &0.2181 &84.2(2.6)&13.0(2.5)&2.83(55)  & 2.07(51)  & 3.66(89)  & 2.58(63) &10.71(70)\\
4.839 & 7 & 0.583 &(GTS-FPS)&1.191 &0.2250 &83.6(2.5)&13.4(2.5)&3.01(56)  & 2.27(54)  & 3.81(91)  & 2.76(66) & 4.89(30)\\
5.164 & 8 & 0.667 &(GTS-FPS)&1.114 &0.2308 &83.5(2.5)&13.4(2.5)&3.09(57)  & 2.30(55)  & 3.68(88)  & 2.74(65) & 7.49(52)\\
5.477 & 9 & 0.750 & GTS-FPS &1.051 &0.2357 &83.0(4.4)&13.8(4.3)&3.25(1.01)& 2.34(81)  & 3.57(1.2) & 2.74(94) & 5.11(12)\\
5.774 &10 & 0.833 &(GTS-FPS)&0.997 &0.2400 &83.4(2.6)&13.4(2.5)&3.21(60)  & 2.23(53)  & 3.23(77)  & 2.55(61) & 6.05(13)\\
6.055 &11 & 0.917 & (FTS)   &2.003&0.1722 &83.0(4.2)&14.5(4.1)&2.50(71) &   -       &    -      &     -    & 6.41(5) \\
6.325 &12 & 1.00  & FTS     &1.918&0.1771 &79.3(4.3)&17.6(4.3)&3.11(76) & 2.78(79)  & 3.67(1.1) & 3.06(88) & 6.21(5) \\
6.708 &13.5&1.125 & (FTS)   &1.808&0.1836 &82.8(4.2)&14.5(4.1)&2.67(76) &   -       &   -       &    -     & 8.10(6)\\
7.071 &15  &1.250 & FTS     &1.716&0.1893 &86.3(7.2)&11.5(7.1)&2.18(1.34)& 1.67(1.06)& 1.97(1.25)& 1.77(38) & 4.91(4)\\
7.745&18  &1.500  &(FTS)    &1.566&0.1989 &82.6(4.2)&14.5(4.1)&2.89(82) & 1.93(62)  & 2.06(66)  & 1.97(64) & 4.62(3)  \\
\hline\\[-3mm]
&&&&&&&&&\multicolumn{4}{l}{\underline{$2s2p\,^1\!P$ (${\xi}=0.9948$\CORnew{\footnotemark[9]}):}} \\
4.472 & 6 & 0.500 & GTS-FPS &1.287 &0.2181 &84.2(2.6)&13.0(2.5)&2.83(55)  & 0.65(12) & 1.27(21)  &  & 12.35(72)\\
4.839 & 7 & 0.583 &(GTS-FPS)&1.191 &0.2250 &83.6(2.5)&13.4(2.5)&3.01(56)  & 0.69(12) & 1.26(21)  &  & 3.12(27)\\
5.164 & 8 & 0.667 &(GTS-FPS)&1.114 &0.2308 &83.5(2.5)&13.4(2.5)&3.09(57)  & 0.69(13) & 1.20(21)  &  & 5.32(51)\\
5.477 & 9 & 0.750 & GTS-FPS &1.051 &0.2357 &83.0(4.4)&13.8(4.3)&3.25(1.01)& 0.69(19) & 1.14(30)  &  & 2.29(12)\\
5.774 &10 & 0.833 &(GTS-FPS)&0.997 &0.2400 &83.4(2.6)&13.4(2.5)&3.21(60)  & 0.66(13) & 1.04(19)  &  & 1.57(11)\\
6.055 &11 & 0.917 & (FTS)   &2.003&0.1722 &83.0(4.2)&14.5(4.1)&2.50(71)   &    -     &      -    &  & 1.23(06) \\
6.325 &12 & 1.00  & FTS     &1.918&0.1771 &79.3(4.3)&17.6(4.3)&3.11(76)   &0.60(14)  & 0.859(190)&  & 1.13(04) \\
6.708 &13.5&1.125 & (FTS)   &1.808&0.1836 &82.8(4.2)&14.5(4.1)&2.67(76)   &   -      &   -       &  & 1.18(04)\\
7.071 &15  &1.250 & FTS     &1.716&0.1893 &86.3(7.2)&11.5(7.1)&2.18(1.34) & 0.42(20) &0.545(240) &  & 0.675(26)\\
7.745&18  &1.500  &(FTS)    &1.566&0.1989 &82.6(4.2)&14.5(4.1)&2.89(82)   & 0.47(13) & 0.545(140)&  & 0.460(20)\\
\hline\hline
\end{tabular}
\normalsize
\footnotetext[1]{Both $2s2p\,^3\!P$ and $2s2p\,^1\!P$ Auger lines were measured in the same spectrum so ion beam
conditions were the same for both.}
\footnotetext[2]{Normalized  yields, ${\frac{dY_A}{d\Omega^\prime}}^\text{tot}$,  given by Eq.~(\ref{eq:sumfisdcsi}) for each of the three conditions expressed by Eq.~(\ref{eq:sdcsAugertheta0isotropic}) (isotropic), Eq.~(\ref{eq:sdcsAugertheta0}) ($D_2=1$) and Eq.~(\ref{eq:sdcsAugertheta0D2}) ($D_2=0.321$). For the $2s2p\,^1\!P$, only $D_2=1$ is possible since there is no fine structure in this state.}
\footnotetext[3]{GTS: gas terminal stripper, GPS: gas post-stripper, FTS: foil terminal stripper, FPS: Foil post-stripper. Parentheses [e.g. (FTS)] indicate that the ion beam fractions for this $E_p$ energy were interpolated from the measured values (no parentheses) which were experimentally determined using the \COR{two spectra measurement} technique~\cite{ben16b,ben18b}.}
\footnotetext[4]{Time-of-flight of ion from last post-stripper to the target - see {\color{black}Eq.~(19)} in supplement of Ref.~\cite{zou25a}.}
\footnotetext[5]{Ground state fraction $f[1s^2] = 1 - f[^3\!S] - f[^1\!S]$ with uncertainty $\Delta f[1s^2] = \sqrt{(\Delta f[^3\!S])^2+(\Delta f[^1\!S])^2}$.}
\footnotetext[6]{{\color{black}$1s2s\,^3\!S$ metastable fraction, $f[^3\!S]$, with an uncertainty $\Delta f[^3\!S]$ determined from the statistical uncertainties [see Eq.~(22) in supplement of Ref.~\cite{zou25a}] in the values of the experimentally determined ratios $p$ and $d$ defined in Eq.~(\ref{eq:pdsigma4P2D}).}}
\footnotetext[7]{{\color{black}$1s2s\,^1\!S$ metastable fraction, $f[^1\!S]$, determined from $f[^3\!S]$  according to Eq.~(\ref{eq:f1Si_3c}) for $\beta^{[i]}=0$ with an uncertainty $\Delta f[^1\!S]=\alpha\, \Delta f[^3\!S]$.}}
\footnotetext[8]{Mean Auger yield $\overline{\xi}$ computed from values given in M\"{u}ller \textit{et al.}  ~\cite{mul18b}.}
\footnotetext[9]{Auger yield $\xi=1-K$ computed from values of $K=0.00524$, the radiative branching ratio given in Goryaev \textit{et al.}  ~\cite{gor17b}. A similar value of $K=0.0052$ is also given by van der Hart and Hansen~\cite{van93a}.}
\end{table*}

\begin{table*}[tbh]
\centering
\caption{\label{tb:O6He2s2pNY} Same as Table~\ref{tb:C4He2s2pNY}, but for $^{16}\text{O}^{6+}(1s^2,1s2s\,^{3,1}\!\!S)$ ion beams. Footnotes same as in Table~\ref{tb:C4He2s2pNY}, except where noted.}
\small
\begin{tabular}{cccccc|ccc||c|c|c||c}
\hline\hline\\[-3mm]
\multicolumn{9}{c||}{$\text{O}^{6+}(1s^2,1s2s\,^{3,1}\!\!S) + \text{He} \rightarrow \text{O}^{6+}(2s2p\,^{3,1}\!P)+ \text{He(All)}$}
&\multicolumn{4}{c}{$0^\circ$ Normalized Auger yields}\\\hline\\[-3mm]
&&&&&&\multicolumn{3}{c||}{Ion beam fractions\footnotemark[1]}&\multicolumn{3}{c||}{Theory\footnotemark[2] - ${\frac{dY_A^{\text{tot}}}{d\Omega^\prime}(\theta=0^\circ)}$}& \multicolumn{1}{c}{Exp.}\\\hline\\[-3mm]
$V_p$ & \multicolumn{2}{c}{$E_p$} & Stripping\footnotemark[3] &$\Delta t_0$\footnotemark[4] & $\alpha$
& $f[1s^2]$\footnotemark[5]& $f[^3\!S]$\footnotemark[6]& $f[^1\!S]$\footnotemark[7]
&\multicolumn{1}{c|}{Isotropic} &\multicolumn{1}{c|}{$D_2$=1}& \multicolumn{1}{c||}{$D_2$=0.283}
& {${\frac{dY_A^\text{exp}}{d\Omega^\prime}}$}\\[0.3mm]
\cline{1-1}\cline{2-3}\cline{5-5}\cline{6-8}\cline{9-13}\\[-2.6mm]
(a.u.) & (MeV) & (MeV/u)             &    Method               & [$\times10^{-6}$s] & (Eq.~\ref{eq:ai}) & \multicolumn{3}{c||}{(\%)}  &\multicolumn{4}{c}{[$\times10^{-21}$~cm$^2$/sr]}\\ \hline \\[-3mm]
&&&&&&&&&\multicolumn{4}{l}{\underline{$2s2p\,^3\!P$ ($\overline{\xi}=0.8504$\footnotemark[8]):}} \\
4.472&8 &0.500 &(GTS-FPS)&1.287 & 0.0171 &80.7(5.2) &19.0(5.2) &0.33(09) & 0.48(15) &0.97(30)  &0.62(19) & - \\
5.000&10&0.625 &FTS-FPS  &1.151 & 0.0234 &79.9(9.4) &19.6(9.4) &0.46(22) & 0.71(36) &1.35(68)  &0.89(45) &1.64(16)\\
5.244&11&0.688 &GTS-FPS  &1.097 & 0.0265 &79.2(9.7) &20.3(9.7) &0.54(26) & 0.86(43) &1.59(80)  &1.07(53) &2.76(26)\\
5.477&12&0.750 &GTS-FPS  &1.051 & 0.0295 &79.4(11.4)&20.0(11.4)&0.59(34) & 0.97(57) &1.76(1.03)&1.19(70) &2.70(16)\\
5.701&13&0.813 &(GTS-FPS)&1.009 & 0.0325 &80.4(5.2) &19.0(5.2) &0.62(17) &    -     &    -     &    -    &3.99(17)\\
5.916&14&0.875 &FTS-FPS  &0.9728& 0.0353 &83.4(11.1)&16.1(11.1)&0.57(39) & 0.94(66) &1.64(1.15)&1.13(80) &2.97(12)\\
6.124&15&0.938 &(FTS-FPS)&0.9398& 0.0381 &80.3(5.2) &19.0(5.2) &0.73(20) &     -    &    -     &    -    &4.50(21)\\
6.325&16&1.000 &(GTS-FPS)&0.9099& 0.0409 &80.2(5.2) &19.0(5.2) &0.78(21) & 1.23(38) &2.07(65)  &1.47(46) &2.70(19)\\
6.708&18&1.125 &(GTS-FPS)&0.8579& 0.0461 &80.1(5.2) &19.0(5.2) &0.88(24) & 1.30(41) &2.12(66)  &1.53(77) & 3.09(13)\\
7.071&20&1.250 & FTS     &1.652 & 0.0074 &85.4(9.5) &14.5(9.5) &0.11(07) & 0.99(66) &1.56(1.04)&1.15(77) & 2.40(07)\\
7.416&22&1.375 &(GTS-FPS)&0.7430& 0.0600 &79.9(5.2) & 19.0(5.2)&1.14(31) &      -   &    -     &    -    & 5.40(21) \\
7.745&24&1.500 &FTS      &1.581 & 0.0087 &82.3(9.3) &17.5(9.3) &0.15(08) & 1.19(66) &1.82(1.00) &1.39(77) & 2.65(05)\\
\hline\\[-3mm]
&&&&&&&&&\multicolumn{4}{l}{\underline{$2s2p\,^1\!P$ (${\xi}=0.9848$\CORnew{\footnotemark[9]}):}} \\
4.472&8 &0.500 &(GTS-FPS)&1.287 & 0.0171 &80.7(5.2) &19.0(5.2) &0.33(09) &0.079(14)&0.184(31)&         & - \\
5.000&10&0.625 &FTS-FPS  &1.151 & 0.0234 &79.9(9.4) &19.6(9.4) &0.46(22) &0.082(19)&0.183(42)&         & 1.21(17)\\
5.244&11&0.688 &GTS-FPS  &1.097 & 0.0265 &79.2(9.7) &20.3(9.7) &0.54(26) &0.085(20)&0.186(47)&         & 0.94(22)\\
5.477&12&0.750 &GTS-FPS  &1.051 & 0.0295 &79.4(11.4)&20.0(11.4)&0.59(34) &0.087(24)&0.186(47)&         & 1.31(20)\\
5.701&13&0.813 &(GTS-FPS)&1.009 & 0.0325 &80.4(5.2) &19.0(5.2) &0.62(17) &    -    &    -    &         & 0.89(15)\\
5.916&14&0.875 &FTS-FPS  &0.9728& 0.0353 &83.4(11.1)&16.1(11.1)&0.57(39) &0.086(29)&0.176(54)&         & 1.29(24)\\
6.124&15&0.938 &(FTS-FPS)&0.9398& 0.0381 &80.3(5.2) &19.0(5.2) &0.73(20) &    -    &    -    &         & 0.67(25)\\
6.325&16&1.000 &(GTS-FPS)&0.9099& 0.0409 &80.2(5.2) &19.0(5.2) &0.78(21) &0.101(20)&0.195(35)&         & 1.49(18)\\
6.708&18&1.125 &(GTS-FPS)&0.8579& 0.0461 &80.1(5.2) &19.0(5.2) &0.88(24) &0.108(23)&0.200(39)&         & 0.82(12)\\
7.071&20&1.250 & FTS     &1.652 & 0.0074 &85.4(9.5) &14.5(9.5) &0.11(07) &0.044(09)&0.092(17)&         & 0.69(07)\\
7.416&22&1.375 &(GTS-FPS)&0.7430& 0.0600 &79.9(5.2) &19.0(5.2) &1.14(31) &    -    &    -    &         & 0.78(17) \\
7.745&24&1.500 &FTS      &1.581 & 0.0087 &82.3(9.3) &17.5(9.3) &0.15(08) &0.036(08)&0.066(13)&         & 0.64(05)\\
\hline\hline
\end{tabular}
\normalsize
\footnotetext[2]{Normalized yields, ${\frac{dY_A}{d\Omega^\prime}}^\text{tot}$,  given by Eq.~(\ref{eq:sumfisdcsi}) for each of the three conditions expressed by Eq.~(\ref{eq:sdcsAugertheta0isotropic}) (isotropic), Eq.~(\ref{eq:sdcsAugertheta0}) ($D_2=1$) and Eq.~(\ref{eq:sdcsAugertheta0D2}) ($D_2=0.283$). For the $2s2p\,^1\!P$ only $D_2=1$ is possible since there is no fine structure in this state.}
\footnotetext[8]{Mean Auger yield $\overline{\xi}$ computed from values given in Goryaev \textit{et al.}  ~\cite{gor17b}.}
\footnotetext[9]{Auger yield $\xi=1-K$ computed from values of $K=0.0152$, the radiative branching ratio given in Goryaev \textit{et al.}  ~\cite{gor17b}.}
\end{table*}


\section{Carbon and Oxygen Auger line identification and energy level diagrams}
\label{apx:ecal}
For accurate spectroscopic work when dealing with Auger emission from projectiles with velocities in the MeV/u range it is important to use the special relativistic electron energy transformations from the laboratory to the projectile rest frame and back. For known Auger electron  energy $\varepsilon^\prime$, the laboratory electron energy $\varepsilon_\pm(\theta)$ at the observation of $\theta$ is given in Doukas \textit{et al.}  ~\cite{dou15a}. Evaluating for our needs here at the $\theta=0^\circ$ ($\theta^\prime = 0^\circ$ or $180^\circ$) laboratory observation angle we have:
\begin{align}
\varepsilon_\pm(0^\circ) &= \gamma_p \varepsilon^\prime + t_p \pm \sqrt{(1+\gamma^\prime)\varepsilon^\prime(1+\gamma_p)t_p}\label{eq:elabrel}
\end{align}
where, in the projectile rest frame, the $(+)$ sign corresponds to forward emission ($\theta^\prime=0^\circ$) and the $(-)$ sign to backward emission ($\theta^\prime=180^\circ$). Primed quantities refer to the projectile rest frame, while unprimed to the laboratory frame.
The reverse transformations are also given as:
\begin{align}
\varepsilon^\prime &= \gamma_p\varepsilon_\pm(0^\circ) + t_p - \sqrt{(1+\gamma)\varepsilon_\pm(0^\circ)(1+\gamma_p)t_p}\label{eq:erestrel}
\end{align}
where the three relativistic $\gamma$ factors have the usual definitions:
\begin{align}
\gamma_p \equiv 1+\frac{t_p}{mc^2},\quad
\gamma \equiv 1+ \frac{\varepsilon}{mc^2},\quad
\gamma^\prime \equiv 1+\frac{\varepsilon^\prime}{mc^2}\label{eq:gammas}
\end{align}
{with $t_p$, the reduced projectile energy (also known as the cusp energy), given by}
\begin{align}
t_p = \frac{m}{M_p}E_p\label{eq:tp}
\end{align}
where $m$ and $M_p$ are the masses of the electron and the projectile, respectively.
In the limit of the relativistic factors going to 1 we obtain the well-known classical results.
As an example, we note that for the case of \REFtwo{1.5~MeV/u} carbon ions ($t_p=822.870$~eV)  and a $2s2p\,^3\!P$  Auger energy $\varepsilon^\prime=265.95$~eV, the difference between the relativistic and classical laboratory energies amounts to more than 3~eV and is readily observable with high-resolution spectrometers.

\begin{figure*}[tbh]
\includegraphics[scale=0.57,angle=0]{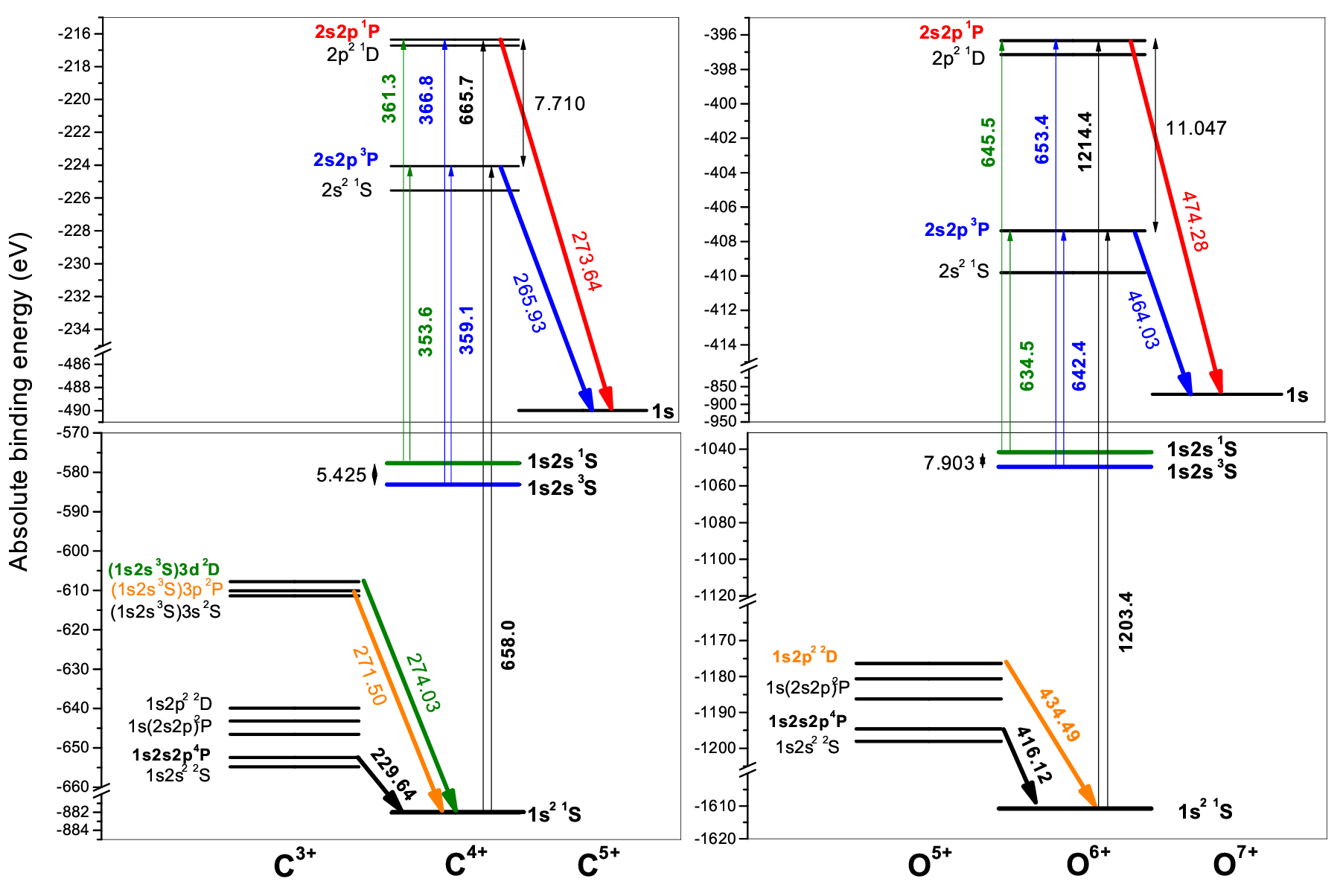}
\caption{\label{fg:Absolute_BE_C3O5_C4O6_Auger}Absolute binding energies of relevant carbon (left) and oxygen (right) levels. Auger transitions and their energies in eV (downward arrows slanted to the right). Bottom panel (left): Black C$^{3+}(1s2s2p\,^4\!P)$ calibration line~\cite{bru85a}, green C$^{3+}[(1s2s\,^3\!S)3d\,^2\!D]$, orange C$^{3+}[(1s2s\,^3\!S)3p\,^2\!P]$. Top panels: Blue  $(2s2p\,^3\!P)$ and red  $(2s2p\,^1\!P)$ Auger transitions. The six excitation energies from each of the three initial ion beam components $1s^2$, $1s2s\,^3\!S$ and $1s2s\,^3\!S$ to the two final states $2s2p^1\!P$ and $2s2p^3\!P$ are also shown (thin upward pointing arrows). The levels $1s2s3l$ are not indicated for oxygen (bottom right) as in the case of carbon because as shown in Table~\ref{tb:energiesOxygen} their Auger energies are quite a bit larger than the $2s2p\,^{3,1}\!P$ and therefore not a problem to identify as in the case of carbon.}
\end{figure*}

Figure~\ref{fg:Absolute_BE_C3O5_C4O6_Auger} presents the relevant energy level diagrams for carbon and oxygen, while the corresponding Auger energies  are listed in Tables~\ref{tb:energiesCarbon} and \ref{tb:energiesOxygen}, respectively.

\begin{table*}[tbh]
\caption{\label{tb:energiesCarbon}Carbon $K$-Auger energies $\varepsilon_A$ listed in increasing energy (eV) resulting from $1s2lnl^\prime \rightarrow 1s^2$ and $2l2l^\prime \rightarrow 1s$ Auger transitions used in the identification of our observed Auger lines (this work). The former are used for energy calibration, while the latter are in the vicinity of the $2s2p\,^{3,1}\!P$ lines. There are no NIST~\CORnew{\cite{kra24a}} recommended values for the energy levels of these doubly excited states. Entries indicated by $-$ denote that no result was acquired. The footnote in the header of each column gives the reference from which the values shown in the column were obtained, unless otherwise indicated. Experimental uncertainties as reported in the corresponding reference. For conversions to eV, we have used the NIST equivalents, 1~a.u. $=27.211386245988(53)$~eV and 1~$cm^{-1} = 1.239842\times10^{-4}$~eV, unless otherwise indicated.}
{\small
\begin{tabular}{r|ccccc|c|cccccc}
\hline\hline
           & \multicolumn{5}{c|}{Experiment} &               & \multicolumn{6}{c}{Theory}\\
           &This &&&&&Calibration& &&&&&\\
State      &work\footnotemark[1] & Rod79\footnotemark[2] & Mac87\footnotemark[3] & Mann87\footnotemark[4] & Kil93\footnotemark[5]
& Values   & Aln02\footnotemark[7]  &Kar09\footnotemark[8] & Gor17\footnotemark[9] & Yer17\footnotemark[10] &Man22\footnotemark[11]&Man22\footnotemark[12]\\
\hline\\[-2.5mm]
$1s2s^2\,^2\!S$   &227.2(6) &227.6(5)&227.06(9)&227.1(2)&-       &227.23(30)\footnotemark[6]&227.1& -       & 227.00  & 227.208 &-& - \\
$1s2s2p\,^4\!P$   &229.6(5)&229.7(5)&\!\!229.639\footnotemark[16]&229.6(2)&-       &\!\!\!\!\!\!\!\!\!\!229.64\footnotemark[6]    &229.5& -       & 229.80  & 229.695 &-& - \\
$1s2s2p\,^2\!P_-$ &235.5(6)&235.5(5)&235.40(4)&235.5(2)&234.3(1)&235.44(20)\footnotemark[6]&235.3& -       & 235.41  & 235.572 &-& - \\
$1s2s2p\,^2\!P_+$ &238.8(6)&238.9(5)&238.92(4)&238.8(2)&237.8(3)&238.86(20)\footnotemark[6]&239.0& -       & 238.97  & 239.024 &-& -\\
$1s2p^2\,^2\!D$   &242.2(6)&242.2(6)&241.98(4)&242.1(2)&241.4(1)&242.15(20)\footnotemark[6]&242.0& -       & 242.18  & 242.099 &-& - \\
\hline\\[-2.5mm]
$2s^2\,^1\!S$     &264.4(6)& -      & -       & -      &-       & -                        &264.2&264.457 &264.30   & 264.45\footnotemark[13] &263.936 &264.417\\
$2s2p\,^3\!P$     &265.9(6) & -     & -       & -      &-       &265.954(1)\footnotemark[14]&265.7& -       &265.94  & 266.02\footnotemark[13] &265.837 &265.962 \\
$(1s2s\,^3\!S)3s\,^2\!S$ &270.6(6)&-&270.70(15)&270.7(2)&-      &-    &271.5 & -       &270.57 &-                       &-       & -\\
$(1s2s\,^3\!S)3p\,^2\!P$ &271.9(6)&-&271.98(10)&271.8(2)&-      &\!\!\!\!\!\!\!\!\!\!\!\!\!\!272.1\footnotemark[6]&272.4 & -       &271.50 &-                       &-       & - \\
$2p^2\,^1\!D$            &273.1(6)&-& -        & -      &-      &-    &272.4 & 273.157 &272.99 &273.27\footnotemark[15] &273.81 &273.141 \\
$2s2p\,^1\!P$            &273.8(7)&-& -        & -      &-      &-    &273.5 & 273.927 &273.64 &273.92\footnotemark[13] &274.289 &273.741 \\
$(1s2s\,^3\!S)3d\,^2\!D$ &274.2(6)&-&274.29(10)&274.2(2)&-      & \!\!\!\!\!\!\!\!\!\!\!\!\!\!274.1\footnotemark[6]& -    & -       &274.02 &-                       &-       & - \\
$2p^2\,^1\!S$            &   -  & - &      -         & -   &-    &-   &-                     &- &283.029 & 281.810  \\
\hline\\[-2.5mm]
$(1s2s\,^1\!S)3s\,^2\!S$ &274.8(6)&-&-         & -      &-      &-    & -      & - &274.59 &-                       &-       & - \\
$(1s2s\,^1\!S)3p\,^2\!P$ &276.9(6)&-&-         & -      &-      &-    & -      & - &276.52 &-                       &-       & - \\
$(1s2s\,^1\!S)3d\,^2\!D$ &278.5(6)&-&  -       &278.9(2)&-      &\!\!\!\!\!\!\!\!\!\!\!\!\!\!278.7\footnotemark[6]& -      & - &278.43 &-                       &-       & - \\
\hline\hline
\end{tabular}
}
\footnotetext[1]{Fitted Auger line peak energies after energy calibration of PSD channels according to the $1s2l2l^\prime$ calibration values proposed by Bruch \textit{et al.}  ~\cite{bru85a} and listed in column seven.}
\footnotetext[2]{Rodbro \textit{et al.}   1979~\cite{rod79a} in 300~keV C$^++$CH$_4$.
}
\footnotetext[3]{Mack 1987~\cite{mack87c} weighted averages (Table 3) calibrated to the $1s2s2p\,^4\!P$ calculation of K.T. Chung~\cite{chu84b}.
}
\footnotetext[4]{Mann 1987~\cite{mann87b}.
}
\footnotetext[5]{Kilgus etl 1993~\cite{kil93a} - Dielectronic Recombination (DR) measurements at the Heidelberg Test Storage Ring (TSR).}
\footnotetext[6]{Bruch \textit{et al.}   1985~\cite{bru85a} - Proposed calibration of carbon $K$-Auger energies based on the measurements by Rodbro \textit{et al.}  ~\cite{rod79a} and theory.}
\footnotetext[7]{Alnaser 2002~\cite{aln02b} (from Tables 1 and 6) using $\varepsilon_A = Z_p^2 E_0 + Z_p E_1 + E_2$ with coefficients for each state $E_0,E_1,E_2$ from Rodbro \textit{et al.}  ~\cite{rod79a}.}
\footnotetext[8]{Kar and Ho 2009~\cite{kar09a} - Stabilization method: $2l2l^\prime\, ^1\!L$ levels in a.u.}
\footnotetext[9]{Goryaev \textit{et al.}   2017~\cite{gor17b} - MZ code with relativistic corrections: $1s2l2l^\prime$ and $2l2l^\prime$ levels in keV.
$\varepsilon_A(1s2snl) = \Delta E_x(1s2snl \rightarrow 1s^22p)+E(1s^22p)-E(1s^2)$,
$\varepsilon_A(2s2p)  = \Delta E_x(2s2p\,^{3,1}\!P_J \rightarrow 1s2s\,^{3,1}\!S)+E(1s2s\,^{3,1}\!S)-E(1s)$,
where $\Delta E_x$ is the value of the x-ray transition energy given in Ref.~\cite{gor17b} and
$E(1s^2), E(1s), E(1s2s\,^{3,1}\!S)$ and $E(1s^22p)$ the energy levels
given in Table~\ref{tb:energy_levels}.}
\footnotetext[10]{Yerokhin \textit{et al.}  2017~\cite{yer17a} - Relativistic configuration-interaction calculation of transition wavelengths. Used $hc=1.23984198\times10^{-4}$~eV-cm to convert wavelength (cm) to energy (eV). See also similar, but older MCDF results by Safronova and Bruch 1994~\cite{saf94a}.}
\footnotetext[11]{Manai \textit{et al.}  2022~\cite{man22a} $E_\text{FAC}$ energy levels computed with respect to the $1s^2$ ground state using the Flexible Atomic code (FAC)~\cite{gu08a}. $\varepsilon_A(2l2l^\prime\,^{3,1}\!L_J) = E_\text{FAC}(2l2l^\prime\,^{3,1}\!L_J) + E(1s^2) - E(1s)$.}
\footnotetext[12]{Manai \textit{et al.}  2022~\cite{man22a} $E_\text{AMBiT}$ energy levels computed with respect to the $1s^2$ ground state using the AMBiT code~\cite{kah19a} (Particle–hole configuration interaction with many-body perturbation theory (CI+MBPT) for fully relativistic calculations of atomic energy levels). $\varepsilon_A(2l2l^\prime\,^{3,1}\!L_J) =
E_\text{AMBiT}(2l2l^\prime\,^{3,1}\!L_J) + E(1s^2) - E(1s)$.}
\footnotetext[13]{Ho 1981~\cite{ho81a} - complex rotation calculations as quoted in  Table 4.5 of Mack 1987~\cite{mack87d}.}
\footnotetext[14]{M\"{u}ller \textit{et al.}  2018~\cite{mul18b} - The $2s2p\,^3\!P_J$ resonance energies were obtained in photoionization measurements of C$^{4+}(1s2s\,^3\!S)$ ions after fitting to theory from which the listed Auger energies were determined (see also Table~\ref{tb:C42s2p3Pfinestructure}). This is probably the most accurate determination to date and should be used for calibration.}
\footnotetext[15]{Peacock \textit{et al.} 1973~\cite{pea73a} - Hartree-Fock type calculations as quoted in Table 4.5 of Mack 1987~\cite{mack87d}.}
\footnotetext[16]{K.T. Chung 1984~\cite{chu84b} - Hartree-Fock with relativistic, Breit-Pauli operator and mass-polarization corrections.}
\end{table*}

\begin{table*}[tbh]
\caption{\label{tb:energiesOxygen}Same as Table~\ref{tb:energiesCarbon}, but for oxygen. Auger energies computed from NIST recommended energy level values are listed in the last column.}
{\small
\begin{tabular}{r|cccc|ccccccc|c}
\hline\hline
 & \multicolumn{4}{c|}{Experiment} & \multicolumn{7}{c|}{Theory}& \\
 & This & & & &&& &&&&\\
State & work\footnotemark[1] & Mac87\footnotemark[2] & Bru87\footnotemark[3] & Kil90\footnotemark[4] & Bru87\footnotemark[3]  & Aln02\footnotemark[5] & Kar09\footnotemark[6] & Gor17\footnotemark[7] &Yer17\footnotemark[8] &Man22\footnotemark[9]& Man22\footnotemark[10]& NIST\footnotemark[11]\\
\hline\\[-2.5mm]
$1s2s^2\,^2\!S$   &412.3(7)&412.67(8)             &412.7(2)& -&412.63&412.4&-&412.50&412.603&- & -     & -\\
$1s2s2p\,^4\!P$   &416.4(7)&\!\!\!\!\!\!\!\!416.08&416.0(2)& -&416.02&415.5&-&416.12&415.973&- &-      &416.124 \\
$1s2s2p\,^2\!P_-$ &425.1(7)&424.81(8)             &425.0(2)& 424.9371(31)\footnotemark[14] &424.99&424.4&-      &424.91&424.945&-       &-      &424.474\\
$1s2s2p\,^2\!P_+$&428.8(7)&\,\,429.38(15) &429.6(2)& -       &429.71&429.4&-      &429.63&429.601&-       &-      &430.094 \\
$1s2p^2\,^2\!D$   &434.6(7)&434.31(8) &434.4(2)& -       &434.38&434.6&-      &434.49&434.313&-       &-      &434.382\\
\hline\\[-2.5mm]
$2s^2\,^1\!S$     &461.5(7)& -        &463(2)\footnotemark[12] &461.9(9) &462.3\footnotemark[12]&-    &462.080&461.60&-      &461.197 &461.924&-\\
$2s2p\,^3\!P$\footnotemark[13]     &463.7(7)& -        &466(2)\footnotemark[12] &463.9(1) &464.4\footnotemark[12] &463.3&-      &463.78&-      &463.677 &464.002&464.029\\
$2p^2\,^1\!D$     &473.8(7)& -        &\!\!\!\!\!\!\!\!471\footnotemark[12]    &474.1(1) &474.8\footnotemark[12] & -    &474.145&473.86&-      &474.583 &474.143&474.275 \\
$2s2p\,^1\!P$\footnotemark[13]     &474.7(7)& -        &477(2)\footnotemark[12] &\,\,476.5(12)&476.0\footnotemark[15] &474.1&475.230&474.79& -     &475.299 &475.034&475.076 \\
$2p^2\,^1\!S$     &485.7(7)& -&-&485.8(1) &-&  -&-      &485.73& -     &487.042 &485.971&480.204 \\
\hline\\[-2.5mm]
$(1s2s\,^3\!S)3s\,^2\!S$   &500.3(7)&500.4(2)&-       & -       &500.5\footnotemark[12]     &500.8&-      & -     & -      & -      &   -   &  - \\
$(1s2s\,^3\!S)3p\,^2\!P$   &501.8(7)&501.9(1)&-       & -       &502.7\footnotemark[12]     &502.2&-      & -     & -      & -      &  -    &   - \\
$(1s2s\,^3\!S)3d\,^2\!D$   &505.5(7)&505.6(1)&-       & -       &506.1\footnotemark[12]     & -   &-      & -     & -      &-       &  -    & - \\
$(1s2s\,^1\!S)3s\,^2\!S$   & -    &  -       &-       & -       &506.2\footnotemark[12]     &-    &-      & -     & -      & -      &   -   &  - \\
$(1s2s\,^1\!S)3p\,^2\!P$   & -    &  -       &-       & -       &509.9\footnotemark[12]     &-    &-      & -     & -      & -      &  -    &   - \\
$(1s2s\,^1\!S)3d\,^2\!D$   & -    &  -       &-       & -       &512.7\footnotemark[12]     & -   &-      & -     & -      &-       &  -    & - \\
\hline\hline
\end{tabular}
}
\footnotetext[1]{Fitted Auger line peak energies after energy calibration of PSD channels according to the values Mack 1987~\cite{mack87c}, listed in the second column.}
\footnotetext[2]{Mack 1987~\cite{mack87c} - see Table~\ref{tb:energiesCarbon}.}
\footnotetext[3]{Bruch \textit{et al.}   1987~\cite{bru87a} - Zero-degree Auger projectile spectroscopy measurements and saddle-point technique with relativistic corrections.}
\footnotetext[4]{Kilgus \textit{et al.}   1990~\cite{kil90a} - Dielectronic Recombination (DR) measurements at the Heidelberg Test Storage Ring (TSR).}
\footnotetext[5]{Alnaser 2002~\cite{aln02b} - see Table~\ref{tb:energiesCarbon}.}
\footnotetext[6]{Kar and Ho 2009~\cite{kar09a} - Stabilization method: $2l2l^\prime\, ^1\!L$ levels in a.u.}
\footnotetext[7]{Goryaev \textit{et al.}   2017~\cite{gor17b} - see Table~\ref{tb:energiesCarbon}.}
\footnotetext[8]{Yerokhin \textit{et al.}   2017~\cite{yer17a} - see Table~\ref{tb:energiesCarbon}.}
\footnotetext[9]{Manai \textit{et al.}  ~\cite{man22a} using FAC code - see Table~\ref{tb:energiesCarbon}.}
\footnotetext[10]{Manai \textit{et al.}  ~\cite{man22a} using AMBiT code - see Table~\ref{tb:energiesCarbon}.}
\footnotetext[11]{$\varepsilon_A(1s2l2l^\prime) = E(1s2l2l^\prime)-E(1s^2)+E(1s^22s)$ or $\varepsilon_A(2l2l^\prime) = E(2l2l^\prime)-E(1s)+E(1s^2)$, where $E(1s2l2l^\prime)$ and $E(2l2l^\prime)$ are the given NIST energies with respect to $1s^22s$ or $1s^2$, respectively~\CORnew{\cite{kra24a}}.}
\footnotetext[12]{Bruch \textit{et al.}   1979~\cite{bru79a} - 23.7$^\circ$ ESCA measurements and semiempirical + \textit{ab initio} theoretical methods.}
\footnotetext[13]{See also Ho~\cite{ho81a} - complex rotation method - which gives Auger energies of 464.25 and 475.23~eV, for $2s2p\,^3\!P$ and $^1\!P$, respectively.}
\footnotetext[14]{Togawa \textit{et al.}~\cite{tog24a} for the most accurate to date experimental values of $1s2s2p\,^2\!P_{1/2-}$ and $1s2s2p\,^2\!P_{3/2-}$ to $1s^22s$ x-ray transition energies from which we obtain a center-of-gravity Auger energy of $\overline{\varepsilon}_A[1s2s2p\,^2\!P_-]=424.9371(31)$~eV using the NIST value for the IP of O$^{5+}(1s^22s)=138.1189(21)$~eV~\cite{kra24a}.}
\footnotetext[15]{Ahmed and Lipsky 1975~\cite{ahm75a} quoted in Bruch \textit{et al.}   1979~\cite{bru79a}.}
\end{table*}

\clearpage
\twocolumngrid

\begin{thebibliography}{97}%
\makeatletter
\providecommand \@ifxundefined [1]{%
 \@ifx{#1\undefined}
}%
\providecommand \@ifnum [1]{%
 \ifnum #1\expandafter \@firstoftwo
 \else \expandafter \@secondoftwo
 \fi
}%
\providecommand \@ifx [1]{%
 \ifx #1\expandafter \@firstoftwo
 \else \expandafter \@secondoftwo
 \fi
}%
\providecommand \natexlab [1]{#1}%
\providecommand \enquote  [1]{``#1''}%
\providecommand \bibnamefont  [1]{#1}%
\providecommand \bibfnamefont [1]{#1}%
\providecommand \citenamefont [1]{#1}%
\providecommand \href@noop [0]{\@secondoftwo}%
\providecommand \href [0]{\begingroup \@sanitize@url \@href}%
\providecommand \@href[1]{\@@startlink{#1}\@@href}%
\providecommand \@@href[1]{\endgroup#1\@@endlink}%
\providecommand \@sanitize@url [0]{\catcode `\\12\catcode `\$12\catcode
  `\&12\catcode `\#12\catcode `\^12\catcode `\_12\catcode `\%12\relax}%
\providecommand \@@startlink[1]{}%
\providecommand \@@endlink[0]{}%
\providecommand \url  [0]{\begingroup\@sanitize@url \@url }%
\providecommand \@url [1]{\endgroup\@href {#1}{\urlprefix }}%
\providecommand \urlprefix  [0]{URL }%
\providecommand \Eprint [0]{\href }%
\providecommand \doibase [0]{https://doi.org/}%
\providecommand \selectlanguage [0]{\@gobble}%
\providecommand \bibinfo  [0]{\@secondoftwo}%
\providecommand \bibfield  [0]{\@secondoftwo}%
\providecommand \translation [1]{[#1]}%
\providecommand \BibitemOpen [0]{}%
\providecommand \bibitemStop [0]{}%
\providecommand \bibitemNoStop [0]{.\EOS\space}%
\providecommand \EOS [0]{\spacefactor3000\relax}%
\providecommand \BibitemShut  [1]{\csname bibitem#1\endcsname}%
\let\auto@bib@innerbib\@empty
\bibitem [{\citenamefont {Laoutaris}\ \emph {et~al.}(2024)\citenamefont
  {Laoutaris}, \citenamefont {Nanos}, \citenamefont {Biniskos}, \citenamefont
  {Passalidis}, \citenamefont {Benis}, \citenamefont {Dubois},\ and\
  \citenamefont {Zouros}}]{lao24a}%
  \BibitemOpen
  \bibfield  {author} {\bibinfo {author} {\bibfnamefont {A.}~\bibnamefont
  {Laoutaris}}, \bibinfo {author} {\bibfnamefont {S.}~\bibnamefont {Nanos}},
  \bibinfo {author} {\bibfnamefont {A.}~\bibnamefont {Biniskos}}, \bibinfo
  {author} {\bibfnamefont {S.}~\bibnamefont {Passalidis}}, \bibinfo {author}
  {\bibfnamefont {E.~P.}\ \bibnamefont {Benis}}, \bibinfo {author}
  {\bibfnamefont {A.}~\bibnamefont {Dubois}},\ and\ \bibinfo {author}
  {\bibfnamefont {T.~J.~M.}\ \bibnamefont {Zouros}},\ }\bibfield  {title}
  {\bibinfo {title} {Projectile excitation to autoionizing states in swift
  collisions of open-shell \textsc{H}e-like ions with helium},\ }\href
  {https://doi.org/10.1103/PhysRevA.109.032825} {\bibfield  {journal} {\bibinfo
   {journal} {Phys. Rev. A}\ }\textbf {\bibinfo {volume} {109}},\ \bibinfo
  {pages} {032825} (\bibinfo {year} {2024})}\BibitemShut {NoStop}%
\bibitem [{\citenamefont {Madesis}\ \emph {et~al.}(2020)\citenamefont
  {Madesis}, \citenamefont {Laoutaris}, \citenamefont {Zouros}, \citenamefont
  {Benis}, \citenamefont {Gao},\ and\ \citenamefont {Dubois}}]{mad20a}%
  \BibitemOpen
  \bibfield  {author} {\bibinfo {author} {\bibfnamefont {I.}~\bibnamefont
  {Madesis}}, \bibinfo {author} {\bibfnamefont {A.}~\bibnamefont {Laoutaris}},
  \bibinfo {author} {\bibfnamefont {T.~J.~M.}\ \bibnamefont {Zouros}}, \bibinfo
  {author} {\bibfnamefont {E.~P.}\ \bibnamefont {Benis}}, \bibinfo {author}
  {\bibfnamefont {J.~W.}\ \bibnamefont {Gao}},\ and\ \bibinfo {author}
  {\bibfnamefont {A.}~\bibnamefont {Dubois}},\ }\bibfield  {title} {\bibinfo
  {title} {\textsc{P}auli shielding and breakdown of spin statistics in
  multielectron multi-open-shell dynamical atomic systems},\ }\href
  {https://doi.org/10.1103/PhysRevLett.124.113401} {\bibfield  {journal}
  {\bibinfo  {journal} {Phys. Rev. Lett.}\ }\textbf {\bibinfo {volume} {124}},\
  \bibinfo {pages} {113401} (\bibinfo {year} {2020})},\ \bibinfo {note} {and
  supplemental material}\BibitemShut {NoStop}%
\bibitem [{\citenamefont {Madesis}\ \emph {et~al.}(2022)\citenamefont
  {Madesis}, \citenamefont {Laoutaris}, \citenamefont {Nanos}, \citenamefont
  {Passalidis}, \citenamefont {Dubois}, \citenamefont {Zouros},\ and\
  \citenamefont {Benis}}]{mad22a}%
  \BibitemOpen
  \bibfield  {author} {\bibinfo {author} {\bibfnamefont {I.}~\bibnamefont
  {Madesis}}, \bibinfo {author} {\bibfnamefont {A.}~\bibnamefont {Laoutaris}},
  \bibinfo {author} {\bibfnamefont {S.}~\bibnamefont {Nanos}}, \bibinfo
  {author} {\bibfnamefont {S.}~\bibnamefont {Passalidis}}, \bibinfo {author}
  {\bibfnamefont {A.}~\bibnamefont {Dubois}}, \bibinfo {author} {\bibfnamefont
  {T.~J.~M.}\ \bibnamefont {Zouros}},\ and\ \bibinfo {author} {\bibfnamefont
  {E.~P.}\ \bibnamefont {Benis}},\ }\bibfield  {title} {\bibinfo {title}
  {State-resolved differential cross sections of single-electron capture in
  swift collisions of \textsc{C}$^{4+}(1s2s\thinspace^3\!\textsc{S})$ ions with
  gas targets},\ }\href {https://doi.org/10.1103/PhysRevA.105.062810}
  {\bibfield  {journal} {\bibinfo  {journal} {Phys. Rev. A}\ }\textbf {\bibinfo
  {volume} {105}},\ \bibinfo {pages} {062810} (\bibinfo {year}
  {2022})}\BibitemShut {NoStop}%
\bibitem [{\citenamefont {Laoutaris}\ \emph {et~al.}(2022)\citenamefont
  {Laoutaris}, \citenamefont {Nanos}, \citenamefont {Madesis}, \citenamefont
  {Passalidis}, \citenamefont {Benis}, \citenamefont {Dubois},\ and\
  \citenamefont {Zouros}}]{lao22a}%
  \BibitemOpen
  \bibfield  {author} {\bibinfo {author} {\bibfnamefont {A.}~\bibnamefont
  {Laoutaris}}, \bibinfo {author} {\bibfnamefont {S.}~\bibnamefont {Nanos}},
  \bibinfo {author} {\bibfnamefont {I.}~\bibnamefont {Madesis}}, \bibinfo
  {author} {\bibfnamefont {S.}~\bibnamefont {Passalidis}}, \bibinfo {author}
  {\bibfnamefont {E.~P.}\ \bibnamefont {Benis}}, \bibinfo {author}
  {\bibfnamefont {A.}~\bibnamefont {Dubois}},\ and\ \bibinfo {author}
  {\bibfnamefont {T.~J.~M.}\ \bibnamefont {Zouros}},\ }\bibfield  {title}
  {\bibinfo {title} {Coherent treatment of transfer-excitation processes in
  swift ion-atom collisions},\ }\href
  {https://doi.org/10.1103/PhysRevA.106.022810} {\bibfield  {journal} {\bibinfo
   {journal} {Phys. Rev. A}\ }\textbf {\bibinfo {volume} {106}},\ \bibinfo
  {pages} {022810} (\bibinfo {year} {2022})}\BibitemShut {NoStop}%
\bibitem [{\citenamefont {Benis}(2024)}]{ben24a}%
  \BibitemOpen
  \bibfield  {author} {\bibinfo {author} {\bibfnamefont {E.~P.}\ \bibnamefont
  {Benis}},\ }\bibinfo {title} {Fast ion-atom collisions: Electron spectroscopy
  of mixed-state beams},\ in\ \href
  {https://doi.org/10.1007/978-981-97-7063-2_4} {\emph {\bibinfo {booktitle}
  {Advances in Atomic Molecular Collisions}}},\ \bibinfo {editor} {edited by\
  \bibinfo {editor} {\bibfnamefont {L.~C.}\ \bibnamefont {Tribedi}}}\ (\bibinfo
   {publisher} {Springer Nature Singapore},\ \bibinfo {address} {Singapore},\
  \bibinfo {year} {2024})\ pp.\ \bibinfo {pages} {71--110}\BibitemShut
  {NoStop}%
\bibitem [{\citenamefont {Sisourat}\ and\ \citenamefont
  {Dubois}(2019)}]{sis19a}%
  \BibitemOpen
  \bibfield  {author} {\bibinfo {author} {\bibfnamefont {N.}~\bibnamefont
  {Sisourat}}\ and\ \bibinfo {author} {\bibfnamefont {A.}~\bibnamefont
  {Dubois}},\ }\bibfield  {title} {\bibinfo {title} {Semiclassical
  close-coupling approaches},\ }in\ \href@noop {} {\emph {\bibinfo {booktitle}
  {Ion-Atom Collision -- The Few-Body Problem in Dynamic Systems}}},\ \bibinfo
  {editor} {edited by\ \bibinfo {editor} {\bibfnamefont {M.}~\bibnamefont
  {Schultz}}}\ (\bibinfo  {publisher} {de Gruyter},\ \bibinfo {address}
  {Berlin/Boston},\ \bibinfo {year} {2019})\ pp.\ \bibinfo {pages}
  {157--178}\BibitemShut {NoStop}%
\bibitem [{\citenamefont {Gabriel}(1972)}]{gab72a}%
  \BibitemOpen
  \bibfield  {author} {\bibinfo {author} {\bibfnamefont {A.~H.}\ \bibnamefont
  {Gabriel}},\ }\bibfield  {title} {\bibinfo {title} {Dielectronic
  \textsc{S}atellite spectra for highly-charged helium-like ion lines},\ }\href
  {https://doi.org/10.1093/mnras/160.1.99} {\bibfield  {journal} {\bibinfo
  {journal} {Mon. Not. R. Astron. Soc.}\ }\textbf {\bibinfo {volume} {160}},\
  \bibinfo {pages} {99} (\bibinfo {year} {1972})}\BibitemShut {NoStop}%
\bibitem [{\citenamefont {Beiersdorfer}(2003)}]{bei03b}%
  \BibitemOpen
  \bibfield  {author} {\bibinfo {author} {\bibfnamefont {P.}~\bibnamefont
  {Beiersdorfer}},\ }\bibfield  {title} {\bibinfo {title} {Laboratory
  \textsc{X}-ray astrophysics},\ }\href
  {https://doi.org/10.1146/annurev.astro.41.011802.094825} {\bibfield
  {journal} {\bibinfo  {journal} {Annu. Rev. Astron. Astrophys.}\ }\textbf
  {\bibinfo {volume} {41}},\ \bibinfo {pages} {343} (\bibinfo {year}
  {2003})}\BibitemShut {NoStop}%
\bibitem [{\citenamefont {Delahaye}\ \emph {et~al.}(2006)\citenamefont
  {Delahaye}, \citenamefont {Pradhan},\ and\ \citenamefont {Zeippen}}]{del06a}%
  \BibitemOpen
  \bibfield  {author} {\bibinfo {author} {\bibfnamefont {F.}~\bibnamefont
  {Delahaye}}, \bibinfo {author} {\bibfnamefont {A.~K.}\ \bibnamefont
  {Pradhan}},\ and\ \bibinfo {author} {\bibfnamefont {C.~J.}\ \bibnamefont
  {Zeippen}},\ }\bibfield  {title} {\bibinfo {title} {Electron impact
  excitation of helium-like ions up to $n = 4$ levels including radiation
  damping},\ }\href {https://doi.org/10.1088/0953-4075/39/17/005} {\bibfield
  {journal} {\bibinfo  {journal} {J. Phys. B}\ }\textbf {\bibinfo {volume}
  {39}},\ \bibinfo {pages} {3465} (\bibinfo {year} {2006})}\BibitemShut
  {NoStop}%
\bibitem [{\citenamefont {Fritsch}(1991)}]{fri91d}%
  \BibitemOpen
  \bibfield  {author} {\bibinfo {author} {\bibfnamefont {W.}~\bibnamefont
  {Fritsch}},\ }\bibfield  {title} {\bibinfo {title} {Electron excitation in
  collisions between protons and excited helium},\ }\href
  {https://doi.org/https://doi.org/10.1016/0375-9601(91)91004-W} {\bibfield
  {journal} {\bibinfo  {journal} {Phys. Lett. A}\ }\textbf {\bibinfo {volume}
  {158}},\ \bibinfo {pages} {227} (\bibinfo {year} {1991})}\BibitemShut
  {NoStop}%
\bibitem [{\citenamefont {Zouros}\ \emph {et~al.}(2025)\citenamefont {Zouros},
  \citenamefont {Laoutaris}, \citenamefont {Nanos},\ and\ \citenamefont
  {Benis}}]{zou25a}%
  \BibitemOpen
  \bibfield  {author} {\bibinfo {author} {\bibfnamefont {T.~J.~M.}\
  \bibnamefont {Zouros}}, \bibinfo {author} {\bibfnamefont {A.}~\bibnamefont
  {Laoutaris}}, \bibinfo {author} {\bibfnamefont {S.}~\bibnamefont {Nanos}},\
  and\ \bibinfo {author} {\bibfnamefont {E.~P.}\ \bibnamefont {Benis}},\
  }\bibfield  {title} {\bibinfo {title} {Determination of the
  $1s2s\,^{3}\!\textsc{S}$ and $1s2s\,^{1}\!\textsc{S}$ metastable fractions in
  swift mixed-state \textsc{H}e-like ion beams},\ }\href
  {https://doi.org/10.1088/1361-6455/adb54d} {\bibfield  {journal} {\bibinfo
  {journal} {J. Phys. B}\ }\textbf {\bibinfo {volume} {58}},\ \bibinfo {pages}
  {055201} (\bibinfo {year} {2025})},\ \bibinfo {note} {and supplemental
  material}\BibitemShut {NoStop}%
\bibitem [{\citenamefont {Benis}\ \emph {et~al.}(2018)\citenamefont {Benis},
  \citenamefont {Madesis}, \citenamefont {Laoutaris}, \citenamefont {Nanos},\
  and\ \citenamefont {Zouros}}]{ben18b}%
  \BibitemOpen
  \bibfield  {author} {\bibinfo {author} {\bibfnamefont {E.~P.}\ \bibnamefont
  {Benis}}, \bibinfo {author} {\bibfnamefont {I.}~\bibnamefont {Madesis}},
  \bibinfo {author} {\bibfnamefont {A.}~\bibnamefont {Laoutaris}}, \bibinfo
  {author} {\bibfnamefont {S.}~\bibnamefont {Nanos}},\ and\ \bibinfo {author}
  {\bibfnamefont {T.~J.~M.}\ \bibnamefont {Zouros}},\ }\bibfield  {title}
  {\bibinfo {title} {Mixed-state ionic beams: An effective tool for collision
  dynamics investigations},\ }\href {https://doi.org/10.3390/atoms6040066}
  {\bibfield  {journal} {\bibinfo  {journal} {Atoms}\ }\textbf {\bibinfo
  {volume} {6}},\ \bibinfo {pages} {66} (\bibinfo {year} {2018})}\BibitemShut
  {NoStop}%
\bibitem [{\citenamefont {Horsdal~Pedersen}(1979)}]{ped79a}%
  \BibitemOpen
  \bibfield  {author} {\bibinfo {author} {\bibfnamefont {E.}~\bibnamefont
  {Horsdal~Pedersen}},\ }\bibfield  {title} {\bibinfo {title} {Metastable-atom
  population of fast, neutral helium beams},\ }\href
  {https://doi.org/10.1103/PhysRevLett.42.440} {\bibfield  {journal} {\bibinfo
  {journal} {Phys. Rev. Lett.}\ }\textbf {\bibinfo {volume} {42}},\ \bibinfo
  {pages} {440} (\bibinfo {year} {1979})}\BibitemShut {NoStop}%
\bibitem [{\citenamefont {Dillingham}\ \emph {et~al.}(1984)\citenamefont
  {Dillingham}, \citenamefont {Newcomb}, \citenamefont {Hall}, \citenamefont
  {Pepmiller},\ and\ \citenamefont {Richard}}]{dil84a}%
  \BibitemOpen
  \bibfield  {author} {\bibinfo {author} {\bibfnamefont {T.~R.}\ \bibnamefont
  {Dillingham}}, \bibinfo {author} {\bibfnamefont {J.}~\bibnamefont {Newcomb}},
  \bibinfo {author} {\bibfnamefont {J.}~\bibnamefont {Hall}}, \bibinfo {author}
  {\bibfnamefont {P.~L.}\ \bibnamefont {Pepmiller}},\ and\ \bibinfo {author}
  {\bibfnamefont {P.}~\bibnamefont {Richard}},\ }\bibfield  {title} {\bibinfo
  {title} {Projectile \textsc{K}-\textsc{A}uger-electron production by bare,
  one-, and two-electron ions},\ }\href@noop {} {\bibfield  {journal} {\bibinfo
   {journal} {Phys. Rev. A}\ }\textbf {\bibinfo {volume} {29}},\ \bibinfo
  {pages} {3029} (\bibinfo {year} {1984})}\BibitemShut {NoStop}%
\bibitem [{\citenamefont {Andersen}\ \emph
  {et~al.}(1992{\natexlab{a}})\citenamefont {Andersen}, \citenamefont {Pan},
  \citenamefont {Schmidt}, \citenamefont {Badnell},\ and\ \citenamefont
  {Pindzola}}]{and92a}%
  \BibitemOpen
  \bibfield  {author} {\bibinfo {author} {\bibfnamefont {L.~H.}\ \bibnamefont
  {Andersen}}, \bibinfo {author} {\bibfnamefont {G.~Y.}\ \bibnamefont {Pan}},
  \bibinfo {author} {\bibfnamefont {H.~T.}\ \bibnamefont {Schmidt}}, \bibinfo
  {author} {\bibfnamefont {N.~R.}\ \bibnamefont {Badnell}},\ and\ \bibinfo
  {author} {\bibfnamefont {M.~S.}\ \bibnamefont {Pindzola}},\ }\bibfield
  {title} {\bibinfo {title} {Absolute measurements and calculations of
  dielectronic recombination with metastable \textsc{H}e-like \textsc{N},
  \textsc{F}, and \textsc{S}i ions},\ }\href@noop {} {\bibfield  {journal}
  {\bibinfo  {journal} {Phys. Rev. A}\ }\textbf {\bibinfo {volume} {45}},\
  \bibinfo {pages} {7868} (\bibinfo {year} {1992}{\natexlab{a}})}\BibitemShut
  {NoStop}%
\bibitem [{\citenamefont {Dinklage}\ \emph {et~al.}(1996)\citenamefont
  {Dinklage}, \citenamefont {Lokajczyk},\ and\ \citenamefont {Kunze}}]{din96a}%
  \BibitemOpen
  \bibfield  {author} {\bibinfo {author} {\bibfnamefont {A.}~\bibnamefont
  {Dinklage}}, \bibinfo {author} {\bibfnamefont {T.}~\bibnamefont
  {Lokajczyk}},\ and\ \bibinfo {author} {\bibfnamefont {H.~J.}\ \bibnamefont
  {Kunze}},\ }\bibfield  {title} {\bibinfo {title} {Measurement of the
  $2\,^3\!\textsc{S}$ population in a $^4\!$\textsc{H}e beam by means of
  laser-induced fluorescence},\ }\href
  {https://doi.org/10.1088/0953-4075/29/9/012} {\bibfield  {journal} {\bibinfo
  {journal} {J. Phys. B}\ }\textbf {\bibinfo {volume} {29}},\ \bibinfo {pages}
  {1655} (\bibinfo {year} {1996})}\BibitemShut {NoStop}%
\bibitem [{\citenamefont {Zamkov}\ \emph
  {et~al.}(2001{\natexlab{a}})\citenamefont {Zamkov}, \citenamefont {Aliabadi},
  \citenamefont {Benis}, \citenamefont {Richard}, \citenamefont {Tawara},\ and\
  \citenamefont {Zouros}}]{zam01a}%
  \BibitemOpen
  \bibfield  {author} {\bibinfo {author} {\bibfnamefont {M.}~\bibnamefont
  {Zamkov}}, \bibinfo {author} {\bibfnamefont {H.}~\bibnamefont {Aliabadi}},
  \bibinfo {author} {\bibfnamefont {E.~P.}\ \bibnamefont {Benis}}, \bibinfo
  {author} {\bibfnamefont {P.}~\bibnamefont {Richard}}, \bibinfo {author}
  {\bibfnamefont {H.}~\bibnamefont {Tawara}},\ and\ \bibinfo {author}
  {\bibfnamefont {T.~J.~M.}\ \bibnamefont {Zouros}},\ }\bibfield  {title}
  {\bibinfo {title} {Energy dependence of the metastable fraction in
  \textsc{B}$^{3+}(1s^2\,^1\!\textsc{S}, 1s2s\,^3\!\textsc{S})$ beams produced
  in collisions with thin-foil and gas targets},\ }\href@noop {} {\bibfield
  {journal} {\bibinfo  {journal} {Phys. Rev. A}\ }\textbf {\bibinfo {volume}
  {64}},\ \bibinfo {pages} {052702} (\bibinfo {year}
  {2001}{\natexlab{a}})}\BibitemShut {NoStop}%
\bibitem [{\citenamefont {Benis}\ \emph {et~al.}(2002)\citenamefont {Benis},
  \citenamefont {Zamkov}, \citenamefont {Richard},\ and\ \citenamefont
  {Zouros}}]{ben02a}%
  \BibitemOpen
  \bibfield  {author} {\bibinfo {author} {\bibfnamefont {E.~P.}\ \bibnamefont
  {Benis}}, \bibinfo {author} {\bibfnamefont {M.}~\bibnamefont {Zamkov}},
  \bibinfo {author} {\bibfnamefont {P.}~\bibnamefont {Richard}},\ and\ \bibinfo
  {author} {\bibfnamefont {T.~J.~M.}\ \bibnamefont {Zouros}},\ }\bibfield
  {title} {\bibinfo {title} {Technique for the determination of the
  $1s2s\,^3\textsc{S}$ metastable fraction in two-electron ion beams},\
  }\href@noop {} {\bibfield  {journal} {\bibinfo  {journal} {Phys. Rev. A}\
  }\textbf {\bibinfo {volume} {65}},\ \bibinfo {pages} {064701} (\bibinfo
  {year} {2002})}\BibitemShut {NoStop}%
\bibitem [{\citenamefont {Dmitriev}\ \emph {et~al.}(2003)\citenamefont
  {Dmitriev}, \citenamefont {Teplova}, \citenamefont {Fainberg},\ and\
  \citenamefont {Belkova}}]{dmi03a}%
  \BibitemOpen
  \bibfield  {author} {\bibinfo {author} {\bibfnamefont {I.~S.}\ \bibnamefont
  {Dmitriev}}, \bibinfo {author} {\bibfnamefont {Y.~A.}\ \bibnamefont
  {Teplova}}, \bibinfo {author} {\bibfnamefont {Y.~A.}\ \bibnamefont
  {Fainberg}},\ and\ \bibinfo {author} {\bibfnamefont {Y.~A.}\ \bibnamefont
  {Belkova}},\ }\bibfield  {title} {\bibinfo {title} {Formation of metastable
  states of light ions in ion–atom collisions},\ }\href
  {https://doi.org/10.1238/Physica.Regular.068a00383} {\bibfield  {journal}
  {\bibinfo  {journal} {Phys. Scr.}\ }\textbf {\bibinfo {volume} {68}},\
  \bibinfo {pages} {383} (\bibinfo {year} {2003})}\BibitemShut {NoStop}%
\bibitem [{\citenamefont {Zamkov}\ \emph
  {et~al.}(2001{\natexlab{b}})\citenamefont {Zamkov}, \citenamefont {Aliabadi},
  \citenamefont {Benis}, \citenamefont {Richard}, \citenamefont {Tawara},\ and\
  \citenamefont {Zouros}}]{zam01b}%
  \BibitemOpen
  \bibfield  {author} {\bibinfo {author} {\bibfnamefont {M.}~\bibnamefont
  {Zamkov}}, \bibinfo {author} {\bibfnamefont {H.}~\bibnamefont {Aliabadi}},
  \bibinfo {author} {\bibfnamefont {E.~P.}\ \bibnamefont {Benis}}, \bibinfo
  {author} {\bibfnamefont {P.}~\bibnamefont {Richard}}, \bibinfo {author}
  {\bibfnamefont {H.}~\bibnamefont {Tawara}},\ and\ \bibinfo {author}
  {\bibfnamefont {T.~J.~M.}\ \bibnamefont {Zouros}},\ }\bibfield  {title}
  {\bibinfo {title} {Stripping energy dependence of
  \textsc{B}$^{3+}(1s2s\,^3\!\textsc{S})$ beam metastable fraction},\
  }\href@noop {} {\bibfield  {journal} {\bibinfo  {journal} {American Institute
  of Physics}\ }\textbf {\bibinfo {volume} {576}},\ \bibinfo {pages} {149}
  (\bibinfo {year} {2001}{\natexlab{b}})}\BibitemShut {NoStop}%
\bibitem [{\citenamefont {Stolterfoht}(1987)}]{sto87a}%
  \BibitemOpen
  \bibfield  {author} {\bibinfo {author} {\bibfnamefont {N.}~\bibnamefont
  {Stolterfoht}},\ }\bibfield  {title} {\bibinfo {title} {High resolution
  \textsc{A}uger spectroscopy in energetic ion atom collisions},\ }\href@noop
  {} {\bibfield  {journal} {\bibinfo  {journal} {Phys. Rep.}\ }\textbf
  {\bibinfo {volume} {146}},\ \bibinfo {pages} {315} (\bibinfo {year}
  {1987})}\BibitemShut {NoStop}%
\bibitem [{\citenamefont {Zouros}\ and\ \citenamefont {Lee}(1997)}]{zou97a}%
  \BibitemOpen
  \bibfield  {author} {\bibinfo {author} {\bibfnamefont {T.~J.~M.}\
  \bibnamefont {Zouros}}\ and\ \bibinfo {author} {\bibfnamefont {D.~H.}\
  \bibnamefont {Lee}},\ }\bibinfo {title} {Zero \textsc{D}egree \textsc{A}uger
  \textsc{E}lectron \textsc{S}pectroscopy of \textsc{P}rojectile
  \textsc{I}ons},\ in\ \href@noop {} {\emph {\bibinfo {booktitle}
  {Accelerator-based Atomic Physics: Techniques and Applications}}},\ \bibinfo
  {editor} {edited by\ \bibinfo {editor} {\bibfnamefont {S.~M.}\ \bibnamefont
  {Shafroth}}\ and\ \bibinfo {editor} {\bibfnamefont {J.~C.}\ \bibnamefont
  {Austin}}}\ (\bibinfo  {publisher} {American Institute of Physics},\ \bibinfo
  {address} {Woodbury, NY},\ \bibinfo {year} {1997})\ Chap.~\bibinfo {chapter}
  {13}, pp.\ \bibinfo {pages} {426--479}\BibitemShut {NoStop}%
\bibitem [{\citenamefont {Benis}\ and\ \citenamefont {Zouros}(2016)}]{ben16b}%
  \BibitemOpen
  \bibfield  {author} {\bibinfo {author} {\bibfnamefont {E.~P.}\ \bibnamefont
  {Benis}}\ and\ \bibinfo {author} {\bibfnamefont {T.~J.~M.}\ \bibnamefont
  {Zouros}},\ }\bibfield  {title} {\bibinfo {title} {Determination of the
  $1s2\ell2\ell^\prime$ state production ratios
  $^4\!\textsc{P}^o/^2\!\textsc{P}$, $^2\!\textsc{D}/^2\!\textsc{P}$ and
  $^2\!\textsc{P}_+/^2\!\textsc{P}_-$ from fast
  ($1s^2,1s2s\thinspace^3\!\textsc{S}$) mixed-state \textsc{H}e-like ion beams
  in collisions with \textsc{H}$_2$ targets},\ }\href
  {http://stacks.iop.org/0953-4075/49/i=23/a=235202} {\bibfield  {journal}
  {\bibinfo  {journal} {J. Phys. B}\ }\textbf {\bibinfo {volume} {49}},\
  \bibinfo {pages} {235202} (\bibinfo {year} {2016})}\BibitemShut {NoStop}%
\bibitem [{\citenamefont {Nanos}\ \emph
  {et~al.}(2023{\natexlab{a}})\citenamefont {Nanos}, \citenamefont {Esponda},
  \citenamefont {Hillenbrand}, \citenamefont {Biniskos}, \citenamefont
  {Laoutaris}, \citenamefont {Quinto}, \citenamefont {Petridis}, \citenamefont
  {Menz}, \citenamefont {Zouros}, \citenamefont {St\"{o}hlker}, \citenamefont
  {Rivarola}, \citenamefont {Monti},\ and\ \citenamefont {Benis}}]{nan23b}%
  \BibitemOpen
  \bibfield  {author} {\bibinfo {author} {\bibfnamefont {S.}~\bibnamefont
  {Nanos}}, \bibinfo {author} {\bibfnamefont {N.~J.}\ \bibnamefont {Esponda}},
  \bibinfo {author} {\bibfnamefont {P.-M.}\ \bibnamefont {Hillenbrand}},
  \bibinfo {author} {\bibfnamefont {A.}~\bibnamefont {Biniskos}}, \bibinfo
  {author} {\bibfnamefont {A.}~\bibnamefont {Laoutaris}}, \bibinfo {author}
  {\bibfnamefont {M.~A.}\ \bibnamefont {Quinto}}, \bibinfo {author}
  {\bibfnamefont {N.}~\bibnamefont {Petridis}}, \bibinfo {author}
  {\bibfnamefont {E.}~\bibnamefont {Menz}}, \bibinfo {author} {\bibfnamefont
  {T.~J.~M.}\ \bibnamefont {Zouros}}, \bibinfo {author} {\bibfnamefont
  {T.}~\bibnamefont {St\"{o}hlker}}, \bibinfo {author} {\bibfnamefont {R.~D.}\
  \bibnamefont {Rivarola}}, \bibinfo {author} {\bibfnamefont {J.~M.}\
  \bibnamefont {Monti}},\ and\ \bibinfo {author} {\bibfnamefont {E.~P.}\
  \bibnamefont {Benis}},\ }\bibfield  {title} {\bibinfo {title} {Cusp-electron
  production in collisions of open-shell \textsc{H}e-like oxygen ions with
  atomic targets},\ }\href {https://doi.org/10.1103/PhysRevA.107.062815}
  {\bibfield  {journal} {\bibinfo  {journal} {Phys. Rev. A}\ }\textbf {\bibinfo
  {volume} {107}},\ \bibinfo {pages} {062815} (\bibinfo {year}
  {2023}{\natexlab{a}})}\BibitemShut {NoStop}%
\bibitem [{\citenamefont {Montenegro}\ \emph {et~al.}(1994)\citenamefont
  {Montenegro}, \citenamefont {Meyerhof},\ and\ \citenamefont
  {McGuire}}]{mon94a}%
  \BibitemOpen
  \bibfield  {author} {\bibinfo {author} {\bibfnamefont {E.}~\bibnamefont
  {Montenegro}}, \bibinfo {author} {\bibfnamefont {W.}~\bibnamefont
  {Meyerhof}},\ and\ \bibinfo {author} {\bibfnamefont {J.}~\bibnamefont
  {McGuire}},\ }\bibfield  {title} {\bibinfo {title} {\textsc{R}ole of
  two-center electron-electron interaction in projectile electron excitation
  and loss},\ }\href
  {https://doi.org/https://doi.org/10.1016/S1049-250X(08)60079-8} {\bibfield
  {journal} {\bibinfo  {journal} {Adv. At. Mol. Opt. Phys.}\ }\textbf {\bibinfo
  {volume} {34}},\ \bibinfo {pages} {249} (\bibinfo {year} {1994})}\BibitemShut
  {NoStop}%
\bibitem [{\citenamefont {Zouros}(1996)}]{zou96b}%
  \BibitemOpen
  \bibfield  {author} {\bibinfo {author} {\bibfnamefont {T.~J.~M.}\
  \bibnamefont {Zouros}},\ }\bibfield  {title} {\bibinfo {title}
  {\textsc{P}rojectile-electron - target-electron interactions:
  \textsc{E}xposing the dynamic role of electrons in fast ion-atom
  collisions},\ }\href@noop {} {\bibfield  {journal} {\bibinfo  {journal}
  {Comments At. Mol. Phys.}\ }\textbf {\bibinfo {volume} {32}},\ \bibinfo
  {pages} {291} (\bibinfo {year} {1996})}\BibitemShut {NoStop}%
\bibitem [{\citenamefont {McGuire}\ \emph {et~al.}(1981)\citenamefont
  {McGuire}, \citenamefont {Stolterfoht},\ and\ \citenamefont
  {Simony}}]{mcg81a}%
  \BibitemOpen
  \bibfield  {author} {\bibinfo {author} {\bibfnamefont {J.~H.}\ \bibnamefont
  {McGuire}}, \bibinfo {author} {\bibfnamefont {N.}~\bibnamefont
  {Stolterfoht}},\ and\ \bibinfo {author} {\bibfnamefont {P.~R.}\ \bibnamefont
  {Simony}},\ }\bibfield  {title} {\bibinfo {title} {Screening and
  antiscreening by projectile electrons in high-velocity atomic collisions},\
  }\href {https://doi.org/10.1103/PhysRevA.24.97} {\bibfield  {journal}
  {\bibinfo  {journal} {Phys. Rev. A}\ }\textbf {\bibinfo {volume} {24}},\
  \bibinfo {pages} {97} (\bibinfo {year} {1981})}\BibitemShut {NoStop}%
\bibitem [{\citenamefont {Macdonald}\ \emph {et~al.}(1973)\citenamefont
  {Macdonald}, \citenamefont {Richard}, \citenamefont {Cocke}, \citenamefont
  {Brown},\ and\ \citenamefont {Sellin}}]{mac73a}%
  \BibitemOpen
  \bibfield  {author} {\bibinfo {author} {\bibfnamefont {J.~R.}\ \bibnamefont
  {Macdonald}}, \bibinfo {author} {\bibfnamefont {P.}~\bibnamefont {Richard}},
  \bibinfo {author} {\bibfnamefont {C.~L.}\ \bibnamefont {Cocke}}, \bibinfo
  {author} {\bibfnamefont {M.}~\bibnamefont {Brown}},\ and\ \bibinfo {author}
  {\bibfnamefont {I.~A.}\ \bibnamefont {Sellin}},\ }\bibfield  {title}
  {\bibinfo {title} {One- and two-electron excited states produced by electron
  exchange, excitation, and electron capture in collisions of fluorine ions in
  argon gas at 34.8 \textsc{M}e\textsc{V}},\ }\href
  {https://doi.org/10.1103/PhysRevLett.31.684} {\bibfield  {journal} {\bibinfo
  {journal} {Phys. Rev. Lett.}\ }\textbf {\bibinfo {volume} {31}},\ \bibinfo
  {pages} {684} (\bibinfo {year} {1973})}\BibitemShut {NoStop}%
\bibitem [{\citenamefont {Hopkins}\ \emph {et~al.}(1974)\citenamefont
  {Hopkins}, \citenamefont {Kauffman}, \citenamefont {Woods},\ and\
  \citenamefont {Richard}}]{hop74a}%
  \BibitemOpen
  \bibfield  {author} {\bibinfo {author} {\bibfnamefont {F.}~\bibnamefont
  {Hopkins}}, \bibinfo {author} {\bibfnamefont {R.~L.}\ \bibnamefont
  {Kauffman}}, \bibinfo {author} {\bibfnamefont {C.~W.}\ \bibnamefont
  {Woods}},\ and\ \bibinfo {author} {\bibfnamefont {P.}~\bibnamefont
  {Richard}},\ }\bibfield  {title} {\bibinfo {title} {$\textsc{K}$
  \textsc{x}-ray transitions in one- and two-electron oxygen and fluorine
  projectiles produced in helium, neon, and argon targets},\ }\href
  {https://doi.org/10.1103/PhysRevA.9.2413} {\bibfield  {journal} {\bibinfo
  {journal} {Phys. Rev. A}\ }\textbf {\bibinfo {volume} {9}},\ \bibinfo {pages}
  {2413} (\bibinfo {year} {1974})}\BibitemShut {NoStop}%
\bibitem [{\citenamefont {Hopkins}\ \emph {et~al.}(1976)\citenamefont
  {Hopkins}, \citenamefont {Little},\ and\ \citenamefont {Cue}}]{hop76a}%
  \BibitemOpen
  \bibfield  {author} {\bibinfo {author} {\bibfnamefont {F.}~\bibnamefont
  {Hopkins}}, \bibinfo {author} {\bibfnamefont {A.}~\bibnamefont {Little}},\
  and\ \bibinfo {author} {\bibfnamefont {N.}~\bibnamefont {Cue}},\ }\bibfield
  {title} {\bibinfo {title} {Inner-shell \textsc{C}oulomb excitation in the
  collisions of few-electron \textsc{F} with \textsc{H}$_{2}$ and
  \textsc{H}e},\ }\href {https://doi.org/10.1103/PhysRevA.14.1634} {\bibfield
  {journal} {\bibinfo  {journal} {Phys. Rev. A}\ }\textbf {\bibinfo {volume}
  {14}},\ \bibinfo {pages} {1634} (\bibinfo {year} {1976})}\BibitemShut
  {NoStop}%
\bibitem [{\citenamefont {Matthews}\ \emph
  {et~al.}(1976{\natexlab{a}})\citenamefont {Matthews}, \citenamefont
  {Fortner}, \citenamefont {Schneider},\ and\ \citenamefont {Moore}}]{matt76a}%
  \BibitemOpen
  \bibfield  {author} {\bibinfo {author} {\bibfnamefont {D.~L.}\ \bibnamefont
  {Matthews}}, \bibinfo {author} {\bibfnamefont {R.~J.}\ \bibnamefont
  {Fortner}}, \bibinfo {author} {\bibfnamefont {D.}~\bibnamefont {Schneider}},\
  and\ \bibinfo {author} {\bibfnamefont {C.~F.}\ \bibnamefont {Moore}},\
  }\bibfield  {title} {\bibinfo {title} {Evidence for nonstatistical population
  of configurations in \textsc{L}i-like neon following \textsc{C}l$^{13+}$
  \ensuremath{\rightarrow} \textsc{N}e collisions},\ }\href
  {https://doi.org/10.1103/PhysRevA.14.1561} {\bibfield  {journal} {\bibinfo
  {journal} {Phys. Rev. A}\ }\textbf {\bibinfo {volume} {14}},\ \bibinfo
  {pages} {1561} (\bibinfo {year} {1976}{\natexlab{a}})}\BibitemShut {NoStop}%
\bibitem [{\citenamefont {Schiebel}\ \emph {et~al.}(1977)\citenamefont
  {Schiebel}, \citenamefont {Doyle}, \citenamefont {Macdonald},\ and\
  \citenamefont {Ellsworth}}]{schi77a}%
  \BibitemOpen
  \bibfield  {author} {\bibinfo {author} {\bibfnamefont {U.}~\bibnamefont
  {Schiebel}}, \bibinfo {author} {\bibfnamefont {B.~L.}\ \bibnamefont {Doyle}},
  \bibinfo {author} {\bibfnamefont {J.~R.}\ \bibnamefont {Macdonald}},\ and\
  \bibinfo {author} {\bibfnamefont {L.~D.}\ \bibnamefont {Ellsworth}},\
  }\bibfield  {title} {\bibinfo {title} {Projectile \textsc{K} x rays from
  \textsc{S}i$^{12+}$ ions in the $1s2s\,^3\!\textsc{S}_1$ metastable state
  incident on helium gas},\ }\href {https://doi.org/10.1103/PhysRevA.16.1089}
  {\bibfield  {journal} {\bibinfo  {journal} {Phys. Rev. A}\ }\textbf {\bibinfo
  {volume} {16}},\ \bibinfo {pages} {1089} (\bibinfo {year}
  {1977})}\BibitemShut {NoStop}%
\bibitem [{\citenamefont {Terasawa}\ \emph {et~al.}(1983)\citenamefont
  {Terasawa}, \citenamefont {Gray}, \citenamefont {Hagmann}, \citenamefont
  {Hall}, \citenamefont {Newcomb}, \citenamefont {Pepmiller},\ and\
  \citenamefont {Richard}}]{ter83a}%
  \BibitemOpen
  \bibfield  {author} {\bibinfo {author} {\bibfnamefont {M.}~\bibnamefont
  {Terasawa}}, \bibinfo {author} {\bibfnamefont {T.~J.}\ \bibnamefont {Gray}},
  \bibinfo {author} {\bibfnamefont {S.}~\bibnamefont {Hagmann}}, \bibinfo
  {author} {\bibfnamefont {J.}~\bibnamefont {Hall}}, \bibinfo {author}
  {\bibfnamefont {J.}~\bibnamefont {Newcomb}}, \bibinfo {author} {\bibfnamefont
  {P.}~\bibnamefont {Pepmiller}},\ and\ \bibinfo {author} {\bibfnamefont
  {P.}~\bibnamefont {Richard}},\ }\bibfield  {title} {\bibinfo {title}
  {Electron capture by and electron excitation of two-electron fluorine ions
  incident on helium},\ }\href@noop {} {\bibfield  {journal} {\bibinfo
  {journal} {Phys. Rev. A}\ }\textbf {\bibinfo {volume} {27}},\ \bibinfo
  {pages} {2868} (\bibinfo {year} {1983})}\BibitemShut {NoStop}%
\bibitem [{\citenamefont {Reymann}\ \emph {et~al.}(1988)\citenamefont
  {Reymann}, \citenamefont {Schartner}, \citenamefont {Sommer},\ and\
  \citenamefont {Tr\"abert}}]{rey88a}%
  \BibitemOpen
  \bibfield  {author} {\bibinfo {author} {\bibfnamefont {K.}~\bibnamefont
  {Reymann}}, \bibinfo {author} {\bibfnamefont {K.-H.}\ \bibnamefont
  {Schartner}}, \bibinfo {author} {\bibfnamefont {B.}~\bibnamefont {Sommer}},\
  and\ \bibinfo {author} {\bibfnamefont {E.}~\bibnamefont {Tr\"abert}},\
  }\bibfield  {title} {\bibinfo {title} {Scaling relation for total excitation
  cross sections of \textsc{H}e in collisions with highly charged ions},\
  }\href {https://doi.org/10.1103/PhysRevA.38.2290} {\bibfield  {journal}
  {\bibinfo  {journal} {Phys. Rev. A}\ }\textbf {\bibinfo {volume} {38}},\
  \bibinfo {pages} {2290} (\bibinfo {year} {1988})}\BibitemShut {NoStop}%
\bibitem [{\citenamefont {Wohrer}\ \emph {et~al.}(1986)\citenamefont {Wohrer},
  \citenamefont {Chetioui}, \citenamefont {Rozet}, \citenamefont {Jolly},
  \citenamefont {Fernandez}, \citenamefont {Stephan}, \citenamefont
  {Brendles},\ and\ \citenamefont {Gayet}}]{woh86a}%
  \BibitemOpen
  \bibfield  {author} {\bibinfo {author} {\bibfnamefont {K.}~\bibnamefont
  {Wohrer}}, \bibinfo {author} {\bibfnamefont {A.}~\bibnamefont {Chetioui}},
  \bibinfo {author} {\bibfnamefont {J.~P.}\ \bibnamefont {Rozet}}, \bibinfo
  {author} {\bibfnamefont {A.}~\bibnamefont {Jolly}}, \bibinfo {author}
  {\bibfnamefont {F.}~\bibnamefont {Fernandez}}, \bibinfo {author}
  {\bibfnamefont {C.}~\bibnamefont {Stephan}}, \bibinfo {author} {\bibfnamefont
  {B.}~\bibnamefont {Brendles}},\ and\ \bibinfo {author} {\bibfnamefont
  {R.}~\bibnamefont {Gayet}},\ }\bibfield  {title} {\bibinfo {title} {Target
  nuclear charge dependence of $1s^2$ to $1s2p$ and $1s^2$ to $1s3p$ excitation
  cross sections of \textsc{F}e$^{24+}$ projectiles in the intermediate
  velocity range},\ }\href {https://doi.org/10.1088/0022-3700/19/13/011}
  {\bibfield  {journal} {\bibinfo  {journal} {J. Phys. B}\ }\textbf {\bibinfo
  {volume} {19}},\ \bibinfo {pages} {1997} (\bibinfo {year}
  {1986})}\BibitemShut {NoStop}%
\bibitem [{\citenamefont {Ali}\ \emph {et~al.}(1991)\citenamefont {Ali},
  \citenamefont {Bhalla}, \citenamefont {Cocke}, \citenamefont {Schulz},\ and\
  \citenamefont {Stockli}}]{ali91a}%
  \BibitemOpen
  \bibfield  {author} {\bibinfo {author} {\bibfnamefont {R.}~\bibnamefont
  {Ali}}, \bibinfo {author} {\bibfnamefont {C.~P.}\ \bibnamefont {Bhalla}},
  \bibinfo {author} {\bibfnamefont {C.~L.}\ \bibnamefont {Cocke}}, \bibinfo
  {author} {\bibfnamefont {M.}~\bibnamefont {Schulz}},\ and\ \bibinfo {author}
  {\bibfnamefont {M.}~\bibnamefont {Stockli}},\ }\bibfield  {title} {\bibinfo
  {title} {Dielectronic recombination on and electron-impact excitation of
  heliumlike argon},\ }\href {https://doi.org/10.1103/PhysRevA.44.223}
  {\bibfield  {journal} {\bibinfo  {journal} {Phys. Rev. A}\ }\textbf {\bibinfo
  {volume} {44}},\ \bibinfo {pages} {223} (\bibinfo {year} {1991})}\BibitemShut
  {NoStop}%
\bibitem [{\citenamefont {Chabot}\ \emph {et~al.}(1994)\citenamefont {Chabot},
  \citenamefont {Wohrer}, \citenamefont {Chetioui}, \citenamefont {Rozet},
  \citenamefont {Touati}, \citenamefont {Vernhet}, \citenamefont {Politis},
  \citenamefont {Stephan}, \citenamefont {Grandin}, \citenamefont {Macias},
  \citenamefont {Martin}, \citenamefont {Riera}, \citenamefont {Sanz},\ and\
  \citenamefont {Gayet}}]{cha94a}%
  \BibitemOpen
  \bibfield  {author} {\bibinfo {author} {\bibfnamefont {M.}~\bibnamefont
  {Chabot}}, \bibinfo {author} {\bibfnamefont {K.}~\bibnamefont {Wohrer}},
  \bibinfo {author} {\bibfnamefont {A.}~\bibnamefont {Chetioui}}, \bibinfo
  {author} {\bibfnamefont {J.~P.}\ \bibnamefont {Rozet}}, \bibinfo {author}
  {\bibfnamefont {A.}~\bibnamefont {Touati}}, \bibinfo {author} {\bibfnamefont
  {D.}~\bibnamefont {Vernhet}}, \bibinfo {author} {\bibfnamefont {M.~F.}\
  \bibnamefont {Politis}}, \bibinfo {author} {\bibfnamefont {C.}~\bibnamefont
  {Stephan}}, \bibinfo {author} {\bibfnamefont {J.~P.}\ \bibnamefont
  {Grandin}}, \bibinfo {author} {\bibfnamefont {A.}~\bibnamefont {Macias}},
  \bibinfo {author} {\bibfnamefont {F.}~\bibnamefont {Martin}}, \bibinfo
  {author} {\bibfnamefont {A.}~\bibnamefont {Riera}}, \bibinfo {author}
  {\bibfnamefont {J.~L.}\ \bibnamefont {Sanz}},\ and\ \bibinfo {author}
  {\bibfnamefont {R.}~\bibnamefont {Gayet}},\ }\bibfield  {title} {\bibinfo
  {title} {New investigation of saturation effect in ion-atom excitation},\
  }\href {http://stacks.iop.org/0953-4075/27/i=1/a=015} {\bibfield  {journal}
  {\bibinfo  {journal} {J. Phys. B}\ }\textbf {\bibinfo {volume} {27}},\
  \bibinfo {pages} {111} (\bibinfo {year} {1994})}\BibitemShut {NoStop}%
\bibitem [{\citenamefont {Adoui}\ \emph {et~al.}(1995)\citenamefont {Adoui},
  \citenamefont {Vernhet}, \citenamefont {Wohrer}, \citenamefont {Plante},
  \citenamefont {Chetioui}, \citenamefont {Rozet}, \citenamefont {Despiney},
  \citenamefont {Stephan}, \citenamefont {Touati}, \citenamefont {Ramillon},
  \citenamefont {Cassimi}, \citenamefont {Grandin},\ and\ \citenamefont
  {Cornille}}]{ado95a}%
  \BibitemOpen
  \bibfield  {author} {\bibinfo {author} {\bibfnamefont {L.}~\bibnamefont
  {Adoui}}, \bibinfo {author} {\bibfnamefont {D.}~\bibnamefont {Vernhet}},
  \bibinfo {author} {\bibfnamefont {K.}~\bibnamefont {Wohrer}}, \bibinfo
  {author} {\bibfnamefont {J.}~\bibnamefont {Plante}}, \bibinfo {author}
  {\bibfnamefont {A.}~\bibnamefont {Chetioui}}, \bibinfo {author}
  {\bibfnamefont {J.}~\bibnamefont {Rozet}}, \bibinfo {author} {\bibfnamefont
  {I.}~\bibnamefont {Despiney}}, \bibinfo {author} {\bibfnamefont
  {C.}~\bibnamefont {Stephan}}, \bibinfo {author} {\bibfnamefont
  {A.}~\bibnamefont {Touati}}, \bibinfo {author} {\bibfnamefont
  {J.}~\bibnamefont {Ramillon}}, \bibinfo {author} {\bibfnamefont
  {A.}~\bibnamefont {Cassimi}}, \bibinfo {author} {\bibfnamefont
  {J.}~\bibnamefont {Grandin}},\ and\ \bibinfo {author} {\bibfnamefont
  {M.}~\bibnamefont {Cornille}},\ }\bibfield  {title} {\bibinfo {title}
  {Excitation of \textsc{A}r$^{16+}$ projectiles in intermediate velocity
  collisions with neutrals},\ }\href
  {https://doi.org/http://dx.doi.org/10.1016/0168-583X(95)00134-4} {\bibfield
  {journal} {\bibinfo  {journal} {Nucl. Instrum. Methods B}\ }\textbf {\bibinfo
  {volume} {98}},\ \bibinfo {pages} {312 } (\bibinfo {year}
  {1995})}\BibitemShut {NoStop}%
\bibitem [{\citenamefont {Matthews}\ \emph
  {et~al.}(1976{\natexlab{b}})\citenamefont {Matthews}, \citenamefont
  {Fortner},\ and\ \citenamefont {Bissinger}}]{matt76b}%
  \BibitemOpen
  \bibfield  {author} {\bibinfo {author} {\bibfnamefont {D.~L.}\ \bibnamefont
  {Matthews}}, \bibinfo {author} {\bibfnamefont {R.~J.}\ \bibnamefont
  {Fortner}},\ and\ \bibinfo {author} {\bibfnamefont {G.}~\bibnamefont
  {Bissinger}},\ }\bibfield  {title} {\bibinfo {title} {Collisional quenching
  of metastable x-ray-emitting states in a fast beam of \textsc{H}e-like
  fluorine},\ }\href {https://doi.org/10.1103/PhysRevLett.36.664} {\bibfield
  {journal} {\bibinfo  {journal} {Phys. Rev. Lett.}\ }\textbf {\bibinfo
  {volume} {36}},\ \bibinfo {pages} {664} (\bibinfo {year}
  {1976}{\natexlab{b}})}\BibitemShut {NoStop}%
\bibitem [{\citenamefont {Zuccatti}\ \emph {et~al.}(1985)\citenamefont
  {Zuccatti}, \citenamefont {Bruch},\ and\ \citenamefont {Rissler}}]{zuc85a}%
  \BibitemOpen
  \bibfield  {author} {\bibinfo {author} {\bibfnamefont {S.}~\bibnamefont
  {Zuccatti}}, \bibinfo {author} {\bibfnamefont {R.}~\bibnamefont {Bruch}},\
  and\ \bibinfo {author} {\bibfnamefont {J.}~\bibnamefont {Rissler}},\
  }\bibfield  {title} {\bibinfo {title} {Excitation cross sections of
  \textsc{L}i$^+(1snl)\,^3\!\textsc{L}$, $^1\!\textsc{L}$ states in collisions
  of \textsc{L}i$^+$ with \textsc{H}e},\ }\href
  {https://doi.org/10.1016/0168-583X(85)90242-3} {\bibfield  {journal}
  {\bibinfo  {journal} {Nucl. Instrum. Methods B}\ }\textbf {\bibinfo {volume}
  {10-11}},\ \bibinfo {pages} {231 } (\bibinfo {year} {1985})}\BibitemShut
  {NoStop}%
\bibitem [{\citenamefont {Newcomb}\ \emph {et~al.}(1984)\citenamefont
  {Newcomb}, \citenamefont {Dillingham}, \citenamefont {Hall}, \citenamefont
  {Varghese}, \citenamefont {Pepmiller},\ and\ \citenamefont
  {Richard}}]{new84b}%
  \BibitemOpen
  \bibfield  {author} {\bibinfo {author} {\bibfnamefont {J.}~\bibnamefont
  {Newcomb}}, \bibinfo {author} {\bibfnamefont {T.~R.}\ \bibnamefont
  {Dillingham}}, \bibinfo {author} {\bibfnamefont {J.}~\bibnamefont {Hall}},
  \bibinfo {author} {\bibfnamefont {S.~L.}\ \bibnamefont {Varghese}}, \bibinfo
  {author} {\bibfnamefont {P.~L.}\ \bibnamefont {Pepmiller}},\ and\ \bibinfo
  {author} {\bibfnamefont {P.}~\bibnamefont {Richard}},\ }\bibfield  {title}
  {\bibinfo {title} {Charge-state dependence of fluorine-projectile \textsc{K}
  \textsc{A}uger-electron production},\ }\href@noop {} {\bibfield  {journal}
  {\bibinfo  {journal} {Phys. Rev. A}\ }\textbf {\bibinfo {volume} {30}},\
  \bibinfo {pages} {106} (\bibinfo {year} {1984})}\BibitemShut {NoStop}%
\bibitem [{\citenamefont {Lee}\ \emph {et~al.}(1991)\citenamefont {Lee},
  \citenamefont {Richard}, \citenamefont {Sanders}, \citenamefont {Zouros},
  \citenamefont {Shinpaugh},\ and\ \citenamefont {Varghese}}]{lee91a}%
  \BibitemOpen
  \bibfield  {author} {\bibinfo {author} {\bibfnamefont {D.~H.}\ \bibnamefont
  {Lee}}, \bibinfo {author} {\bibfnamefont {P.}~\bibnamefont {Richard}},
  \bibinfo {author} {\bibfnamefont {J.~M.}\ \bibnamefont {Sanders}}, \bibinfo
  {author} {\bibfnamefont {T.~J.~M.}\ \bibnamefont {Zouros}}, \bibinfo {author}
  {\bibfnamefont {J.~L.}\ \bibnamefont {Shinpaugh}},\ and\ \bibinfo {author}
  {\bibfnamefont {S.~L.}\ \bibnamefont {Varghese}},\ }\bibfield  {title}
  {\bibinfo {title} {Electron capture and excitation studied by state-resolved
  \textsc{KLL} \textsc{A}uger measurement in 0.25-2 \textsc{M}e\textsc{V}/u
  \textsc{F}$^{7+}(1s^2\thinspace^1\!\textsc{S}, 1s2s\thinspace^3\!\textsc{S})$
  + \textsc{H}$_2$/\textsc{H}e collisions},\ }\href
  {https://doi.org/https://doi.org/10.1016/0168-583X(91)95981-I} {\bibfield
  {journal} {\bibinfo  {journal} {Nucl. Instrum. Methods B}\ }\textbf {\bibinfo
  {volume} {56-57}},\ \bibinfo {pages} {99} (\bibinfo {year}
  {1991})}\BibitemShut {NoStop}%
\bibitem [{\citenamefont {Harissopulos}\ \emph {et~al.}(2021)\citenamefont
  {Harissopulos}, \citenamefont {Andrianis}, \citenamefont {Axiotis},
  \citenamefont {Lagoyannis}, \citenamefont {Karydas}, \citenamefont {Kotsina},
  \citenamefont {Laoutaris}, \citenamefont {Apostolopoulos}, \citenamefont
  {Theodorou}, \citenamefont {Zouros}, \citenamefont {Madesis},\ and\
  \citenamefont {Benis}}]{har21a}%
  \BibitemOpen
  \bibfield  {author} {\bibinfo {author} {\bibfnamefont {S.}~\bibnamefont
  {Harissopulos}}, \bibinfo {author} {\bibfnamefont {M.}~\bibnamefont
  {Andrianis}}, \bibinfo {author} {\bibfnamefont {M.}~\bibnamefont {Axiotis}},
  \bibinfo {author} {\bibfnamefont {A.}~\bibnamefont {Lagoyannis}}, \bibinfo
  {author} {\bibfnamefont {A.~G.}\ \bibnamefont {Karydas}}, \bibinfo {author}
  {\bibfnamefont {Z.}~\bibnamefont {Kotsina}}, \bibinfo {author} {\bibfnamefont
  {A.}~\bibnamefont {Laoutaris}}, \bibinfo {author} {\bibfnamefont
  {G.}~\bibnamefont {Apostolopoulos}}, \bibinfo {author} {\bibfnamefont
  {A.}~\bibnamefont {Theodorou}}, \bibinfo {author} {\bibfnamefont {T.~J.~M.}\
  \bibnamefont {Zouros}}, \bibinfo {author} {\bibfnamefont {I.}~\bibnamefont
  {Madesis}},\ and\ \bibinfo {author} {\bibfnamefont {E.~P.}\ \bibnamefont
  {Benis}},\ }\bibfield  {title} {\bibinfo {title} {The \textsc{T}andem
  \textsc{A}ccelerator \textsc{L}aboratory of \textsc{NCSR}
  ``\textsc{D}emokritos": \textsc{C}urrent status and perspectives},\ }\href
  {https://doi.org/10.1140/epjp/s13360-021-01596-5} {\bibfield  {journal}
  {\bibinfo  {journal} {Eur. Phys. J. Plus}\ }\textbf {\bibinfo {volume}
  {136}},\ \bibinfo {pages} {617} (\bibinfo {year} {2021})}\BibitemShut
  {NoStop}%
\bibitem [{\citenamefont {Tr\"{a}bert}(2024)}]{tra24a}%
  \BibitemOpen
  \bibfield  {author} {\bibinfo {author} {\bibfnamefont {E.}~\bibnamefont
  {Tr\"{a}bert}},\ }\bibfield  {title} {\bibinfo {title} {Long-lived levels in
  multiply and highly charged ions},\ }\href
  {https://www.mdpi.com/2218-2004/12/3/12} {\bibfield  {journal} {\bibinfo
  {journal} {Atoms}\ }\textbf {\bibinfo {volume} {12}},\ \bibinfo {pages} {12}
  (\bibinfo {year} {2024})}\BibitemShut {NoStop}%
\bibitem [{\citenamefont {Madesis}\ \emph {et~al.}(2019)\citenamefont
  {Madesis}, \citenamefont {Laoutaris}, \citenamefont {Zouros}, \citenamefont
  {Nanos},\ and\ \citenamefont {Benis}}]{mad19a}%
  \BibitemOpen
  \bibfield  {author} {\bibinfo {author} {\bibfnamefont {I.}~\bibnamefont
  {Madesis}}, \bibinfo {author} {\bibfnamefont {A.}~\bibnamefont {Laoutaris}},
  \bibinfo {author} {\bibfnamefont {T.~J.~M.}\ \bibnamefont {Zouros}}, \bibinfo
  {author} {\bibfnamefont {S.}~\bibnamefont {Nanos}},\ and\ \bibinfo {author}
  {\bibfnamefont {E.~P.}\ \bibnamefont {Benis}},\ }\bibfield  {title} {\bibinfo
  {title} {Projectile electron spectroscopy and new answers to old questions:
  Latest results at the new atomic physics beamline in \textsc{D}emokritos,
  \textsc{A}thens},\ }in\ \href {https://doi.org/10.1142/11588} {\emph
  {\bibinfo {booktitle} {State-of-the-Art Reviews on Energetic Ion-Atom and
  Ion-Molecule Collisions}}},\ \bibinfo {series} {Interdisciplinary Research on
  Particle Collisions and Quantitative Spectroscopy}, Vol.~\bibinfo {volume}
  {2},\ \bibinfo {editor} {edited by\ \bibinfo {editor} {\bibfnamefont
  {D.}~\bibnamefont {Belki\'{c}}}, \bibinfo {editor} {\bibfnamefont
  {I.}~\bibnamefont {Bray}},\ and\ \bibinfo {editor} {\bibfnamefont
  {A.}~\bibnamefont {Kadyrov}}}\ (\bibinfo  {publisher} {World Scientific},\
  \bibinfo {address} {Singapore},\ \bibinfo {year} {2019})\ Chap.~\bibinfo
  {chapter} {1}, pp.\ \bibinfo {pages} {1--31}\BibitemShut {NoStop}%
\bibitem [{\citenamefont {\emph{SIS} \emph{Scientific Instrument
  Services}}(2014)}]{sim14a}%
  \BibitemOpen
  \bibfield  {author} {\bibinfo {author} {\bibnamefont {\emph{SIS}
  \emph{Scientific Instrument Services}}},\ }\href@noop {} {\emph {\bibinfo
  {title} {SIMION version 8.1.2.20}}},\ \bibinfo {address} {Ringoes, NJ, see
  http://www.simion.com} (\bibinfo {year} {2014})\BibitemShut {NoStop}%
\bibitem [{\citenamefont {Nanos}\ \emph
  {et~al.}(2023{\natexlab{b}})\citenamefont {Nanos}, \citenamefont {Biniskos},
  \citenamefont {Laoutaris}, \citenamefont {Andrianis}, \citenamefont {Zouros},
  \citenamefont {Lagoyannis},\ and\ \citenamefont {Benis}}]{nan23a}%
  \BibitemOpen
  \bibfield  {author} {\bibinfo {author} {\bibfnamefont {S.}~\bibnamefont
  {Nanos}}, \bibinfo {author} {\bibfnamefont {A.}~\bibnamefont {Biniskos}},
  \bibinfo {author} {\bibfnamefont {A.}~\bibnamefont {Laoutaris}}, \bibinfo
  {author} {\bibfnamefont {M.}~\bibnamefont {Andrianis}}, \bibinfo {author}
  {\bibfnamefont {T.~J.~M.}\ \bibnamefont {Zouros}}, \bibinfo {author}
  {\bibfnamefont {A.}~\bibnamefont {Lagoyannis}},\ and\ \bibinfo {author}
  {\bibfnamefont {E.~P.}\ \bibnamefont {Benis}},\ }\bibfield  {title} {\bibinfo
  {title} {Determination of the ion beam energy width in tandem \textsc{V}an de
  \textsc{G}raaff accelerators via \textsc{A}uger projectile spectroscopy},\
  }\href {https://doi.org/https://doi.org/10.1016/j.nimb.2023.05.027}
  {\bibfield  {journal} {\bibinfo  {journal} {Nucl. Instrum. Methods B}\
  }\textbf {\bibinfo {volume} {541}},\ \bibinfo {pages} {93} (\bibinfo {year}
  {2023}{\natexlab{b}})}\BibitemShut {NoStop}%
\bibitem [{\citenamefont {Bruch}\ \emph {et~al.}(1979)\citenamefont {Bruch},
  \citenamefont {Schneider}, \citenamefont {Schwarz}, \citenamefont {Meinhart},
  \citenamefont {Johnson},\ and\ \citenamefont {Taulbjerg}}]{bru79a}%
  \BibitemOpen
  \bibfield  {author} {\bibinfo {author} {\bibfnamefont {R.}~\bibnamefont
  {Bruch}}, \bibinfo {author} {\bibfnamefont {D.}~\bibnamefont {Schneider}},
  \bibinfo {author} {\bibfnamefont {W.~H.~E.}\ \bibnamefont {Schwarz}},
  \bibinfo {author} {\bibfnamefont {M.}~\bibnamefont {Meinhart}}, \bibinfo
  {author} {\bibfnamefont {B.~M.}\ \bibnamefont {Johnson}},\ and\ \bibinfo
  {author} {\bibfnamefont {K.}~\bibnamefont {Taulbjerg}},\ }\bibfield  {title}
  {\bibinfo {title} {Projectile \textsc{A}uger spectra of gas- and foil-excited
  oxygen ions at \textsc{M}e\textsc{V} energies},\ }\href
  {https://doi.org/10.1103/PhysRevA.19.587} {\bibfield  {journal} {\bibinfo
  {journal} {Phys. Rev. A}\ }\textbf {\bibinfo {volume} {19}},\ \bibinfo
  {pages} {587} (\bibinfo {year} {1979})}\BibitemShut {NoStop}%
\bibitem [{\citenamefont {Kilgus}\ \emph {et~al.}(1990)\citenamefont {Kilgus},
  \citenamefont {Berger}, \citenamefont {Blatt}, \citenamefont {Grieser},
  \citenamefont {Habs}, \citenamefont {Hochadel}, \citenamefont {Jaeschke},
  \citenamefont {Kr\"amer}, \citenamefont {Neumann}, \citenamefont
  {Neureither}, \citenamefont {Ott}, \citenamefont {Schwalm}, \citenamefont
  {Steck}, \citenamefont {Stokstad}, \citenamefont {Szmola}, \citenamefont
  {Wolf}, \citenamefont {Schuch}, \citenamefont {M\"uller},\ and\ \citenamefont
  {Wagner}}]{kil90a}%
  \BibitemOpen
  \bibfield  {author} {\bibinfo {author} {\bibfnamefont {G.}~\bibnamefont
  {Kilgus}}, \bibinfo {author} {\bibfnamefont {J.}~\bibnamefont {Berger}},
  \bibinfo {author} {\bibfnamefont {P.}~\bibnamefont {Blatt}}, \bibinfo
  {author} {\bibfnamefont {M.}~\bibnamefont {Grieser}}, \bibinfo {author}
  {\bibfnamefont {D.}~\bibnamefont {Habs}}, \bibinfo {author} {\bibfnamefont
  {B.}~\bibnamefont {Hochadel}}, \bibinfo {author} {\bibfnamefont
  {E.}~\bibnamefont {Jaeschke}}, \bibinfo {author} {\bibfnamefont
  {D.}~\bibnamefont {Kr\"amer}}, \bibinfo {author} {\bibfnamefont
  {R.}~\bibnamefont {Neumann}}, \bibinfo {author} {\bibfnamefont
  {G.}~\bibnamefont {Neureither}}, \bibinfo {author} {\bibfnamefont
  {W.}~\bibnamefont {Ott}}, \bibinfo {author} {\bibfnamefont {D.}~\bibnamefont
  {Schwalm}}, \bibinfo {author} {\bibfnamefont {M.}~\bibnamefont {Steck}},
  \bibinfo {author} {\bibfnamefont {R.}~\bibnamefont {Stokstad}}, \bibinfo
  {author} {\bibfnamefont {E.}~\bibnamefont {Szmola}}, \bibinfo {author}
  {\bibfnamefont {A.}~\bibnamefont {Wolf}}, \bibinfo {author} {\bibfnamefont
  {R.}~\bibnamefont {Schuch}}, \bibinfo {author} {\bibfnamefont
  {A.}~\bibnamefont {M\"uller}},\ and\ \bibinfo {author} {\bibfnamefont
  {M.}~\bibnamefont {Wagner}},\ }\bibfield  {title} {\bibinfo {title}
  {\textsc{D}ielectronic recombination of hydrogenlike oxygen in a heavy-ion
  storage ring},\ }\href {https://doi.org/10.1103/PhysRevLett.64.737}
  {\bibfield  {journal} {\bibinfo  {journal} {Phys. Rev. Lett.}\ }\textbf
  {\bibinfo {volume} {64}},\ \bibinfo {pages} {737} (\bibinfo {year}
  {1990})}\BibitemShut {NoStop}%
\bibitem [{\citenamefont {Azarov}\ \emph {et~al.}(2023)\citenamefont {Azarov},
  \citenamefont {Kramida},\ and\ \citenamefont {Ralchenko}}]{aza23a}%
  \BibitemOpen
  \bibfield  {author} {\bibinfo {author} {\bibfnamefont {V.}~\bibnamefont
  {Azarov}}, \bibinfo {author} {\bibfnamefont {A.}~\bibnamefont {Kramida}},\
  and\ \bibinfo {author} {\bibfnamefont {Y.}~\bibnamefont {Ralchenko}},\
  }\bibfield  {title} {\bibinfo {title} {A critical compilation of experimental
  data on the $1s2l2l^\prime$ core-excited states of \textsc{L}i-like ions from
  carbon to uranium},\ }\href
  {https://doi.org/https://doi.org/10.1016/j.adt.2022.101548} {\bibfield
  {journal} {\bibinfo  {journal} {At. Data Nucl. Data Tables}\ }\textbf
  {\bibinfo {volume} {149}},\ \bibinfo {pages} {101548} (\bibinfo {year}
  {2023})}\BibitemShut {NoStop}%
\bibitem [{\citenamefont {Sisourat}\ \emph {et~al.}(2011)\citenamefont
  {Sisourat}, \citenamefont {Pilskog},\ and\ \citenamefont {Dubois}}]{sis11a}%
  \BibitemOpen
  \bibfield  {author} {\bibinfo {author} {\bibfnamefont {N.}~\bibnamefont
  {Sisourat}}, \bibinfo {author} {\bibfnamefont {I.}~\bibnamefont {Pilskog}},\
  and\ \bibinfo {author} {\bibfnamefont {A.}~\bibnamefont {Dubois}},\
  }\bibfield  {title} {\bibinfo {title} {Non perturbative treatment of
  multielectron processes in ion-molecule scattering: Application to
  \textsc{H}e${}^{2+}$-\textsc{H}${}_{2}$ collisions},\ }\href
  {https://doi.org/10.1103/PhysRevA.84.052722} {\bibfield  {journal} {\bibinfo
  {journal} {Phys. Rev. A}\ }\textbf {\bibinfo {volume} {84}},\ \bibinfo
  {pages} {052722} (\bibinfo {year} {2011})}\BibitemShut {NoStop}%
\bibitem [{\citenamefont {Gao}\ \emph {et~al.}(2018)\citenamefont {Gao},
  \citenamefont {Wu}, \citenamefont {Wang}, \citenamefont {Sisourat},\ and\
  \citenamefont {Dubois}}]{gao18a}%
  \BibitemOpen
  \bibfield  {author} {\bibinfo {author} {\bibfnamefont {J.~W.}\ \bibnamefont
  {Gao}}, \bibinfo {author} {\bibfnamefont {Y.}~\bibnamefont {Wu}}, \bibinfo
  {author} {\bibfnamefont {J.~G.}\ \bibnamefont {Wang}}, \bibinfo {author}
  {\bibfnamefont {N.}~\bibnamefont {Sisourat}},\ and\ \bibinfo {author}
  {\bibfnamefont {A.}~\bibnamefont {Dubois}},\ }\bibfield  {title} {\bibinfo
  {title} {State-selective electron transfer in \textsc{H}e$^{+}$ + \textsc{H}e
  collisions at intermediate energies},\ }\href
  {https://doi.org/10.1103/PhysRevA.97.052709} {\bibfield  {journal} {\bibinfo
  {journal} {Phys. Rev. A}\ }\textbf {\bibinfo {volume} {97}},\ \bibinfo
  {pages} {052709} (\bibinfo {year} {2018})}\BibitemShut {NoStop}%
\bibitem [{\citenamefont {Igarashi}\ and\ \citenamefont {Kato}(2024)}]{iga24b}%
  \BibitemOpen
  \bibfield  {author} {\bibinfo {author} {\bibfnamefont {A.}~\bibnamefont
  {Igarashi}}\ and\ \bibinfo {author} {\bibfnamefont {D.}~\bibnamefont
  {Kato}},\ }\bibfield  {title} {\bibinfo {title} {Cross sections in
  \textsc{H}e$^+$ + \textsc{H}e collision at intermediate energies},\ }\href
  {https://doi.org/10.1140/epjd/s10053-024-00875-x} {\bibfield  {journal}
  {\bibinfo  {journal} {Eur. Phys. J. D}\ }\textbf {\bibinfo {volume} {78}},\
  \bibinfo {pages} {79} (\bibinfo {year} {2024})}\BibitemShut {NoStop}%
\bibitem [{\citenamefont {Madesis}(2021)}]{mad21a}%
  \BibitemOpen
  \bibfield  {author} {\bibinfo {author} {\bibfnamefont {I.}~\bibnamefont
  {Madesis}},\ }\emph {\bibinfo {title} {Investigation of Electron Capture in
  Swift \textsc{C}$^{4+}(1s2s\,^3\!\textsc{S})$ Collisions with Gas Targets
  Using a Zero-degree \textsc{A}uger Projectile Spectroscopy Apparatus Built
  with the \textsc{L}45 Beam Line at the ``Demokritos" 5.5~\textsc{MV} Tandem
  Accelerator}},\ \href@noop {} {Ph.D. thesis},\ \bibinfo  {school} {University
  of Crete} (\bibinfo {year} {2021})\BibitemShut {NoStop}%
\bibitem [{\citenamefont {Kramida}\ \emph {et~al.}(2024)\citenamefont
  {Kramida}, \citenamefont {{Yu.~Ralchenko}},\ and\ \citenamefont {{and NIST
  ASD Team}}}]{kra24a}%
  \BibitemOpen
  \bibfield  {author} {\bibinfo {author} {\bibfnamefont {A.}~\bibnamefont
  {Kramida}}, \bibinfo {author} {\bibnamefont {{Yu.~Ralchenko}}},  
  \bibinfo {author} {\bibfnamefont {J.}~\bibnamefont
  {Reader}}\ and\ \bibinfo {author}{\bibnamefont{NIST ASD
  Team}},\ }\href {https://doi.org/https://doi.org/10.18434/T4W30F} {\bibinfo
  {title} {\textit{{NIST} atomic spectra database (v5.11)}, [Online] Available:
  https://physics.nist.gov/asd}} (\bibinfo {year} {2024}),\ \bibinfo {note}
  {\textsc{G}aithersburg, \textsc{MD}: \textsc{N}ational Institute of Standards
  and Technology}\BibitemShut {NoStop}%
\bibitem [{\citenamefont {M\"uller}\ \emph {et~al.}(2018)\citenamefont
  {M\"uller}, \citenamefont {Lindroth}, \citenamefont {Bari}, \citenamefont
  {Borovik}, \citenamefont {Hillenbrand}, \citenamefont {Holste}, \citenamefont
  {Indelicato}, \citenamefont {Kilcoyne}, \citenamefont {Klumpp}, \citenamefont
  {Martins}, \citenamefont {Viefhaus}, \citenamefont {Wilhelm},\ and\
  \citenamefont {Schippers}}]{mul18b}%
  \BibitemOpen
  \bibfield  {author} {\bibinfo {author} {\bibfnamefont {A.}~\bibnamefont
  {M\"uller}}, \bibinfo {author} {\bibfnamefont {E.}~\bibnamefont {Lindroth}},
  \bibinfo {author} {\bibfnamefont {S.}~\bibnamefont {Bari}}, \bibinfo {author}
  {\bibfnamefont {A.}~\bibnamefont {Borovik}}, \bibinfo {author} {\bibfnamefont
  {P.-M.}\ \bibnamefont {Hillenbrand}}, \bibinfo {author} {\bibfnamefont
  {K.}~\bibnamefont {Holste}}, \bibinfo {author} {\bibfnamefont
  {P.}~\bibnamefont {Indelicato}}, \bibinfo {author} {\bibfnamefont {A.~L.~D.}\
  \bibnamefont {Kilcoyne}}, \bibinfo {author} {\bibfnamefont {S.}~\bibnamefont
  {Klumpp}}, \bibinfo {author} {\bibfnamefont {M.}~\bibnamefont {Martins}},
  \bibinfo {author} {\bibfnamefont {J.}~\bibnamefont {Viefhaus}}, \bibinfo
  {author} {\bibfnamefont {P.}~\bibnamefont {Wilhelm}},\ and\ \bibinfo {author}
  {\bibfnamefont {S.}~\bibnamefont {Schippers}},\ }\bibfield  {title} {\bibinfo
  {title} {Photoionization of metastable heliumlike
  \textsc{C}$^{4+}(1s2s\,^{3}\textsc{S}_{1})$ ions: Precision study of
  intermediate doubly excited states},\ }\href
  {https://doi.org/10.1103/PhysRevA.98.033416} {\bibfield  {journal} {\bibinfo
  {journal} {Phys. Rev. A}\ }\textbf {\bibinfo {volume} {98}},\ \bibinfo
  {pages} {033416} (\bibinfo {year} {2018})}\BibitemShut {NoStop}%
\bibitem [{\citenamefont {van~der Hart}\ and\ \citenamefont
  {Hansen}(1993)}]{van93a}%
  \BibitemOpen
  \bibfield  {author} {\bibinfo {author} {\bibfnamefont {H.~W.}\ \bibnamefont
  {van~der Hart}}\ and\ \bibinfo {author} {\bibfnamefont {J.~E.}\ \bibnamefont
  {Hansen}},\ }\bibfield  {title} {\bibinfo {title} {Competition between
  radiative and non-radiative decay for doubly excited $2lnl^\prime$ and
  $3lnl^\prime$ states in \textsc{C}$^{4+}$},\ }\href
  {http://stacks.iop.org/0953-4075/26/i=4/a=007} {\bibfield  {journal}
  {\bibinfo  {journal} {J. Phys. B}\ }\textbf {\bibinfo {volume} {26}},\
  \bibinfo {pages} {641} (\bibinfo {year} {1993})}\BibitemShut {NoStop}%
\bibitem [{\citenamefont {Chen}(1992)}]{che92a}%
  \BibitemOpen
  \bibfield  {author} {\bibinfo {author} {\bibfnamefont {M.~H.}\ \bibnamefont
  {Chen}},\ }\bibfield  {title} {\bibinfo {title} {Effect of intermediate
  coupling on angular distribution of \textsc{A}uger electrons},\ }\href
  {https://doi.org/10.1103/PhysRevA.45.1684} {\bibfield  {journal} {\bibinfo
  {journal} {Phys. Rev. A}\ }\textbf {\bibinfo {volume} {45}},\ \bibinfo
  {pages} {1684} (\bibinfo {year} {1992})}\BibitemShut {NoStop}%
\bibitem [{\citenamefont {Surzhykov}\ \emph {et~al.}(2008)\citenamefont
  {Surzhykov}, \citenamefont {Jentschura}, \citenamefont {St\"ohlker},
  \citenamefont {Gumberidze},\ and\ \citenamefont {Fritzsche}}]{sur08a}%
  \BibitemOpen
  \bibfield  {author} {\bibinfo {author} {\bibfnamefont {A.}~\bibnamefont
  {Surzhykov}}, \bibinfo {author} {\bibfnamefont {U.~D.}\ \bibnamefont
  {Jentschura}}, \bibinfo {author} {\bibfnamefont {T.}~\bibnamefont
  {St\"ohlker}}, \bibinfo {author} {\bibfnamefont {A.}~\bibnamefont
  {Gumberidze}},\ and\ \bibinfo {author} {\bibfnamefont {S.}~\bibnamefont
  {Fritzsche}},\ }\bibfield  {title} {\bibinfo {title} {Alignment of heavy
  few-electron ions following excitation by relativistic \textsc{C}oulomb
  collisions},\ }\href {https://doi.org/10.1103/PhysRevA.77.042722} {\bibfield
  {journal} {\bibinfo  {journal} {Phys. Rev. A}\ }\textbf {\bibinfo {volume}
  {77}},\ \bibinfo {pages} {042722} (\bibinfo {year} {2008})}\BibitemShut
  {NoStop}%
\bibitem [{\citenamefont {Fritzsche}\ \emph {et~al.}(2012)\citenamefont
  {Fritzsche}, \citenamefont {Surzhykov}, \citenamefont {Gumberidze},\ and\
  \citenamefont {St\"{o}hlker}}]{fri12a}%
  \BibitemOpen
  \bibfield  {author} {\bibinfo {author} {\bibfnamefont {S.}~\bibnamefont
  {Fritzsche}}, \bibinfo {author} {\bibfnamefont {A.}~\bibnamefont
  {Surzhykov}}, \bibinfo {author} {\bibfnamefont {A.}~\bibnamefont
  {Gumberidze}},\ and\ \bibinfo {author} {\bibfnamefont {T.}~\bibnamefont
  {St\"{o}hlker}},\ }\bibfield  {title} {\bibinfo {title} {Electron emission
  from highly charged ions: Signatures of magnetic interactions and retardation
  in strong fields},\ }\href {https://doi.org/10.1088/1367-2630/14/8/083018}
  {\bibfield  {journal} {\bibinfo  {journal} {New J. Phys.}\ }\textbf {\bibinfo
  {volume} {14}},\ \bibinfo {pages} {083018} (\bibinfo {year}
  {2012})}\BibitemShut {NoStop}%
\bibitem [{\citenamefont {Kabachnik}\ \emph {et~al.}(1994)\citenamefont
  {Kabachnik}, \citenamefont {Tulkki}, \citenamefont {Aksela},\ and\
  \citenamefont {Ricz}}]{kab94a}%
  \BibitemOpen
  \bibfield  {author} {\bibinfo {author} {\bibfnamefont {N.~M.}\ \bibnamefont
  {Kabachnik}}, \bibinfo {author} {\bibfnamefont {J.}~\bibnamefont {Tulkki}},
  \bibinfo {author} {\bibfnamefont {H.}~\bibnamefont {Aksela}},\ and\ \bibinfo
  {author} {\bibfnamefont {S.}~\bibnamefont {Ricz}},\ }\bibfield  {title}
  {\bibinfo {title} {Coherence and correlation in the anisotropy of \textsc{N}e
  \textsc{KL-LLL} satellite \textsc{A}uger decay},\ }\href
  {https://doi.org/10.1103/PhysRevA.49.4653} {\bibfield  {journal} {\bibinfo
  {journal} {Phys. Rev. A}\ }\textbf {\bibinfo {volume} {49}},\ \bibinfo
  {pages} {4653} (\bibinfo {year} {1994})}\BibitemShut {NoStop}%
\bibitem [{\citenamefont {Mehlhorn}\ and\ \citenamefont
  {Taulbjerg}(1980)}]{mehl80a}%
  \BibitemOpen
  \bibfield  {author} {\bibinfo {author} {\bibfnamefont {W.}~\bibnamefont
  {Mehlhorn}}\ and\ \bibinfo {author} {\bibfnamefont {K.}~\bibnamefont
  {Taulbjerg}},\ }\bibfield  {title} {\bibinfo {title} {Angular distribution of
  electrons from autoionising states with unresolved fine structure},\ }\href
  {http://stacks.iop.org/0022-3700/13/i=3/a=007} {\bibfield  {journal}
  {\bibinfo  {journal} {J. Phys. B}\ }\textbf {\bibinfo {volume} {13}},\
  \bibinfo {pages} {445} (\bibinfo {year} {1980})}\BibitemShut {NoStop}%
\bibitem [{\citenamefont {M\"uller}\ \emph {et~al.}(2025)\citenamefont
  {M\"uller}, \citenamefont {Hillenbrand}, \citenamefont {Wang}, \citenamefont
  {Schippers}, \citenamefont {Bray}, \citenamefont {Kheifets}, \citenamefont
  {Lindroth}, \citenamefont {Reinwardt}, \citenamefont {Martins}, \citenamefont
  {Seltmann},\ and\ \citenamefont {Trinter}}]{mul25a}%
  \BibitemOpen
  \bibfield  {author} {\bibinfo {author} {\bibfnamefont {A.}~\bibnamefont
  {M\"uller}}, \bibinfo {author} {\bibfnamefont {P.-M.}\ \bibnamefont
  {Hillenbrand}}, \bibinfo {author} {\bibfnamefont {S.-X.}\ \bibnamefont
  {Wang}}, \bibinfo {author} {\bibfnamefont {S.}~\bibnamefont {Schippers}},
  \bibinfo {author} {\bibfnamefont {I.}~\bibnamefont {Bray}}, \bibinfo {author}
  {\bibfnamefont {A.~S.}\ \bibnamefont {Kheifets}}, \bibinfo {author}
  {\bibfnamefont {E.}~\bibnamefont {Lindroth}}, \bibinfo {author}
  {\bibfnamefont {S.}~\bibnamefont {Reinwardt}}, \bibinfo {author}
  {\bibfnamefont {M.}~\bibnamefont {Martins}}, \bibinfo {author} {\bibfnamefont
  {J.}~\bibnamefont {Seltmann}},\ and\ \bibinfo {author} {\bibfnamefont
  {F.}~\bibnamefont {Trinter}},\ }\bibfield  {title} {\bibinfo {title} {Direct
  double ionization of the \textsc{H}e-like \textsc{B}$^{3+}$ ion by a single
  photon},\ }\href {https://doi.org/10.1103/PhysRevA.111.023115} {\bibfield
  {journal} {\bibinfo  {journal} {Phys. Rev. A}\ }\textbf {\bibinfo {volume}
  {111}},\ \bibinfo {pages} {023115} (\bibinfo {year} {2025})}\BibitemShut
  {NoStop}%
\bibitem [{\citenamefont {Zouros}\ \emph {et~al.}(1989)\citenamefont {Zouros},
  \citenamefont {Lee},\ and\ \citenamefont {Richard}}]{zou89b}%
  \BibitemOpen
  \bibfield  {author} {\bibinfo {author} {\bibfnamefont {T.~J.~M.}\
  \bibnamefont {Zouros}}, \bibinfo {author} {\bibfnamefont {D.~H.}\
  \bibnamefont {Lee}},\ and\ \bibinfo {author} {\bibfnamefont {P.}~\bibnamefont
  {Richard}},\ }\bibfield  {title} {\bibinfo {title} {Projectile 1s$\to$2p
  excitation due to electron-electron interaction in collisions of
  \textsc{O}$^{5+}$ and \textsc{F}$^{6+}$ ions {w}ith \textsc{H}$_2$ and
  \textsc{H}e targets},\ }\href {https://doi.org/10.1103/PhysRevLett.62.2261}
  {\bibfield  {journal} {\bibinfo  {journal} {Phys. Rev. Lett.}\ }\textbf
  {\bibinfo {volume} {62}},\ \bibinfo {pages} {2261} (\bibinfo {year}
  {1989})}\BibitemShut {NoStop}%
\bibitem [{\citenamefont {Montenegro}\ \emph {et~al.}(1993)\citenamefont
  {Montenegro}, \citenamefont {Belkacem}, \citenamefont {Spooner},
  \citenamefont {Meyerhof},\ and\ \citenamefont {Shah}}]{mon93b}%
  \BibitemOpen
  \bibfield  {author} {\bibinfo {author} {\bibfnamefont {E.~C.}\ \bibnamefont
  {Montenegro}}, \bibinfo {author} {\bibfnamefont {A.}~\bibnamefont
  {Belkacem}}, \bibinfo {author} {\bibfnamefont {D.~W.}\ \bibnamefont
  {Spooner}}, \bibinfo {author} {\bibfnamefont {W.~E.}\ \bibnamefont
  {Meyerhof}},\ and\ \bibinfo {author} {\bibfnamefont {M.~B.}\ \bibnamefont
  {Shah}},\ }\bibfield  {title} {\bibinfo {title} {Impact-parameter dependence
  of the electron-electron interaction in (\textsc{L}i$^{2+}$,
  \textsc{C}$^{5+}$)+(\textsc{H}$_2$,\textsc{H}e) electron loss},\ }\href
  {https://doi.org/10.1103/PhysRevA.47.1045} {\bibfield  {journal} {\bibinfo
  {journal} {Phys. Rev. A}\ }\textbf {\bibinfo {volume} {47}},\ \bibinfo
  {pages} {1045} (\bibinfo {year} {1993})}\BibitemShut {NoStop}%
\bibitem [{\citenamefont {Montenegro}\ \emph {et~al.}(1992)\citenamefont
  {Montenegro}, \citenamefont {Melo}, \citenamefont {Meyerhof},\ and\
  \citenamefont {de~Pinho}}]{mon92a}%
  \BibitemOpen
  \bibfield  {author} {\bibinfo {author} {\bibfnamefont {E.~C.}\ \bibnamefont
  {Montenegro}}, \bibinfo {author} {\bibfnamefont {W.~S.}\ \bibnamefont
  {Melo}}, \bibinfo {author} {\bibfnamefont {W.~E.}\ \bibnamefont {Meyerhof}},\
  and\ \bibinfo {author} {\bibfnamefont {A.~G.}\ \bibnamefont {de~Pinho}},\
  }\bibfield  {title} {\bibinfo {title} {Separation of the screening and
  antiscreening effects in the electron loss of \textsc{H}e$^+$ on
  \textsc{H}$_2$ and \textsc{H}e},\ }\href
  {https://doi.org/10.1103/PhysRevLett.69.3033} {\bibfield  {journal} {\bibinfo
   {journal} {Phys. Rev. Lett.}\ }\textbf {\bibinfo {volume} {69}},\ \bibinfo
  {pages} {3033} (\bibinfo {year} {1992})}\BibitemShut {NoStop}%
\bibitem [{\citenamefont {Olson}\ and\ \citenamefont {Smith}(1973)}]{ols73a}%
  \BibitemOpen
  \bibfield  {author} {\bibinfo {author} {\bibfnamefont {R.~E.}\ \bibnamefont
  {Olson}}\ and\ \bibinfo {author} {\bibfnamefont {F.~T.}\ \bibnamefont
  {Smith}},\ }\bibfield  {title} {\bibinfo {title} {Effect of long-range forces
  in near-resonant charge transfer: Application to \textsc{H}e$^{+}$ +
  \textsc{K}, \textsc{R}b, and \textsc{C}s},\ }\href
  {https://doi.org/10.1103/PhysRevA.7.1529} {\bibfield  {journal} {\bibinfo
  {journal} {Phys. Rev. A}\ }\textbf {\bibinfo {volume} {7}},\ \bibinfo {pages}
  {1529} (\bibinfo {year} {1973})}\BibitemShut {NoStop}%
\bibitem [{\citenamefont {McCullough}\ \emph {et~al.}(1978)\citenamefont
  {McCullough}, \citenamefont {Goffe},\ and\ \citenamefont {Gilbody}}]{mcc78a}%
  \BibitemOpen
  \bibfield  {author} {\bibinfo {author} {\bibfnamefont {R.~W.}\ \bibnamefont
  {McCullough}}, \bibinfo {author} {\bibfnamefont {T.~V.}\ \bibnamefont
  {Goffe}},\ and\ \bibinfo {author} {\bibfnamefont {H.~B.}\ \bibnamefont
  {Gilbody}},\ }\bibfield  {title} {\bibinfo {title} {Formation of fast
  metastable helium atoms in electron capture by \textsc{H}e$^+$ ions in
  alkali-metal vapours},\ }\href {https://doi.org/10.1088/0022-3700/11/13/017}
  {\bibfield  {journal} {\bibinfo  {journal} {J. Phys. B}\ }\textbf {\bibinfo
  {volume} {11}},\ \bibinfo {pages} {2333} (\bibinfo {year}
  {1978})}\BibitemShut {NoStop}%
\bibitem [{\citenamefont {Lee}\ \emph {et~al.}(1992)\citenamefont {Lee},
  \citenamefont {Zouros}, \citenamefont {Sanders}, \citenamefont {Richard},
  \citenamefont {Anthony}, \citenamefont {Wang},\ and\ \citenamefont
  {McGuire}}]{lee92a}%
  \BibitemOpen
  \bibfield  {author} {\bibinfo {author} {\bibfnamefont {D.~H.}\ \bibnamefont
  {Lee}}, \bibinfo {author} {\bibfnamefont {T.~J.~M.}\ \bibnamefont {Zouros}},
  \bibinfo {author} {\bibfnamefont {J.~M.}\ \bibnamefont {Sanders}}, \bibinfo
  {author} {\bibfnamefont {P.}~\bibnamefont {Richard}}, \bibinfo {author}
  {\bibfnamefont {J.~M.}\ \bibnamefont {Anthony}}, \bibinfo {author}
  {\bibfnamefont {Y.~D.}\ \bibnamefont {Wang}},\ and\ \bibinfo {author}
  {\bibfnamefont {J.~H.}\ \bibnamefont {McGuire}},\ }\bibfield  {title}
  {\bibinfo {title} {$\textsc{K}$-shell ionization of \textsc{O}$^{4+}$ and
  \textsc{C}$^{2+}$ ions in fast collisions with \textsc{H}$_2$ and \textsc{H}e
  gas targets},\ }\href {https://doi.org/10.1103/PhysRevA.46.1374} {\bibfield
  {journal} {\bibinfo  {journal} {Phys. Rev. A}\ }\textbf {\bibinfo {volume}
  {46}},\ \bibinfo {pages} {1374} (\bibinfo {year} {1992})}\BibitemShut
  {NoStop}%
\bibitem [{\citenamefont {Rzadkiewicz}\ \emph {et~al.}(2006)\citenamefont
  {Rzadkiewicz}, \citenamefont {St\"ohlker}, \citenamefont
  {Bana\ifmmode~\acute{s}\else \'{s}\fi{}}, \citenamefont {Beyer},
  \citenamefont {Bosch}, \citenamefont {Brandau}, \citenamefont {Dong},
  \citenamefont {Fritzsche}, \citenamefont {Gojska}, \citenamefont
  {Gumberidze}, \citenamefont {Hagmann}, \citenamefont {Ionescu}, \citenamefont
  {Kozhuharov}, \citenamefont {Nandi}, \citenamefont {Reuschl}, \citenamefont
  {Sierpowski}, \citenamefont {Spillmann}, \citenamefont {Surzhykov},
  \citenamefont {Tashenov}, \citenamefont {Trassinelli},\ and\ \citenamefont
  {Trotsenko}}]{rza06a}%
  \BibitemOpen
  \bibfield  {author} {\bibinfo {author} {\bibfnamefont {J.}~\bibnamefont
  {Rzadkiewicz}}, \bibinfo {author} {\bibfnamefont {T.}~\bibnamefont
  {St\"ohlker}}, \bibinfo {author} {\bibfnamefont {D.}~\bibnamefont
  {Bana\ifmmode~\acute{s}\else \'{s}\fi{}}}, \bibinfo {author} {\bibfnamefont
  {H.~F.}\ \bibnamefont {Beyer}}, \bibinfo {author} {\bibfnamefont
  {F.}~\bibnamefont {Bosch}}, \bibinfo {author} {\bibfnamefont
  {C.}~\bibnamefont {Brandau}}, \bibinfo {author} {\bibfnamefont {C.~Z.}\
  \bibnamefont {Dong}}, \bibinfo {author} {\bibfnamefont {S.}~\bibnamefont
  {Fritzsche}}, \bibinfo {author} {\bibfnamefont {A.}~\bibnamefont {Gojska}},
  \bibinfo {author} {\bibfnamefont {A.}~\bibnamefont {Gumberidze}}, \bibinfo
  {author} {\bibfnamefont {S.}~\bibnamefont {Hagmann}}, \bibinfo {author}
  {\bibfnamefont {D.~C.}\ \bibnamefont {Ionescu}}, \bibinfo {author}
  {\bibfnamefont {C.}~\bibnamefont {Kozhuharov}}, \bibinfo {author}
  {\bibfnamefont {T.}~\bibnamefont {Nandi}}, \bibinfo {author} {\bibfnamefont
  {R.}~\bibnamefont {Reuschl}}, \bibinfo {author} {\bibfnamefont
  {D.}~\bibnamefont {Sierpowski}}, \bibinfo {author} {\bibfnamefont
  {U.}~\bibnamefont {Spillmann}}, \bibinfo {author} {\bibfnamefont
  {A.}~\bibnamefont {Surzhykov}}, \bibinfo {author} {\bibfnamefont
  {S.}~\bibnamefont {Tashenov}}, \bibinfo {author} {\bibfnamefont
  {M.}~\bibnamefont {Trassinelli}},\ and\ \bibinfo {author} {\bibfnamefont
  {S.}~\bibnamefont {Trotsenko}},\ }\bibfield  {title} {\bibinfo {title}
  {Selective population of the $[1s2s]$ $^{1}\textsc{S}_{0}$ and $[1s2s]$
  $^{3}\textsc{S}_{1}$ states of \textsc{H}e-like uranium},\ }\href
  {https://doi.org/10.1103/PhysRevA.74.012511} {\bibfield  {journal} {\bibinfo
  {journal} {Phys. Rev. A}\ }\textbf {\bibinfo {volume} {74}},\ \bibinfo
  {pages} {012511} (\bibinfo {year} {2006})}\BibitemShut {NoStop}%
\bibitem [{\citenamefont {Zhu}\ \emph {et~al.}(2024)\citenamefont {Zhu},
  \citenamefont {Zhang}, \citenamefont {Gao}, \citenamefont {Guo},
  \citenamefont {Xu}, \citenamefont {Zhang}, \citenamefont {Zhao},
  \citenamefont {Lin}, \citenamefont {Zhu}, \citenamefont {Xing}, \citenamefont
  {Cui}, \citenamefont {Passalidis}, \citenamefont {Dubois},\ and\
  \citenamefont {Ma}}]{zhu24a}%
  \BibitemOpen
  \bibfield  {author} {\bibinfo {author} {\bibfnamefont {X.}~\bibnamefont
  {Zhu}}, \bibinfo {author} {\bibfnamefont {S.}~\bibnamefont {Zhang}}, \bibinfo
  {author} {\bibfnamefont {Y.}~\bibnamefont {Gao}}, \bibinfo {author}
  {\bibfnamefont {D.}~\bibnamefont {Guo}}, \bibinfo {author} {\bibfnamefont
  {J.}~\bibnamefont {Xu}}, \bibinfo {author} {\bibfnamefont {R.}~\bibnamefont
  {Zhang}}, \bibinfo {author} {\bibfnamefont {D.}~\bibnamefont {Zhao}},
  \bibinfo {author} {\bibfnamefont {K.}~\bibnamefont {Lin}}, \bibinfo {author}
  {\bibfnamefont {X.}~\bibnamefont {Zhu}}, \bibinfo {author} {\bibfnamefont
  {D.}~\bibnamefont {Xing}}, \bibinfo {author} {\bibfnamefont {S.}~\bibnamefont
  {Cui}}, \bibinfo {author} {\bibfnamefont {S.}~\bibnamefont {Passalidis}},
  \bibinfo {author} {\bibfnamefont {A.}~\bibnamefont {Dubois}},\ and\ \bibinfo
  {author} {\bibfnamefont {X.}~\bibnamefont {Ma}},\ }\bibfield  {title}
  {\bibinfo {title} {Direct evidence of breakdown of spin statistics in
  ion-atom charge exchange collisions},\ }\href
  {https://doi.org/10.1103/PhysRevLett.133.173002} {\bibfield  {journal}
  {\bibinfo  {journal} {Phys. Rev. Lett.}\ }\textbf {\bibinfo {volume} {133}},\
  \bibinfo {pages} {173002} (\bibinfo {year} {2024})}\BibitemShut {NoStop}%
\bibitem [{\citenamefont {Zouros}(1995)}]{zou94b}%
  \BibitemOpen
  \bibfield  {author} {\bibinfo {author} {\bibfnamefont {T.~J.~M.}\
  \bibnamefont {Zouros}},\ }\bibfield  {title} {\bibinfo {title} {Excitation
  and ionization in fast ion-atom collisions due to projectile electron--target
  electron interactions},\ }in\ \href@noop {} {\emph {\bibinfo {booktitle}
  {Applications of Particle and Laser Beams in Materials Technology}}},\ Vol.\
  \bibinfo {volume} {283},\ \bibinfo {editor} {edited by\ \bibinfo {editor}
  {\bibfnamefont {P.}~\bibnamefont {Misailides}}},\ \bibinfo {organization}
  {NATO Advanced Study Institute Series E: Applied Sciences}\ (\bibinfo
  {publisher} {Kluwer Academic},\ \bibinfo {address} {Netherlands},\ \bibinfo
  {year} {1995})\ pp.\ \bibinfo {pages} {37--52}\BibitemShut {NoStop}%
\bibitem [{\citenamefont {Gumberidze}\ \emph {et~al.}(2019)\citenamefont
  {Gumberidze}, \citenamefont {Thorn}, \citenamefont {Surzhykov}, \citenamefont
  {Fontes}, \citenamefont {Najjari}, \citenamefont {Voitkiv}, \citenamefont
  {Fritzsche}, \citenamefont {Bana\ifmmode~\acute{s}\else \'{s}\fi{}},
  \citenamefont {Beyer}, \citenamefont {Chen}, \citenamefont {Grisenti},
  \citenamefont {Hagmann}, \citenamefont {Hess}, \citenamefont {Hillenbrand},
  \citenamefont {Indelicato}, \citenamefont {Kozhuharov}, \citenamefont
  {Lestinsky}, \citenamefont {M\"artin}, \citenamefont {Petridis},
  \citenamefont {Popov}, \citenamefont {Schuch}, \citenamefont {Spillmann},
  \citenamefont {Tashenov}, \citenamefont {Trotsenko}, \citenamefont {Warczak},
  \citenamefont {Weber}, \citenamefont {Wen}, \citenamefont {Winters},
  \citenamefont {Winters}, \citenamefont {Yin},\ and\ \citenamefont
  {St\"ohlker}}]{gum19a}%
  \BibitemOpen
  \bibfield  {author} {\bibinfo {author} {\bibfnamefont {A.}~\bibnamefont
  {Gumberidze}}, \bibinfo {author} {\bibfnamefont {D.~B.}\ \bibnamefont
  {Thorn}}, \bibinfo {author} {\bibfnamefont {A.}~\bibnamefont {Surzhykov}},
  \bibinfo {author} {\bibfnamefont {C.~J.}\ \bibnamefont {Fontes}}, \bibinfo
  {author} {\bibfnamefont {B.}~\bibnamefont {Najjari}}, \bibinfo {author}
  {\bibfnamefont {A.}~\bibnamefont {Voitkiv}}, \bibinfo {author} {\bibfnamefont
  {S.}~\bibnamefont {Fritzsche}}, \bibinfo {author} {\bibfnamefont
  {D.}~\bibnamefont {Bana\ifmmode~\acute{s}\else \'{s}\fi{}}}, \bibinfo
  {author} {\bibfnamefont {H.~F.}\ \bibnamefont {Beyer}}, \bibinfo {author}
  {\bibfnamefont {W.}~\bibnamefont {Chen}}, \bibinfo {author} {\bibfnamefont
  {R.~E.}\ \bibnamefont {Grisenti}}, \bibinfo {author} {\bibfnamefont
  {S.}~\bibnamefont {Hagmann}}, \bibinfo {author} {\bibfnamefont
  {R.}~\bibnamefont {Hess}}, \bibinfo {author} {\bibfnamefont {P.-M.}\
  \bibnamefont {Hillenbrand}}, \bibinfo {author} {\bibfnamefont
  {P.}~\bibnamefont {Indelicato}}, \bibinfo {author} {\bibfnamefont
  {C.}~\bibnamefont {Kozhuharov}}, \bibinfo {author} {\bibfnamefont
  {M.}~\bibnamefont {Lestinsky}}, \bibinfo {author} {\bibfnamefont
  {R.}~\bibnamefont {M\"artin}}, \bibinfo {author} {\bibfnamefont
  {N.}~\bibnamefont {Petridis}}, \bibinfo {author} {\bibfnamefont {R.~V.}\
  \bibnamefont {Popov}}, \bibinfo {author} {\bibfnamefont {R.}~\bibnamefont
  {Schuch}}, \bibinfo {author} {\bibfnamefont {U.}~\bibnamefont {Spillmann}},
  \bibinfo {author} {\bibfnamefont {S.}~\bibnamefont {Tashenov}}, \bibinfo
  {author} {\bibfnamefont {S.}~\bibnamefont {Trotsenko}}, \bibinfo {author}
  {\bibfnamefont {A.}~\bibnamefont {Warczak}}, \bibinfo {author} {\bibfnamefont
  {G.}~\bibnamefont {Weber}}, \bibinfo {author} {\bibfnamefont
  {W.}~\bibnamefont {Wen}}, \bibinfo {author} {\bibfnamefont {D.~F.~A.}\
  \bibnamefont {Winters}}, \bibinfo {author} {\bibfnamefont {N.}~\bibnamefont
  {Winters}}, \bibinfo {author} {\bibfnamefont {Z.}~\bibnamefont {Yin}},\ and\
  \bibinfo {author} {\bibfnamefont {T.}~\bibnamefont {St\"ohlker}},\ }\bibfield
   {title} {\bibinfo {title} {Electron- and proton-impact excitation of
  heliumlike uranium in relativistic collisions},\ }\href
  {https://doi.org/10.1103/PhysRevA.99.032706} {\bibfield  {journal} {\bibinfo
  {journal} {Phys. Rev. A}\ }\textbf {\bibinfo {volume} {99}},\ \bibinfo
  {pages} {032706} (\bibinfo {year} {2019})}\BibitemShut {NoStop}%
\bibitem [{\citenamefont {Andersen}\ \emph
  {et~al.}(1992{\natexlab{b}})\citenamefont {Andersen}, \citenamefont {Pan},
  \citenamefont {Schmidt}, \citenamefont {Pindzola},\ and\ \citenamefont
  {Badnell}}]{and92b}%
  \BibitemOpen
  \bibfield  {author} {\bibinfo {author} {\bibfnamefont {L.~H.}\ \bibnamefont
  {Andersen}}, \bibinfo {author} {\bibfnamefont {G.-Y.}\ \bibnamefont {Pan}},
  \bibinfo {author} {\bibfnamefont {H.~T.}\ \bibnamefont {Schmidt}}, \bibinfo
  {author} {\bibfnamefont {M.~S.}\ \bibnamefont {Pindzola}},\ and\ \bibinfo
  {author} {\bibfnamefont {N.~R.}\ \bibnamefont {Badnell}},\ }\bibfield
  {title} {\bibinfo {title} {State-selective measurements and calculations of
  dielectronic recombination with \textsc{L}i-like \textsc{N}$^{4+}$,
  \textsc{F}$^{6+}$, and \textsc{S}i$^{11+}$ ions},\ }\href
  {https://doi.org/10.1103/PhysRevA.45.6332} {\bibfield  {journal} {\bibinfo
  {journal} {Phys. Rev. A}\ }\textbf {\bibinfo {volume} {45}},\ \bibinfo
  {pages} {6332} (\bibinfo {year} {1992}{\natexlab{b}})}\BibitemShut {NoStop}%
\bibitem [{\citenamefont {Deveney}\ \emph {et~al.}(1993)\citenamefont
  {Deveney}, \citenamefont {Kessel}, \citenamefont {Fuller}, \citenamefont
  {Reaves}, \citenamefont {Bellantone}, \citenamefont {Shafroth},\ and\
  \citenamefont {Jones}}]{dev93b}%
  \BibitemOpen
  \bibfield  {author} {\bibinfo {author} {\bibfnamefont {E.~F.}\ \bibnamefont
  {Deveney}}, \bibinfo {author} {\bibfnamefont {Q.~C.}\ \bibnamefont {Kessel}},
  \bibinfo {author} {\bibfnamefont {R.~J.}\ \bibnamefont {Fuller}}, \bibinfo
  {author} {\bibfnamefont {M.~P.}\ \bibnamefont {Reaves}}, \bibinfo {author}
  {\bibfnamefont {R.~A.}\ \bibnamefont {Bellantone}}, \bibinfo {author}
  {\bibfnamefont {S.~M.}\ \bibnamefont {Shafroth}},\ and\ \bibinfo {author}
  {\bibfnamefont {N.}~\bibnamefont {Jones}},\ }\bibfield  {title} {\bibinfo
  {title} {Projectile \textsc{A}uger electron spectra of \textsc{C}$^{3+}$
  following 12-\textsc{M}e\textsc{V} collisions with \textsc{H}e targets},\
  }\href {https://doi.org/10.1103/PhysRevA.48.2926} {\bibfield  {journal}
  {\bibinfo  {journal} {Phys. Rev. A}\ }\textbf {\bibinfo {volume} {48}},\
  \bibinfo {pages} {2926} (\bibinfo {year} {1993})}\BibitemShut {NoStop}%
\bibitem [{\citenamefont {Zaytsev}\ \emph {et~al.}(2019)\citenamefont
  {Zaytsev}, \citenamefont {Maltsev}, \citenamefont {Tupitsyn},\ and\
  \citenamefont {Shabaev}}]{zay19a}%
  \BibitemOpen
  \bibfield  {author} {\bibinfo {author} {\bibfnamefont {V.~A.}\ \bibnamefont
  {Zaytsev}}, \bibinfo {author} {\bibfnamefont {I.~A.}\ \bibnamefont
  {Maltsev}}, \bibinfo {author} {\bibfnamefont {I.~I.}\ \bibnamefont
  {Tupitsyn}},\ and\ \bibinfo {author} {\bibfnamefont {V.~M.}\ \bibnamefont
  {Shabaev}},\ }\bibfield  {title} {\bibinfo {title} {Complex-scaled
  relativistic configuration-interaction study of the $\textsc{LL}$ resonances
  in heliumlike ions: From boron to argon},\ }\href
  {https://doi.org/10.1103/PhysRevA.100.052504} {\bibfield  {journal} {\bibinfo
   {journal} {Phys. Rev. A}\ }\textbf {\bibinfo {volume} {100}},\ \bibinfo
  {pages} {052504} (\bibinfo {year} {2019})}\BibitemShut {NoStop}%
\bibitem [{\citenamefont {Manai}\ \emph {et~al.}(2022)\citenamefont {Manai},
  \citenamefont {Salhi}, \citenamefont {Nasr},\ and\ \citenamefont
  {Jelassi}}]{man22a}%
  \BibitemOpen
  \bibfield  {author} {\bibinfo {author} {\bibfnamefont {S.}~\bibnamefont
  {Manai}}, \bibinfo {author} {\bibfnamefont {D.~E.}\ \bibnamefont {Salhi}},
  \bibinfo {author} {\bibfnamefont {S.~B.}\ \bibnamefont {Nasr}},\ and\
  \bibinfo {author} {\bibfnamefont {H.}~\bibnamefont {Jelassi}},\ }\bibfield
  {title} {\bibinfo {title} {Relativistic theoretical calculations of singly
  and doubly energy levels, lifetimes, wavelengths, weighted oscillator
  strengths and radiative rates for helium-like ions with $Z=5-9$},\ }\href
  {https://doi.org/https://doi.org/10.1016/j.rinp.2022.105487} {\bibfield
  {journal} {\bibinfo  {journal} {Results in Phys.}\ }\textbf {\bibinfo
  {volume} {37}},\ \bibinfo {pages} {105487} (\bibinfo {year}
  {2022})}\BibitemShut {NoStop}%
\bibitem [{\citenamefont {Kahl}\ and\ \citenamefont {Berengut}(2019)}]{kah19a}%
  \BibitemOpen
  \bibfield  {author} {\bibinfo {author} {\bibfnamefont {E.}~\bibnamefont
  {Kahl}}\ and\ \bibinfo {author} {\bibfnamefont {J.}~\bibnamefont
  {Berengut}},\ }\bibfield  {title} {\bibinfo {title} {\textsc{AMB}i\textsc{T}:
  A programme for high-precision relativistic atomic structure calculations},\
  }\href {https://doi.org/https://doi.org/10.1016/j.cpc.2018.12.014} {\bibfield
   {journal} {\bibinfo  {journal} {Comp. Phys. Com.}\ }\textbf {\bibinfo
  {volume} {238}},\ \bibinfo {pages} {232} (\bibinfo {year}
  {2019})}\BibitemShut {NoStop}%
\bibitem [{\citenamefont {Goryaev}\ \emph {et~al.}(2017)\citenamefont
  {Goryaev}, \citenamefont {Vainshtein},\ and\ \citenamefont {Urnov}}]{gor17b}%
  \BibitemOpen
  \bibfield  {author} {\bibinfo {author} {\bibfnamefont {F.~F.}\ \bibnamefont
  {Goryaev}}, \bibinfo {author} {\bibfnamefont {L.~A.}\ \bibnamefont
  {Vainshtein}},\ and\ \bibinfo {author} {\bibfnamefont {A.~M.}\ \bibnamefont
  {Urnov}},\ }\bibfield  {title} {\bibinfo {title} {Atomic data for
  doubly-excited states $2lnl^\prime$ of \textsc{H}e-like ions and
  $1s2lnl^\prime$ of \textsc{L}i-like ions with $\textsc{Z}=6-36$ and
  $n=2,3$},\ }\href
  {https://doi.org/http://dx.doi.org/10.1016/j.adt.2016.04.002} {\bibfield
  {journal} {\bibinfo  {journal} {At. Data Nucl. Data Tables}\ }\textbf
  {\bibinfo {volume} {113}},\ \bibinfo {pages} {117 } (\bibinfo {year}
  {2017})}\BibitemShut {NoStop}%
\bibitem [{\citenamefont {Doukas}\ \emph {et~al.}(2015)\citenamefont {Doukas},
  \citenamefont {Madesis}, \citenamefont {Dimitriou}, \citenamefont
  {Laoutaris}, \citenamefont {Zouros},\ and\ \citenamefont {Benis}}]{dou15a}%
  \BibitemOpen
  \bibfield  {author} {\bibinfo {author} {\bibfnamefont {S.}~\bibnamefont
  {Doukas}}, \bibinfo {author} {\bibfnamefont {I.}~\bibnamefont {Madesis}},
  \bibinfo {author} {\bibfnamefont {A.}~\bibnamefont {Dimitriou}}, \bibinfo
  {author} {\bibfnamefont {A.}~\bibnamefont {Laoutaris}}, \bibinfo {author}
  {\bibfnamefont {T.~J.~M.}\ \bibnamefont {Zouros}},\ and\ \bibinfo {author}
  {\bibfnamefont {E.~P.}\ \bibnamefont {Benis}},\ }\bibfield  {title} {\bibinfo
  {title} {Determination of the solid angle and response function of a
  hemispherical spectrograph with injection lens for \textsc{A}uger electrons
  emitted from long lived projectile states},\ }\href
  {https://doi.org/10.1063/1.4917274} {\bibfield  {journal} {\bibinfo
  {journal} {Rev. Sci. Instrum.}\ }\textbf {\bibinfo {volume} {86}},\ \bibinfo
  {pages} {043111} (\bibinfo {year} {2015})}\BibitemShut {NoStop}%
\bibitem [{\citenamefont {Bruch}\ \emph {et~al.}(1985)\citenamefont {Bruch},
  \citenamefont {Chung}, \citenamefont {Luken},\ and\ \citenamefont
  {Culberson}}]{bru85a}%
  \BibitemOpen
  \bibfield  {author} {\bibinfo {author} {\bibfnamefont {R.}~\bibnamefont
  {Bruch}}, \bibinfo {author} {\bibfnamefont {K.~T.}\ \bibnamefont {Chung}},
  \bibinfo {author} {\bibfnamefont {W.~L.}\ \bibnamefont {Luken}},\ and\
  \bibinfo {author} {\bibfnamefont {J.~C.}\ \bibnamefont {Culberson}},\
  }\bibfield  {title} {\bibinfo {title} {Recalibration of the \textsc{KLL}
  \textsc{A}uger spectrum of carbon},\ }\href@noop {} {\bibfield  {journal}
  {\bibinfo  {journal} {Phys. Rev. A}\ }\textbf {\bibinfo {volume} {31}},\
  \bibinfo {pages} {310} (\bibinfo {year} {1985})}\BibitemShut {NoStop}%
\bibitem [{\citenamefont {R{\o}dbro}\ \emph {et~al.}(1979)\citenamefont
  {R{\o}dbro}, \citenamefont {Bruch},\ and\ \citenamefont {Bisgaard}}]{rod79a}%
  \BibitemOpen
  \bibfield  {author} {\bibinfo {author} {\bibfnamefont {M.}~\bibnamefont
  {R{\o}dbro}}, \bibinfo {author} {\bibfnamefont {R.}~\bibnamefont {Bruch}},\
  and\ \bibinfo {author} {\bibfnamefont {P.}~\bibnamefont {Bisgaard}},\
  }\bibfield  {title} {\bibinfo {title} {High-resolution projectile
  \textsc{A}uger spectroscopy for \textsc{L}i, \textsc{B}e, \textsc{B} and
  \textsc{C} excited in single gas collisions \textsc{I}. \textsc{L}ine
  energies for prompt decays},\ }\href@noop {} {\bibfield  {journal} {\bibinfo
  {journal} {J. Phys. B}\ }\textbf {\bibinfo {volume} {12}},\ \bibinfo {pages}
  {2413} (\bibinfo {year} {1979})}\BibitemShut {NoStop}%
\bibitem [{\citenamefont {Mack}\ and\ \citenamefont {Niehaus}(1987)}]{mack87c}%
  \BibitemOpen
  \bibfield  {author} {\bibinfo {author} {\bibfnamefont {M.}~\bibnamefont
  {Mack}}\ and\ \bibinfo {author} {\bibfnamefont {A.}~\bibnamefont {Niehaus}},\
  }\bibfield  {title} {\bibinfo {title} {\textsc{K}-shell excited
  \textsc{L}i-like ions: Electron spectroscopy of the doublet term system},\
  }\href@noop {} {\bibfield  {journal} {\bibinfo  {journal} {Nucl. Instrum.
  Methods B}\ }\textbf {\bibinfo {volume} {23}},\ \bibinfo {pages} {291 }
  (\bibinfo {year} {1987})}\BibitemShut {NoStop}%
\bibitem [{\citenamefont {Chung}(1984)}]{chu84b}%
  \BibitemOpen
  \bibfield  {author} {\bibinfo {author} {\bibfnamefont {K.~T.}\ \bibnamefont
  {Chung}},\ }\bibfield  {title} {\bibinfo {title} {Fine structures and
  transition wavelengths for $1s2s2p\,^4\!\textsc{P}$ and
  $1s2p2p\,^4\!\textsc{P}$ of lithiumlike ions},\ }\href@noop {} {\bibfield
  {journal} {\bibinfo  {journal} {Phys. Rev. A}\ }\textbf {\bibinfo {volume}
  {29}},\ \bibinfo {pages} {682} (\bibinfo {year} {1984})}\BibitemShut
  {NoStop}%
\bibitem [{\citenamefont {Mann}(1987)}]{mann87b}%
  \BibitemOpen
  \bibfield  {author} {\bibinfo {author} {\bibfnamefont {R.}~\bibnamefont
  {Mann}},\ }\bibfield  {title} {\bibinfo {title} {High-resolution \textsc{K}
  and \textsc{L} \textsc{A}uger electron spectra induced by single- and
  double-electron capture from \textsc{H}$_{2}$, \textsc{H}e, and \textsc{X}e
  atoms to \textsc{C}$^{4+}$ and \textsc{C}$^{5+}$ ions at 10--100-ke\textsc{V}
  energies},\ }\href {https://doi.org/10.1103/PhysRevA.35.4988} {\bibfield
  {journal} {\bibinfo  {journal} {Phys. Rev. A}\ }\textbf {\bibinfo {volume}
  {35}},\ \bibinfo {pages} {4988} (\bibinfo {year} {1987})}\BibitemShut
  {NoStop}%
\bibitem [{\citenamefont {Kilgus}\ \emph {et~al.}(1993)\citenamefont {Kilgus},
  \citenamefont {Habs}, \citenamefont {Schwalm}, \citenamefont {Wolf},
  \citenamefont {Schuch},\ and\ \citenamefont {Badnell}}]{kil93a}%
  \BibitemOpen
  \bibfield  {author} {\bibinfo {author} {\bibfnamefont {G.}~\bibnamefont
  {Kilgus}}, \bibinfo {author} {\bibfnamefont {D.}~\bibnamefont {Habs}},
  \bibinfo {author} {\bibfnamefont {D.}~\bibnamefont {Schwalm}}, \bibinfo
  {author} {\bibfnamefont {A.}~\bibnamefont {Wolf}}, \bibinfo {author}
  {\bibfnamefont {R.}~\bibnamefont {Schuch}},\ and\ \bibinfo {author}
  {\bibfnamefont {N.~R.}\ \bibnamefont {Badnell}},\ }\bibfield  {title}
  {\bibinfo {title} {Dielectronic recombination from ground state of heliumlike
  carbon ions},\ }\href@noop {} {\bibfield  {journal} {\bibinfo  {journal}
  {Phys. Rev. A}\ }\textbf {\bibinfo {volume} {47}},\ \bibinfo {pages} {4859}
  (\bibinfo {year} {1993})}\BibitemShut {NoStop}%
\bibitem [{\citenamefont {Alnaser}(2002)}]{aln02b}%
  \BibitemOpen
  \bibfield  {author} {\bibinfo {author} {\bibfnamefont {A.~S.}\ \bibnamefont
  {Alnaser}},\ }\emph {\bibinfo {title} {Electron Correlation Leading to
  Double-K-Shell Vacancy Production in Li-Like Ions Colliding with Helium}},\
  \href@noop {} {Ph.D. thesis},\ \bibinfo  {school} {Western Michigan
  University, Kalamazoo} (\bibinfo {year} {2002})\BibitemShut {NoStop}%
\bibitem [{\citenamefont {Kar}\ and\ \citenamefont {Ho}(2009)}]{kar09a}%
  \BibitemOpen
  \bibfield  {author} {\bibinfo {author} {\bibfnamefont {S.}~\bibnamefont
  {Kar}}\ and\ \bibinfo {author} {\bibfnamefont {Y.~K.}\ \bibnamefont {Ho}},\
  }\bibfield  {title} {\bibinfo {title} {Effect of screened \textsc{C}oulomb
  potentials on the resonance states of two-electron highly stripped atoms
  using the stabilization method},\ }\href
  {https://doi.org/10.1088/0953-4075/42/4/044007} {\bibfield  {journal}
  {\bibinfo  {journal} {J. Phys. B}\ }\textbf {\bibinfo {volume} {42}},\
  \bibinfo {pages} {044007} (\bibinfo {year} {2009})}\BibitemShut {NoStop}%
\bibitem [{\citenamefont {Yerokhin}\ \emph {et~al.}(2017)\citenamefont
  {Yerokhin}, \citenamefont {Surzhykov},\ and\ \citenamefont
  {M\"uller}}]{yer17a}%
  \BibitemOpen
  \bibfield  {author} {\bibinfo {author} {\bibfnamefont {V.~A.}\ \bibnamefont
  {Yerokhin}}, \bibinfo {author} {\bibfnamefont {A.}~\bibnamefont
  {Surzhykov}},\ and\ \bibinfo {author} {\bibfnamefont {A.}~\bibnamefont
  {M\"uller}},\ }\bibfield  {title} {\bibinfo {title} {Relativistic
  configuration-interaction calculations of the energy levels of the
  $1{s}^{2}2l$ and $1s2l2{l}^\prime$ states in lithiumlike ions: Carbon through
  chlorine},\ }\href {https://doi.org/10.1103/PhysRevA.96.042505} {\bibfield
  {journal} {\bibinfo  {journal} {Phys. Rev. A}\ }\textbf {\bibinfo {volume}
  {96}},\ \bibinfo {pages} {042505} (\bibinfo {year} {2017})}\BibitemShut
  {NoStop}%
\bibitem [{\citenamefont {Safronova}\ and\ \citenamefont
  {Bruch}(1994)}]{saf94a}%
  \BibitemOpen
  \bibfield  {author} {\bibinfo {author} {\bibfnamefont {U.~I.}\ \bibnamefont
  {Safronova}}\ and\ \bibinfo {author} {\bibfnamefont {R.}~\bibnamefont
  {Bruch}},\ }\bibfield  {title} {\bibinfo {title} {Transition and
  \textsc{A}uger energies of \textsc{L}i-like ions ($1s2lnl^\prime$
  configurations)},\ }\href@noop {} {\bibfield  {journal} {\bibinfo  {journal}
  {Phys. Scr.}\ }\textbf {\bibinfo {volume} {50}},\ \bibinfo {pages} {45}
  (\bibinfo {year} {1994})}\BibitemShut {NoStop}%
\bibitem [{\citenamefont {Gu}(2008)}]{gu08a}%
  \BibitemOpen
  \bibfield  {author} {\bibinfo {author} {\bibfnamefont {M.~F.}\ \bibnamefont
  {Gu}},\ }\bibfield  {title} {\bibinfo {title} {The flexible atomic code},\
  }\href {https://doi.org/10.1139/p07-197} {\bibfield  {journal} {\bibinfo
  {journal} {Can. J. Phys.}\ }\textbf {\bibinfo {volume} {86}},\ \bibinfo
  {pages} {675} (\bibinfo {year} {2008})}\BibitemShut {NoStop}%
\bibitem [{\citenamefont {Ho}(1981)}]{ho81a}%
  \BibitemOpen
  \bibfield  {author} {\bibinfo {author} {\bibfnamefont {Y.~K.}\ \bibnamefont
  {Ho}},\ }\bibfield  {title} {\bibinfo {title} {Complex-coordinate
  calculations for doubly excited states of two-electron atoms},\ }\href
  {https://doi.org/10.1103/PhysRevA.23.2137} {\bibfield  {journal} {\bibinfo
  {journal} {Phys. Rev. A}\ }\textbf {\bibinfo {volume} {23}},\ \bibinfo
  {pages} {2137} (\bibinfo {year} {1981})}\BibitemShut {NoStop}%
\bibitem [{\citenamefont {Mack}(1987)}]{mack87d}%
  \BibitemOpen
  \bibfield  {author} {\bibinfo {author} {\bibfnamefont {E.~M.}\ \bibnamefont
  {Mack}},\ }\emph {\bibinfo {title} {Electron Capture to Autoionizing States
  of Multiply Charged Ions}},\ \href@noop {} {Ph.D. thesis},\ \bibinfo
  {school} {Utrecht University} (\bibinfo {year} {1987})\BibitemShut {NoStop}%
\bibitem [{\citenamefont {Peacock}\ \emph {et~al.}(1973)\citenamefont
  {Peacock}, \citenamefont {Hobby},\ and\ \citenamefont {Galanti}}]{pea73a}%
  \BibitemOpen
  \bibfield  {author} {\bibinfo {author} {\bibfnamefont {N.~J.}\ \bibnamefont
  {Peacock}}, \bibinfo {author} {\bibfnamefont {M.~G.}\ \bibnamefont {Hobby}},\
  and\ \bibinfo {author} {\bibfnamefont {M.}~\bibnamefont {Galanti}},\
  }\bibfield  {title} {\bibinfo {title} {Satellite spectra for helium-like ions
  in laser-produced plasmas},\ }\href
  {http://stacks.iop.org/0022-3700/6/i=10/a=007} {\bibfield  {journal}
  {\bibinfo  {journal} {J. Phys. B}\ }\textbf {\bibinfo {volume} {6}},\
  \bibinfo {pages} {L298} (\bibinfo {year} {1973})}\BibitemShut {NoStop}%
\bibitem [{\citenamefont {Bruch}\ \emph {et~al.}(1987)\citenamefont {Bruch},
  \citenamefont {Stolterfoht}, \citenamefont {Datz}, \citenamefont {Miller},
  \citenamefont {Pepmiller}, \citenamefont {Yamazaki}, \citenamefont {Krause},
  \citenamefont {Swenson}, \citenamefont {Chung},\ and\ \citenamefont
  {Davis}}]{bru87a}%
  \BibitemOpen
  \bibfield  {author} {\bibinfo {author} {\bibfnamefont {R.}~\bibnamefont
  {Bruch}}, \bibinfo {author} {\bibfnamefont {N.}~\bibnamefont {Stolterfoht}},
  \bibinfo {author} {\bibfnamefont {S.}~\bibnamefont {Datz}}, \bibinfo {author}
  {\bibfnamefont {P.~D.}\ \bibnamefont {Miller}}, \bibinfo {author}
  {\bibfnamefont {P.~L.}\ \bibnamefont {Pepmiller}}, \bibinfo {author}
  {\bibfnamefont {Y.}~\bibnamefont {Yamazaki}}, \bibinfo {author}
  {\bibfnamefont {H.~F.}\ \bibnamefont {Krause}}, \bibinfo {author}
  {\bibfnamefont {J.~K.}\ \bibnamefont {Swenson}}, \bibinfo {author}
  {\bibfnamefont {K.~T.}\ \bibnamefont {Chung}},\ and\ \bibinfo {author}
  {\bibfnamefont {B.~F.}\ \bibnamefont {Davis}},\ }\bibfield  {title} {\bibinfo
  {title} {High-resolution \textsc{KLL} \textsc{A}uger spectra of multiply
  ionized oxygen projectiles studied by zero-degree electron spectroscopy},\
  }\href {https://doi.org/10.1103/PhysRevA.35.4114} {\bibfield  {journal}
  {\bibinfo  {journal} {Phys. Rev. A}\ }\textbf {\bibinfo {volume} {35}},\
  \bibinfo {pages} {4114} (\bibinfo {year} {1987})}\BibitemShut {NoStop}%
\bibitem [{\citenamefont {Togawa}\ \emph {et~al.}(2024)\citenamefont {Togawa},
  \citenamefont {K\"uhn}, \citenamefont {Shah}, \citenamefont {Zaytsev},
  \citenamefont {Oreshkina}, \citenamefont {Buck}, \citenamefont {Bernitt},
  \citenamefont {Steinbr\"ugge}, \citenamefont {Seltmann}, \citenamefont
  {Hoesch}, \citenamefont {Keitel}, \citenamefont {Pfeifer}, \citenamefont
  {Leutenegger},\ and\ \citenamefont {L\'opez-Urrutia}}]{tog24a}%
  \BibitemOpen
  \bibfield  {author} {\bibinfo {author} {\bibfnamefont {M.}~\bibnamefont
  {Togawa}}, \bibinfo {author} {\bibfnamefont {S.}~\bibnamefont {K\"uhn}},
  \bibinfo {author} {\bibfnamefont {C.}~\bibnamefont {Shah}}, \bibinfo {author}
  {\bibfnamefont {V.~A.}\ \bibnamefont {Zaytsev}}, \bibinfo {author}
  {\bibfnamefont {N.~S.}\ \bibnamefont {Oreshkina}}, \bibinfo {author}
  {\bibfnamefont {J.}~\bibnamefont {Buck}}, \bibinfo {author} {\bibfnamefont
  {S.}~\bibnamefont {Bernitt}}, \bibinfo {author} {\bibfnamefont
  {R.}~\bibnamefont {Steinbr\"ugge}}, \bibinfo {author} {\bibfnamefont
  {J.}~\bibnamefont {Seltmann}}, \bibinfo {author} {\bibfnamefont
  {M.}~\bibnamefont {Hoesch}}, \bibinfo {author} {\bibfnamefont {C.~H.}\
  \bibnamefont {Keitel}}, \bibinfo {author} {\bibfnamefont {T.}~\bibnamefont
  {Pfeifer}}, \bibinfo {author} {\bibfnamefont {M.~A.}\ \bibnamefont
  {Leutenegger}},\ and\ \bibinfo {author} {\bibfnamefont {J.~R.~C.}\
  \bibnamefont {L\'opez-Urrutia}},\ }\bibfield  {title} {\bibinfo {title}
  {High-accuracy measurements of core-excited transitions in light
  \textsc{L}i-like ions},\ }\href
  {https://doi.org/10.1103/PhysRevA.110.L030802} {\bibfield  {journal}
  {\bibinfo  {journal} {Phys. Rev. A}\ }\textbf {\bibinfo {volume} {110}},\
  \bibinfo {pages} {L030802} (\bibinfo {year} {2024})}\BibitemShut {NoStop}%
\bibitem [{\citenamefont {Ahmed}\ and\ \citenamefont {Lipsky}(1975)}]{ahm75a}%
  \BibitemOpen
  \bibfield  {author} {\bibinfo {author} {\bibfnamefont {M.}~\bibnamefont
  {Ahmed}}\ and\ \bibinfo {author} {\bibfnamefont {L.}~\bibnamefont {Lipsky}},\
  }\bibfield  {title} {\bibinfo {title} {Triply excited states of
  three-electron atomic systems},\ }\href
  {https://doi.org/10.1103/PhysRevA.12.1176} {\bibfield  {journal} {\bibinfo
  {journal} {Phys. Rev. A}\ }\textbf {\bibinfo {volume} {12}},\ \bibinfo
  {pages} {1176} (\bibinfo {year} {1975})}\BibitemShut {NoStop}%
\end{thebibliography}
%
\end{document}